\documentclass[%
 reprint,
 amsmath,amssymb,
 aps,
]{revtex4-2}

\usepackage{graphicx}
\usepackage{dcolumn}
\usepackage{bm}
\usepackage{multirow}
\usepackage{xcolor}
\usepackage{hyperref}

\begin{document}

\title{Radial Uncertainty Product for Spherically Symmetric Potential in Position Space}

\author{Avoy Jana}
\affiliation{%
 Master in Science, Physics\\
 Indian Institute of Technology, Delhi
}%

\date{\today}

\begin{abstract}
\textbf{Abstract} This paper presents a detailed analysis of the radial uncertainty product  for quantum systems with spherically symmetric potentials. Using the principles of quantum mechanics, the study derives the radial uncertainty relation analogous to the Cartesian form and investigates its implications for three key spherically symmetric potentials: the Hydrogen atom, the infinite spherical potential well, and the spherical harmonic oscillator, all within the non-relativistic regime. Employing the Schrödinger equation in spherical coordinates, the paper rigorously evaluates the normalized radial wave functions, expectation values, and uncertainties associated with both position and momentum. Analytical derivations and numerical computations highlight the dependence of the uncertainty product on quantum numbers and system-specific parameters.\\

\end{abstract}

\maketitle

\section{Introduction}
In many papers, the uncertainty relation is shown using the product of the expectation values of the squared position operator, $\langle \hat r^2 \rangle$, and the squared momentum operator, $\langle \hat p^2 \rangle$, or their uncertainties, $\Delta \hat{r}$ and $\Delta \hat p_r$, respectively, is expressed as an inequality in two and three dimensions, (Refs.~\cite{khelashvili2022,bracher2011}) and even in multidimensional spaces,(Refs.~\cite{aljaber2016,Dehesa_2021}) where $p$ represents the total momentum of the system and $r$ denotes the radial position. In this paper, we want to show radial uncertainty product $\Delta \hat{r}\Delta \hat p_r$ for a spherical quantum system as we will see ${[\hat{r},\hat{p}_{r}]=i\hbar}$ is analogous to ${[\hat{x},\hat{p}_{x}]=i\hbar}$. Consequently the radial uncertainty relation becomes $\Delta \hat{r}\Delta \hat p_r\geq \frac{\hbar}{2}$ like $\Delta x\Delta p_{x}\geq \frac{\hbar}{2}$. And both uncertainty relations are examples of general type, $\Delta A \Delta B \geq \sqrt{\frac{1}{4} \langle i[A,B] \rangle^2}$ (Refs.~\cite{Li_2020}) for two observables $A$ and $B$.
We will conduct a comprehensive investigation into the radial uncertainty product [$\Delta \hat{r} \Delta \hat p_r $]  within the context of three distinct spherically symmetric potentials, the Hydrogen atom [$V(r)\propto \frac{1}{r}$], the infinite spherical well (ISW) [$V(r)=0$], and the spherical harmonic oscillator (SHO) [$V(r)\propto r^2$] only for the non-relativistic case.

\section{Formulation for Radial uncertainty product}
The time-independent Schrodinger equation in spherical coordinate ($r,\theta,\phi$) is given by
\begin{equation*}
    - \frac{\hbar^2}{2m} \nabla^2 \psi(r,\theta,\phi)+V(r)\psi(r,\theta,\phi)=E\psi(r,\theta,\phi)
\end{equation*}
Here for our works, $V(r)$ is the spherically symmetric potential and the Laplacian in spherical coordinate $(r,\theta,\phi)$ is given by
\begin{equation*}
    \nabla^2=\frac{1}{r^2}\frac{\partial}{\partial{r}}\left(r^2\frac{\partial}{\partial{r}}\right)+\frac{1}{r^2\sin\theta}\frac{\partial}{\partial{\theta}}\left(\sin\theta\frac{\partial}{\partial{\theta}}\right)+\frac{1}{r^2\sin^2\theta}\frac{\partial{^2}}{\partial{^2\phi}}
\end{equation*}
We know for spherically symmetric potential the solution can be written as 
\begin{equation*}
    \psi(r,\theta,\phi)=R(r)\times Y(\theta,\phi)
\end{equation*}
where $R(r)$ is the radial wave function and $Y(\theta,\phi)$ is Spherical harmonics.
The probability density function for the radial part is given by $P(r)=r^2|R(r)|^2$ from for d-dimensional position space, radial distribution is given by $P(r)=r^{d-1}|R(r)|^2$. (Refs.~\cite{AlJaber_1998}) Now see how to normalize radial wave function. The normalization condition is given by
\begin{equation*}
    \int_{0}^{\infty} P(r) dr = 1 \implies\int_{0}^{\infty} r^2 |R(r)|^2 dr = 1
\end{equation*}
The expectation value of an operator $\hat{A}$ in position space is given by
\begin{equation*}
    \langle\hat{A}\rangle=\int_{0}^{\infty} r^2 R^*(r)\hat{A}R(r) dr
\end{equation*}
So expectation of $\hat{r}$, $\hat{r}^2$ are
\begin{equation}
    \langle \hat{r} \rangle=\int_{0}^{\infty} r^2 R^*(r)\hat{r}R(r) dr=\int_{0}^{\infty} r^3 |R(r)|^2 dr 
\end{equation}
\begin{equation}
    \langle \hat{r}^2 \rangle=\int_{0}^{\infty} r^2 R^*(r^2)\hat{r}^2R(r) dr=\int_{0}^{\infty} r^4 |R(r)|^2 dr
\end{equation}
Now what will be the radial momentum operator?
In classical mechanics, the radial momentum operator is defined by,
\begin{equation*}
    \hat{p}_{rc}=\frac{1}{r}\hat{r}.\hat{p}
\end{equation*}
where $\frac{\hat{r}}{r}$ is unit radial vector and $\hat{p}$ is momentum vector.
In quantum mechanics, this definition becomes ambiguous, since the components of $r$ and $p$ don't commute. So, $\hat{p_{r}}$ is not Hermitian, so it's not observable. Since $\hat{p_{r}}$ should be a Hermitian operator; (Refs.~\cite{Paz_2001}) we need to define newly a symmetric operator given by
\begin{equation*}
    \hat{p_{r}}=\frac{1}{2}\frac{1}{r}(\hat{r}.\hat{p}+\hat{p}.\hat{r})
\end{equation*}
Now $\hat{e_{r}}=\frac{\hat{r}}{r}$ and $\hat{p}$ is momentum, in quantum mechanics for spherical coordinates, it is defined by $\hat{p}=-i\hbar\Vec{\nabla}$.
So 
\begin{equation*}
    \hat{p_{r}}=\frac{1}{2}(\hat{e_{r}}.\hat{p}+\hat{p}.\hat{e_{r}})
\end{equation*}
For an arbitrary wave function $\psi$,
\begin{equation}
    \hat{p_{r}}\psi= \frac{1}{2}(-i\hbar)[\hat{e_{r}}.\Vec{\nabla}\psi+\Vec{\nabla}.(\hat{e_{r}}\psi)] \label{eq:equ3}
\end{equation}
The first part of equation,
\begin{equation*}
    \hat{e_{r}}.\Vec{\nabla}\psi=\hat{e_{r}}.\left(\hat{e_{r}}\frac{\partial{\psi}}{\partial{r}}+\hat{e_{\theta}}\frac{1}{r}\frac{\partial{\psi}}{\partial{\theta}}+\hat{e_{\phi}}\frac{1}{r\sin\theta}\frac{\partial{\psi}}{\partial{\phi}}\right)=\frac{\partial{\psi}}{\partial{r}}
\end{equation*}
The second part of equation,
\begin{equation*}
\begin{alignedat}{2}
    &\Vec{\nabla}.(\hat{e_{r}}\psi)\\
    &=\left(\hat{e_{r}}\frac{\partial}{\partial{r}}+\hat{e_{\theta}}\frac{1}{r}\frac{\partial}{\partial{\theta}}+\hat{e_{\phi}}\frac{1}{r\sin\theta}\frac{\partial}{\partial{\phi}}\right).(\hat{e_{r}}\psi)\\
    &=\hat{e_{r}}.\frac{\partial{\hat{e_{r}}}}{\partial{r}}\psi+\frac{\partial{\psi}}{\partial{r}}+\hat{e_{\theta}}.\frac{\partial{\hat{e_{r}}}}{\partial{\theta}}\frac{1}{r}\psi+\hat{e_{\phi}}.\frac{\partial{\hat{e_{r}}}}{\partial{\phi}}\frac{1}{r\sin\theta}\psi
\end{alignedat}
\end{equation*}
The unit vectors and their derivatives in spherical co-ordinate system given by,
\begin{equation*}
\begin{alignedat}{2}
    &\hat{e_{r}}=(\sin\theta\cos\phi,\sin\theta\sin\phi,\cos\theta)\\ 
    &\hat{e_{\theta}}=(\cos\theta\cos\phi,\cos\theta\sin\phi,-\sin\theta)\\ 
    &\hat{e_{\phi}}=(-\sin\phi,\cos\phi,0)\\
\end{alignedat}
\end{equation*}
\begin{equation*}
\begin{alignedat}{2}
    &\frac{\partial{\hat{e_{r}}}}{\partial{r}}=0\\ 
    &\frac{\partial\hat{e_{r}}}{\partial{\theta}}=(\cos\theta\cos\phi,\cos\theta\sin\phi,-\sin\theta)=\hat{e_{\theta}}\\
    &\frac{\partial\hat{e_{r}}}{\partial{\phi}}=(-\sin\theta\sin\phi,\sin\theta\cos\phi,0)=\sin\theta\,\,\hat{e_{\phi}}\\
\end{alignedat}
\end{equation*}
Finally, 
\begin{equation*}
    \vec\nabla.(\hat{e_{r}}\psi)=\frac{\partial{\psi}}{\partial{r}}+\frac{2}{r}\psi
\end{equation*}
From equation, we get,
\begin{equation*}
\begin{alignedat}{2}
    &\hat{p_{r}}\psi=-\frac{i\hbar}{2}\left(\frac{\partial{\psi}}{\partial{r}}+\frac{\partial{\psi}}{\partial{r}}+\frac{2}{r}\psi\right)=-i\hbar\left(\frac{\partial{\psi}}{\partial{r}}+\frac{1}{r}\psi\right)
\end{alignedat}
\end{equation*}
Hence the radial momentum operator can be written as 
\begin{equation}
    \hat{p_{r}}=-i\hbar\left(\frac{\partial}{\partial{r}}+\frac{1}{r}\right)
\end{equation}
In a compact form it can represented as $\hat{p_{r}}=-i\hbar\frac{1}{r}\frac{\partial}{\partial{r}}[r]$. In a d-dimensional position space, the radial momentum operator is given by $ \hat{p_{r}}=-i\hbar(\frac{\partial}{\partial{r}}+\frac{d-1}{2r})$ (Refs.~\cite{Paz_2001}) and for $d=3$, it's our result.
Lets check whether $\hat{r}$ and $\hat{p_{r}}$ commute or not.
The commutator of  $\hat{r}$ and $\hat{p_{r}}$ is given by $[\hat{r},\hat{p_{r}}]=(\hat{r}\hat{p_{r}}-\hat{p_{r}}\hat{r})$ and considering an arbitrary radial wave function $R(r)$,
\begin{equation*}
[\hat{r},\hat{p_{r}}]R(r)=(-i\hbar)\left[\frac{\partial}{\partial{r}}[rR(r)]-\frac{1}{r}\frac{\partial}{\partial{r}}[r^2R(r)]\right] =i\hbar R(r)
\end{equation*}
\begin{equation}
    [\hat{r},\hat{p_{r}}]=i\hbar
\end{equation}
We see that for ${[\hat{x},\hat{p}_{x}]=i\hbar}$, we calculate uncertainty product $\Delta x\Delta p_{x}$ as mentioned in introduction. Hence as $[\hat{r},\hat{p_{r}}]=i\hbar$, we will calculate radial uncertainty product $\Delta \hat{r}\Delta p_{r}$ for spherical quantum system is of simple form than $\Delta \hat{r}\Delta p$.
It can be shown also that the momentum square operator can be written as 
\begin{equation}
    \hat{p}^2_{r}=-\hbar^2\left(\frac{\partial^2}{\partial{r^2}}+\frac{2}{r}\frac{\partial}{\partial{r}}\right)
\end{equation}
where in d-dimensional position space $\hat{p}^2_{r}$ is given by $\hat{p}^2_{r}=-\hbar^2\left(\frac{\partial^2}{\partial{r^2}}+\frac{d-1}{r}\frac{\partial}{\partial{r}}+\frac{(d-1)(d-3)}{4r^2}\right)$ (Refs.~\cite{Paz_2001}) and for $d=3$, it's our result. Now expectation of $\hat{p_{r}}$ and $\hat{p^2_{r}}$ are
\begin{equation}
    \langle \hat{p_{r}} \rangle=\int_{0}^{\infty} r^2R^*(r)(-i\hbar)\left(\frac{\partial}{\partial{r}}+\frac{1}{r}\right)R(r) dr \label{eq:equ7}
\end{equation}
\begin{equation}
    \langle   \hat p_r^2  \rangle=\int_{0}^{\infty} r^2R^*(r)(-\hbar^2)\left(\frac{\partial^2}{\partial{r^2}}+\frac{2}{r}\frac{\partial}{\partial{r}}\right)R(r) dr \label{eq:equ8}
\end{equation}
But there is a fantastic way to find $\langle\hat{p}^2_{r}\rangle$ if we know about the total energy of the system as total energy is related to total momentum i.e. also to radial momentum.
We see this approach when calculating it later in H-atom, spherical harmonic oscillator, and infinite spherical well problem respectively. By the definition of uncertainty, the uncertainties of radial position and radial momentum are given by 
\begin{equation}
\begin{aligned}
    \Delta \hat{r}=\sqrt{\langle \hat{r}^2 \rangle-\langle \hat{r} \rangle^2}\\
    \Delta \hat p_r=\sqrt{\langle\hat{p}^2_{r}\rangle-\langle\hat{p}_r\rangle^2}
\end{aligned}
\end{equation}
\section{Hydro-genic Atom}
\subsection{Radial wave function}
Here we go for general calculation for one electron system (hydrogenic atoms) like H-atom, He$^+$, Li$^{2+}$ and Be$^{3+}$ respectively.
The time-independent Schrodinger equation for this system is given by
\begin{equation}
   -\frac{\hbar^2}{2\mu}\nabla^2\psi(r,\theta,\phi)-\frac{Zke^2}{r}\psi(r,\theta,\phi)=E\psi(r,\theta,\phi)
\end{equation}
where $Z$, $\mu$ is the atomic number and reduced mass of one electron system, $k=\frac{1}{4\pi\epsilon_{0}}$ and $e$ is the electronic charge. It is important to say that this is a non-relativistic case.\\
Let's see how different the Schrodinger equation is in the sense of reduced mass and atomic number.
\begin{table}[htb]
    \centering
    \begin{ruledtabular}
    \begin{tabular}{lcc}
        \textrm{System} & 
        \textrm{Atomic number ($Z$)} & 
        \textrm{Reduced Mass (\(\mu\))} \\
        \colrule
        H & 1 & \(9.104878 \times 10^{-31}\) kg \\
        He\(^+\) & 2 & \(9.108597 \times 10^{-31}\) kg \\
        Li\(^{2+}\) & 3 & \(9.109010 \times 10^{-31}\) kg \\
        Be\(^{3+}\) & 4 & \(9.109272 \times 10^{-31}\) kg \\
    \end{tabular}
    \end{ruledtabular}
    \caption{Reduced Mass (\(\mu\)) and atomic number ($Z$) for one-electron systems}
\end{table}
The Laplacian can be be written as $\nabla^2=\frac{1}{r^2}\frac{d}{dr}(r^2\frac{d}{dr})-\frac{\hat{L}^2}{r^2\hbar^2}$ and the angular momentum operator $\hat{L}^2$ operates on a state gives eigenvalue $\ell(\ell+1)\hbar^2$ i.e. $\hat{L}^2\psi=\ell(\ell+1)\hbar^2\psi$, then the well-known radial Schrodinger equation (Refs.~\cite{Griffiths_2018}) will be 
\begin{equation}
\begin{alignedat}{2}
    &\left[-\frac{\hbar^2}{2\mu} \left(\frac{1}{r^2}\frac{d}{dr}\left(r^2\frac{d}{dr}\right)\right) -\frac{Zke^2}{r} + \frac{\ell(\ell + 1)\hbar^2}{2\mu r^2}\right]R(r)\\
    &= E R(r)\\
\end{alignedat}
\end{equation}
where $\ell$ is the azimuthal quantum number (angular momentum quantum number). And it comes from effective potential defined by $V_{eff}(r)=V(r)+\frac{\ell(\ell+1){\hbar}^2}{2\mu r^2}$, here $\ell$ can take values $0,1,2,3,....,n-1$ where $n$ is the principal quantum number.
The solution (Refs.~\cite{AlJaber_1998}) of it is given by
\begin{equation*}
\begin{alignedat}{2}
    &R_{n_r\ell}(r)=N_{n_r\ell}e^{-{\frac{Zr}{(n_r+\ell+1)a_{0}}}}\\
    &\left(\frac{2Z}{(n_r+\ell+1)a_{0}}r\right)^\ell L^{2\ell+1}_{n_r}\left({\frac{2Z}{(n_r+\ell+1)a_{0}}r}\right)\\
\end{alignedat}
\end{equation*}
And we know for the Hydrogen atom, $n=n_r+\ell+1$ (Refs.~\cite{Bransden_Joachain_1989}) where $n_r$, $n$, $\ell$ are radial, principal, and azimuthal quantum numbers. The radial quantum number $n_r$ is specifically used for the radial solution of the Schrodinger equation and the number of nodes in the radial wave function. So, the revised radial wave function (Refs.~\cite{Supriadi_2019}) can be written as,
\begin{equation}
    R_{n\ell}(r)=N_{n\ell}e^{-{\frac{Zr}{na_{0}}}}\left(\frac{2Z}{na_{0}}r\right)^\ell L^{2\ell+1}_{n-\ell-1}\left({\frac{2Z}{na_{0}}r}\right)
\end{equation}
where $N_{n\ell}$ is the normalization constant, it can be calculated by normalizing condition and $a_{0}$ is Bohr radius.
And $L^{2\ell+1}_{n-\ell-1}$ is associated Laguerre function (Generalized Laguerre function),(Refs.~\cite{Abramowitz_Stegun_1972}) is defined as 
\begin{equation*}
    L^a_b(x)=\frac{e^x x^{-a}}{b!}\frac{d^b}{dx^b}(e^{-x} x^{b+a})
\end{equation*}
Some examples are shown in table II.
\begin{table}[htb]
    \centering
    \begin{ruledtabular}
    \begin{tabular}{ccc}
        \textrm{$b$} & \textrm{$a$} & \textrm{Associated Laguerre Function $L_{b}^{a}(x)$} \\
        \colrule
        $k$ & $0$ & $1$ \\
        $k$ & $1$ & $-x + k + 1$ \\
        $k$ & $2$ & $\frac{1}{2} \left[ x^2 - 2(k+2)x + (k+1)(k+2) \right]$ \\
    \end{tabular}
    \end{ruledtabular}
    \caption{Some Associated Laguerre Functions}
\end{table}
The three properties of the associated Laguerre function, those have an important role in the context of calculation are\\
1. Orthonormal property 
\begin{equation}
    \int_{0}^{\infty}z^a e^{-z}L^a_b(z)L^a_c(z)dz=\frac{\Gamma{(a+b+1)}}{\Gamma{(b+1)}}\delta_{bc}
\end{equation}
Many authors like Griffiths (Refs.~\cite{Griffiths_2018}) uses the orthogonality property as $\int_{0}^{\infty}z^a e^{-z}L^a_b(z)L^a_c(z)dz=\frac{\Gamma{(a+b+1)}}{[\Gamma{(b+1)}]^3}\delta_{bc}$, but this will not affect in our calculation, we see.\\
2. Recursive property
\begin{equation}
\begin{alignedat}{2}
    &zL^a_b(z)=(a+2b+1)L^a_b(z)\\
    &-\frac{b+1}{a+b+1}L^a_{b+1}(z)-(a+b)^2L^a_{b-1}(z)\\
\end{alignedat}
\end{equation} 
3. Derivative property
\begin{equation}
    \frac{d}{dz}L^a_b(z)=L^{a+1}_b(z)=\frac{1}{z}[bL^a_b(z)-(b+a)L^a_{b-1}(z)]
\end{equation}
Let's introduce a dimensionless parameter $\rho$ as 
\begin{equation*}
    \rho=\frac{2Z}{na_{0}}r
\end{equation*}
For that, the radial wave function form looks like
\begin{equation}
    R_{n\ell}(\rho)=N_{n\ell}e^{-\frac{\rho}{2}}{\rho}^{\ell}L^{2\ell+1}_{n-\ell-1}(\rho)
\end{equation}
Let make normalize.
\begin{equation}
    P(r)=r^2|R(r)|^2\implies P(\rho)=\left({\frac{na_{0}}{2Z}}\right)^2{\rho}^2|R_{n\ell}(\rho)|^2  
\end{equation}
As $r=\frac{na_{0}}{2Z}\rho$, then $dr=\frac{na_{0}}{2Z}d\rho$ and 
\begin{equation}
\begin{alignedat}{2}
    &\int_0^{\infty}P(r)dr\\ 
    &={\left(\frac{na_{0}}{2Z}\right)}^3 \int_0^{\infty}{\rho}^2|R_{n\ell}(\rho)|^2d\rho\\
    &={\left(\frac{na_{0}}{2Z}\right)}^3|N_{n\ell}|^2 \int_0^{\infty}{\rho}e^{-\rho}{\rho}^{2\ell+1}[L^{2\ell+1}_{n-\ell-1}(\rho)]^2d\rho\\
    &={\left(\frac{na_{0}}{2Z}\right)}^3|N_{n\ell}|^2 I_1 \label{eq:equ34}
\end{alignedat}
\end{equation}
Let naming $2\ell+1=a$, $n-\ell-1=b$ and $\rho=z$, the integral $I_1$ becomes,
\begin{equation*}
\begin{alignedat}{2}
    &I_1\\
    &=\int_0^{\infty}z.e^{-z}z^a[L^a_b(z)]^2dz\\
    &=\int_0^{\infty}z^ae^{-z}\left[(a+2b+1)L^a_b(z)\right.\\
    &\left.-\frac{b+1}{a+b+1}L^a_{b+1}(z)-(a+b)^2L^a_{b-1}(z)\right]L^a_b(z)dz\\  
    &=(a+2b+1)\int_0^{\infty}z^ae^{-z}[L^a_b(z)]^2dz\text{ (using equ. 13 )}\\
    &=(a+2b+1)\frac{\Gamma{(a+b+1)}}{\Gamma{(b+1)}}\\
\end{alignedat}
\end{equation*}
Now, \begin{center}
    $a+2b+1=2n$,
    $a+b+1=n+\ell+1$,
    $b+1=n-\ell$
\end{center}
Finally,
\begin{equation*}
I_1=(2n)\frac{\Gamma{(n+\ell+1)}}{\Gamma(n-\ell)}=(2n)\frac{(n+\ell)!}{(n-\ell-1)!}  
\end{equation*}
From equ (18),
\begin{equation*}
    {\left(\frac{na_{0}}{2Z}\right)}^3|N_{n\ell}|^2 I_1={\left(\frac{na_{0}}{2Z}\right)}^3|N_{n\ell}|^2(2n)\frac{(n+\ell)!}{(n-\ell-1)!}=1
\end{equation*}\\
The normalization constant is given by
\begin{equation}
N_{n\ell}=\sqrt{\left(\frac{2Z}{na_{0}}\right)^3\frac{(n-\ell-1)!}{2n(n+\ell)!}}
\end{equation}

\subsection{Uncertainty in Radial position}
As per equ (1) ,
\begin{equation}
\begin{alignedat}{2}
    &\langle \hat{r} \rangle\\
    &=\left(\frac{na_{0}}{2Z}\right)^4\int_0^{\infty}{\rho}^3|R_{n\ell}(\rho)|^2d\rho\\ 
    &=\left(\frac{na_{0}}{2Z}\right)^4|N_{n\ell}|^2 \int_0^{\infty}{\rho}^2e^{-\rho}{\rho}^{2\ell+1}[L_{n-\ell-1}^{2\ell+1}(\rho)]^2d\rho\\
    &=\left(\frac{na_{0}}{2Z}\right)^4|N_{n\ell}|^2 I_2 \label{eq:equ38}
\end{alignedat}
\end{equation}\\
Using previous nomenclature, the integral $I_2$ becomes,
\begin{equation*}
I_2=\int_0^{\infty}z^2.e^{-z}z^a[L^a_b(z)]^2dz\\  
\end{equation*}
Using the recursive property of equ (14) ,\\
\begin{equation*}
\begin{alignedat}{2}
    &zL^a_b(z)=(a+2b+1)L^a_b(z)-\frac{b+1}{a+b+1}L^a_{b+1}(z)\\
    &-(a+b)^2L^a_{b-1}(z)\\
\end{alignedat}
\end{equation*}
\begin{equation}
\begin{alignedat}{2}
    &z^2L^a_b(z)=\left[(a+2b+1)^2+(b+1)(a+b+1)+b(a+b)\right]\\
    &L_b^a(z)+\text{other terms contain } L_c^a(z) \text{ where } c\neq b \\
\end{alignedat}
\end{equation}
Finally, 
\begin{equation*}
\begin{alignedat}{2}
    &I_2=\int_{0}^{\infty}[(a+2b+1)^2+(b+1)(a+b+1)\\
    &+b(a+b)]z^a e^{-z}[L^a_b(z)]^2dz\text{ (using equ. (13)) }\\
    &=[(a+2b+1)^2+(b+1)(a+b+1)+b(a+b)]\\
    &\frac{\Gamma{(a+b+1)}}{\Gamma{(b+1)}}\\  
\end{alignedat}
\end{equation*}
Now \begin{center}
    $(a+2b+1)^2=(2n)^2=4n^2$
\end{center}
\begin{center}
    $(b+1)(a+b+1)=(n-\ell)(n+\ell+1)$
\end{center}
\begin{center}
    $b(a+b)=(n-\ell-1)(n+\ell)$
\end{center}
Hence
\begin{equation*}
    (a+2b+1)^2+(b+1)(a+b+1)+b(a+b)=2[3n^2-\ell(\ell+1)]
\end{equation*}
\begin{equation*}
    I_2=2[3n^2-\ell(\ell+1)]\frac{(n+\ell)!}{(n-\ell-1)!}
\end{equation*}
From equ. (20),
\begin{equation}
    \langle \hat{r} \rangle= \frac{1}{2}\frac{a_{0}}{Z}[3n^2-\ell(\ell+1)]
\end{equation}
Now, our aim is to evaluate $\langle \hat{r}^2 \rangle$, as from equ. (2), 
\begin{equation}
\begin{alignedat}{2}
    &\langle \hat{r}^2 \rangle\\
    &=\left(\frac{na_{0}}{2Z}\right)^5\int_0^{\infty}{\rho}^4|R_{n\ell}(\rho)|^2d\rho\\ 
    &=\left(\frac{na_{0}}{2Z}\right)^5|N_{n\ell}|^2 \int_0^{\infty}{\rho}^3e^{-\rho}{\rho}^{2\ell+1}[L_{n-\ell-1}^{2\ell+1}(\rho)]^2d\rho\\
    &=\left(\frac{na_{0}}{2Z}\right)^5|N_{n\ell}|^2 I_3  \label{eq:equ21}
\end{alignedat}
\end{equation}
Using previous nomenclature, the integral $I_3$ becomes,
\begin{equation*}
    I_3=\int_0^{\infty}z^3.e^{-z}z^a[L^a_b(z)]^2dz
\end{equation*}\\
From equ. (21), we get,
\begin{equation*}
\begin{alignedat}{2}
    &z^2L^a_b(z)\\ 
    &=[(a+2b+1)^2+(b+1)(a+b+1)+b(a+b)]L_b^a(z)\\
    &+\left[-\frac{(a+2b+1)(b+1)}{a+b+1}-\frac{(b+1)(a+2b+3)}{(a+b+1)}\right]L_{b+1}^a(z)\\
    &+[-(a+2b+1)(a+b)^2-(a+2b-1)(a+b)^2]L_{b-1}^a(z)\\
    &+\frac{(b+1)(b+2)}{(a+b+1)(a+b+2)}L_{b+2}^a(z)+(a+b)^2(a+b-1)^2L_{b-2}^a(z)\\
    &=U(a,b)L_b^a(z)+V(a,b)L_{b+1}^a(z)+W(a,b)L_{b-1}^a(z)\\
    &+X(a,b)L_{b+2}^a(z)+Y(a,b)L_{b-2}^a(z)\\
\end{alignedat}
\end{equation*}
where $U,V,W,X,Y$ are shorthand symbols.
\begin{widetext}
\begin{equation*}
\begin{alignedat}{2}
    &z^3L^a_b(z)\\
    &=z\left[U(a,b)L_b^a(z)+V(a,b)L_{b+1}^a(z)+W(a,b)L_{b-1}^a(z)+X(a,b)L_{b+2}^a(z)+Y(a,b)L_{b-2}^a(z)\right]\\  
    &=U(a,b)\left[(a+2b+1)L_b^a-\frac{(b+1)}{(a+b+1}L_{b+1}^a(z)-(a+b)^2L_{b-1}^a(z)\right]\\
    &  +V(a,b)\left[(a+2b+3)L_{b+1}^a(z)-\frac{b+2}{a+b+2}L_{b+2}^a(z)-(a+b+1)^2L_b^a(z)\right]\\
    &  +W(a,b)\left[(a+2b-1)L_{b-1}^a(z)-\frac{b}{(a+b)}L_b^a(z)-(a+b-1)^2L_{b-2}^a(z)\right]\\
    &  +X(a,b)\left[(a+2b+5)L_{b+2}^a(z)+\frac{b+3}{a+b+3}L_{b+3}^a(z)-(a+b+2)^2L_{b+1}^a(z)\right]\\
    &  +Y(a,b)\left[(a+2b-3)L_{b-2}^a(z)-\frac{b-1}{a+b-1}L_{b-1}^a(z)-(a+b-2)^2L_{b-3}^a(z)\right]
\end{alignedat}
\end{equation*}
\end{widetext}
\begin{equation*}
\begin{alignedat}{2}
    &z^3L_b^a(z)\\
    &=U(a,b)[(a+2b+1)]+V(a,b)[-(a+b+1)^2]\\
    &+W(a,b)\left[-\frac{b}{a+b}\right]L_b^a(z) \\
    &+\text{other terms not contain $L_c^a(z)$ where $c \neq b$}\\
    &=[U'(a,b)+V'(a,b)+W'(a,b)]L_b^a(z)\\
    &+ \text{other terms not contain $L_c^a(z)$ where $c \neq b$}\\
\end{alignedat}
\end{equation*}
\begin{equation*}
    U(a,b)=(a+2b+1)^2+(b+1)(a+b+1)+b(a+b)
\end{equation*}
\begin{equation*}
    U'(a,b)=U(a,b)(a+2b+1)=2[6n^3-2n{\ell}^2-2n\ell]
\end{equation*}
\begin{equation*}
    V(a,b)=-\frac{(a+2b+1)(b+1)}{(a+b+1}-\frac{(b+1)(a+2b+3)}{(a+b+1)}
\end{equation*}
\begin{equation*}
\begin{alignedat}{2}
    &V'(a,b)=V(a,b)[-(a+b+1)^2]\\
    &=2(2n^3+3n^2+n-{\ell}^2-\ell-2n{\ell}^2-2n\ell)\\
\end{alignedat}
\end{equation*}
\begin{equation*}
    W(a,b)=-(a+2b+1)(a+b)^2-(a+2b-1)(a+b)^2
\end{equation*}
\begin{equation*}
\begin{alignedat}{2}
    &W'(a,b)=W(a,b)\left[-\frac{b}{a+b}\right]\\
    &=2(2n^3-3n^2+n+{\ell}^2+\ell-2n{\ell}^2-2n\ell)\\
\end{alignedat}
\end{equation*}
\begin{equation*}
   U'(a,b)+V'(a,b)+W'(a,b)=2[10n^3+2n-6n{\ell}^2-6n\ell]  
\end{equation*}
Finally, $z^3L_b^a(z)=2[10n^3+2n-6n{\ell}^2-6n\ell]L_b^a(z)+ \text{other terms contain $L_c^a(z)$ where $c\neq b$}$.\\
\begin{equation*}
\begin{alignedat}{2}
    &I_3=\int_0^{\infty}\left[[U'(a,b)+V'(a,b)+W'(a,b)]L_b^a(z)\right.\\
    &\left.+ \text{other terms contain $L_c^a(z)$ where $c\neq b$}\right]e^{-z}z^a[L^a_b]dz\\  
    &=[U'(a,b)+V'(a,b)+W'(a,b)]\frac{\Gamma{(a+b+1)}}{\Gamma{(b+1)}}\\
    &=2[10n^3+2n-6n{\ell}^2-6n\ell].\frac{(n+\ell)!}{(n-\ell-1)!}
\end{alignedat}
\end{equation*}
From equ. (23), 
\begin{equation}
   \langle \hat{r}^2 \rangle=\frac{1}{2}.\frac{n^2a_0^2}{Z^2}[5n^2-3\ell(\ell+1)+1]  
\end{equation}
Previously we have in equ. (22),
\begin{equation*}
    \langle \hat{r} \rangle= \frac{1}{2}\frac{a_{0}}{Z}[3n^2-\ell(\ell+1)]
\end{equation*}
And the uncertainty in radial position will be,
\begin{equation}
    \Delta \hat{r} =\frac{1}{2}\frac{a_0}{Z}\sqrt{n^2(n^2+2)-[\ell(\ell+1)]^2}
\end{equation}

\subsection{Uncertainty in Radial momentum}
We see that the radial momentum operator is given by,
\begin{equation*}
    \hat{p}_{r}=-i\hbar\frac{1}{r}\frac{\partial}{\partial{r}}[r]=-i\hbar\left(\frac{\partial}{\partial{r}}+\frac{1}{r}\right)
\end{equation*}
As per equ. (7),\\
\begin{equation}
\begin{alignedat}{2}
    &\langle \hat{p_{r}} \rangle\\
    &=\int_{0}^{\infty} r^2R_{n\ell}^*(r)(-i\hbar)\left(\frac{\partial}{\partial{r}}+\frac{1}{r}\right)R_{n\ell}(r) dr\\
    &=-i\hbar\left[\int_0^{\infty}r^2R_{n\ell}\frac{\partial{R_{n\ell}}}{\partial{r}}dr+\int_0^{\infty}r^2R_{n\ell}.\frac{1}{r}R_{n\ell}\,dr\right]\\
    &=-i\hbar[I_4+I_5] \label{eq:equ25}
\end{alignedat}
\end{equation}\\
First integral of equ. (24) becomes, 
\begin{equation*}
    I_4=\int_0^{\infty}r^2R_{n\ell}\,dR_{n\ell}=\int_0^{\infty}\left(\frac{na_0}{2Z}\right)^2{\rho}^2R_{n\ell}(\rho)\,dR_{n\ell}(\rho)
\end{equation*}
From equ. (16),
$R_{n\ell}(\rho)=N_{n\ell}e^{-\frac{\rho}{2}}{\rho}^{\ell}L^{2\ell+1}_{n-\ell-1}(\rho)$\\
Then 
\begin{equation*}
\begin{alignedat}{2}
    &dR_{n\ell}(\rho)=N_{n\ell}\left[-\frac{1}{2}e^{-\frac{\rho}{2}}{\rho}^{\ell}L_{n-\ell-1}^{2\ell+1}(\rho\right.\\
    &\left.+e^{-\frac{\rho}{2}}{\ell}{\rho}^{\ell-1}L_{n-\ell-1}^{2\ell+1}(\rho)+e^{-\frac{\rho}{2}}{\rho}^{\ell}\frac{dL_{n-\ell-1}^{2\ell+1}(\rho)}{d\rho}\right]d\rho\\
\end{alignedat}
\end{equation*}
Using nomenclature,
\begin{equation}
\begin{alignedat}{2}
    &dR_{n\ell}(\rho)=N_{n\ell}\left[-\frac{1}{2}e^{-\frac{\rho}{2}}{\rho}^{\ell}L_b^a(\rho)\right.\\
    &\left.+e^{-\frac{\rho}{2}}\ell{\rho}^{\ell-1}L_b^a(\rho)+e^{-\frac{\rho}{2}}{\rho}^{\ell}\frac{dL_b^a(\rho)}{d\rho}\right]d\rho\\
    &=N_{n\ell}\left[-\frac{1}{2}e^{-\frac{\rho}{2}}{\rho}^{\ell}L_b^a(\rho)+(b+\ell)e^{-\frac{\rho}{2}}{\rho}^{\ell-1}L_b^a(\rho)\right.\\
    &\left.-(b+a)e^{-\frac{\rho}{2}}{\rho}^{\ell-1}L_{b+1}^a(\rho)\right]d\rho
\end{alignedat}
\end{equation}
And as a result, $I_4$ becomes,
\begin{equation}
\begin{alignedat}{2}
    &I_4=\left(\frac{na_0}{2Z}\right)^2|N_{n\ell}|^2\left[-\frac{1}{2}\int_0^{\infty}\rho.{\rho}^{a}e^{-\rho}[L_b^a(\rho)]^2d\rho\right.\\
    &+(b+\ell)\int_0^{\infty}{\rho}^{a}e^{-\rho}[L_b^a(\rho)]^2d\rho\\
    &\left.-(b+a)\int_0^{\infty}{\rho}^{a}e^{-\rho}L_b^a(\rho)L_{b-1}^a(\rho)d\rho\right]\\
    &=\left(\frac{na_0}{2Z}\right)^2|N_{n\ell}|^2\left[-\frac{1}{2}I_6+(b+\ell)I_7\right]\\
\end{alignedat}
\end{equation}
Now, 
\begin{equation}
\begin{alignedat}{2}
    &I_6=\int_0^{\infty}\rho.{\rho}^{a}e^{-\rho}[L_b^a(\rho)]^2d\rho\\
.   &=\int_0^{\infty} 
    \left[(a+2b+1)L^a_b(z)-\frac{b+1}{a+b+1}L^a_{b+1}(z)\right.\\
    &\left.-(a+b)^2L^a_{b-1}(z)\right]{\rho}^{a}e^{-\rho}L_b^a(\rho)d\rho\\  
    &=(a+2b+1)\int_0^{\infty}{\rho}^{a}e^{-\rho}[L_b^a(\rho)]^2d\rho=2nI_7\\
\end{alignedat}
\end{equation}
So, $I_4=-(\frac{na_0}{2Z})^2|N_{n\ell}|^2 I_7$ and from equ. (13), $I_7=\frac{(n+\ell)!}{(n-\ell-1)!}$ gives $I_4=-\frac{Z}{n^2a_0}$.
First integral of equ. (24) becomes, 
\begin{equation*}
\begin{alignedat}{2}
    &I_5=\int_0^{\infty}r|R_{n\ell}|^2dr\\
    &=\left(\frac{na_0}{2Z}\right)^2|N_{n\ell}|^2\int_0^{\infty}e^{-\rho}{\rho}^{2\ell+1}[L_{n-\ell-1}^{2\ell+1}(\rho)]^2d\rho\\
    &=\frac{Z}{n^2a_0}\\
\end{alignedat}
\end{equation*}
There is also a fascinating way to find integral $I_5=\int_0^{\infty}r|R_{n\ell}|^2dr=\langle \frac{1}{\hat{r}} \rangle$ by energy calculation. Virial theorem states that for $V(r)\propto r^n$, $\langle T  \rangle=\frac{n}{2}\langle V  \rangle$. For Coulomb potential, $V(r)\propto \frac{1}{r}$ and $\langle T  \rangle=-\frac{1}{2}\langle V  \rangle$ that states that $\langle E  \rangle=\langle T  \rangle+\langle V  \rangle=\frac{1}{2}\langle V  \rangle$. For Hydrogen or Hydrogen like atom, $V=-\frac{Zke^2}{r}$ and $E=-\frac{Z^2{\hbar}^2}{2\mu n^2a_0^2}$ where $a_0=\frac{{\hbar}^2}{\mu ke^2}$. After a little bit simplification, we get, $\langle \frac{1}{\hat{r}} \rangle=\frac{Z}{n^2a_0}$. It's clear from equ. (26), 
\begin{equation}
    \langle   \hat p_r \rangle=0
\end{equation}
Now our aim is $\langle   \hat p_r^2  \rangle$, for that we can use $\langle   \hat p_r^2  \rangle=\int_{0}^{\infty} r^2R^*(r)(-\hbar^2)(\frac{\partial^2}{\partial{r^2}}+\frac{2}{r}\frac{\partial}{\partial{r}})R(r) dr$, in a similar way as we calculate $\langle \hat{p_{r}} \rangle$. But here to avoid similar long calculation, we do it by energy method. Total energy of H-atom or Hydrogen like atom, $E_{n\ell m}=E_n=-\frac{Z^2{\hbar}^2}{2\mu n^2a_0^2}$. Total momentum is defined as $p_{n\ell m}=\sqrt{p_r^2+p_{\ell}^2}$ where $p_r$ is radial momentum and $p_\ell$ is orbital momentum. We know that, orbital angular momentum is defined as $L=p_\ell r=\sqrt{\ell(\ell+1)}\hbar$, from this we easily can say, $p_{\ell}=\frac{L}{r}=\frac{\sqrt{\ell(\ell+1)}}{r}\hbar$. The momentum-energy relation $\langle T=-E_{n\ell m}=\frac{p_{n\ell m}^2}{2\mu}\rangle$ gives $\langle\frac{Z^2{\hbar}^2}{n^2a_0^2}=p_r^2+\frac{\ell(\ell+1)}{r^2}{\hbar}^2\rangle$. Hence, $\langle   \hat{p}_r^2  \rangle=\frac{Z^2{\hbar}^2}{n^2a_0^2}-\ell(\ell+1){\hbar}^2\langle   \frac{1}{\hat{r}^2}  \rangle$, now we have to calculate $\langle   \frac{1}{\hat{r}^2}  \rangle$. Now, 
\begin{equation}
\begin{alignedat}{2}
    &\langle   \frac{1}{\hat{r}^2}  \rangle=\int_0^{\infty}|R_{n\ell}|^2dr\\
    &=\frac{na_0}{2Z}|N_{n\ell}|^2\int_0^{\infty}\frac{1}{\rho}e^{-\rho}{\rho}^{2\ell+1}[L_{n-\ell-1}^{2\ell+1}(\rho)]^2d\rho\\
\end{alignedat}
\end{equation}
It's clear that it can't be done this integral using orthogonality condition and recursive formula as we applied before for calculation. For this type of situation, the radial Schrodinger equation can be used. (Refs.~\cite{Zettili_2013}) Let $f(r)=rR(r)$, $\frac{d^2}{dr^2}(\frac{f}{r})+\frac{2}{r}\frac{d}{dr}(\frac{f}{r})=\frac{1}{r}f''$. The radial Schrodinger equation (equ. (11)) for Hydro-genic atom will be
\begin{equation}
    -\frac{{\hbar}^2}{2\mu}\frac{d^2f(r)}{dr^2}+\left[\frac{\ell(\ell+1){\hbar}^2}{2\mu r^2}-\frac{Zke^2}{r}\right]f(r)=Ef(r) 
\end{equation}
Arranging the equation as
\begin{equation}
    \frac{f''}{f}=\frac{\ell(\ell+1)}{r^2}-\frac{2\mu Zke^2}{{\hbar}^2r}+\frac{Z^2}{n^2a_0^2} \text{ substituting $E_n=-\frac{ Z^2{\hbar}^2}{2\mu n^2a_0^2}$}
\end{equation}
Now differentiating the equation with respect to $\ell$,
\begin{equation}
    \frac{\partial}{\partial{\ell}}\left[\frac{f''}{f}\right]=\frac{2(\ell+1)}{r^2}-\frac{2Z^2}{n^3a_0^2}
\end{equation}
For $n=n(\ell)=n_r+\ell+1$, $\frac{dn}{d\ell}=1$. Now multiplying both sides of equ. (34) by $f^2(r)$,
\begin{equation}
    f^2(r)\frac{\partial}{\partial{\ell}}\left[\frac{f''(r)}{f(r)}\right]=(2\ell+1)f^2(r)\frac{1}{r^2}-\frac{2Z^2}{n^3a_0^2}f^2(r)
\end{equation}
Now integrating the equ. (35) with respect to $r$ from $0$ to $\infty$,
\begin{equation}
\begin{alignedat}{2}
    &\int_0^{\infty}\frac{\partial}{\partial{\ell}}\left[\frac{f''(r)}{f(r)}\right]dr=(2\ell+1)\int_0^{\infty}f^2(r)\frac{1}{r^2}dr\\
    &-\frac{2Z^2}{n^3a_0^2}\int_0^{\infty}f^2(r)dr\\
\end{alignedat}
\end{equation}
The normalization condition is given by $\int_0^{\infty}r^2|R(r)|^2dr=\int_0^{\infty}f^2(r)dr=1$ and $\langle   \frac{1}{\hat{r}^2}  \rangle=\int_0^{\infty}f^2(r)\frac{1}{\hat{r}^2}dr$. So equ. (36) becomes,
\begin{equation}
    \int_0^{\infty}\frac{\partial}{\partial{\ell}}\left[\frac{f''(r)}{f(r)}\right]dr=(2\ell+1)\langle   \frac{1}{\hat{r}^2}  \rangle-\frac{2Z^2}{n^3a_0^2}
\end{equation}
After a little bit manipulation, got $\int_0^{\infty}\frac{\partial}{\partial{\ell}}\left[\frac{f''(r)}{f(r)}\right]dr=0$.\\
Finally, from equ. (37),
\begin{equation}
    \langle   \frac{1}{\hat{r}^2}  \rangle=\frac{2Z^2}{(2\ell+1)n^3a_0^2}
\end{equation}
Then
\begin{equation} 
 \langle   \hat{p}_r^2  \rangle=\frac{Z^2{\hbar}^2}{n^2a_0^2}\left[1-\frac{2\ell(\ell+1)}{n(2\ell+1)}\right]
\end{equation}
Finally the uncertainty in radial momentum will be,
\begin{equation}
    \Delta \hat p_r=\frac{Z\hbar}{na_0}\sqrt{1-\frac{2\ell(\ell+1)}{n(2\ell+1)}}
\end{equation}
\subsection{Radial Uncertainty Product}
From equ. (25),
\begin{equation*}
    \Delta \hat{r} =\frac{1}{2}\frac{a_0}{Z}\sqrt{n^2(n^2+2)-[\ell(\ell+1)]^2}
\end{equation*}
Hence the radial uncertainty product will be,
\begin{equation}
\begin{alignedat}{2}
    &\Delta \hat{r} \Delta \hat p_r\\
    &=\frac{1}{2n}\hbar\sqrt{n^2(n^2+2)-[\ell(\ell+1)]^2}\sqrt{1-\frac{2\ell(\ell+1)}{n(2\ell+1)}}\\  
\end{alignedat}
\end{equation}
\subsection{Results and Graphical analysis}
\subsubsection{Expectation value of Radial position}
We have already derived the generalized formula for expectation or average of radial position is given by,
\begin{equation*}
\langle \hat{r} \rangle=\frac{1}{2}\frac{a_{0}}{Z}[3n^2-\ell(\ell+1)]
\end{equation*}
for all principal quantum numbers ($n$) and azimuthal quantum numbers ($\ell$). They are tabulated in Table III.
If $\ell=0$, then $\langle \hat{r} \rangle=\frac{1}{2}\frac{a_{0}}{Z}3n^2$ and for ground state wave function ($n=1,\ell=0$), it becomes $\langle \hat{r} \rangle=\frac{3}{2}\frac{a_{0}}{Z}$, that is a well-known result, we see in basic level of Hydrogen atom. Now Fig. (1) and (2) are about $\langle \hat{r} \rangle$ vs $n$ for a specific $\ell$ and $\langle \hat{r} \rangle$ vs $\ell$ for a specific $n$. In these cases, the locus of $\langle \hat{r} \rangle$ is an asymmetric downward and upward parabolic respectively and as well all points on the locus will be in first quadrant only as $n\geq1$ and $\ell\geq0$.
\begin{figure}[h!]
    \centering
    \includegraphics[width=0.5\textwidth]{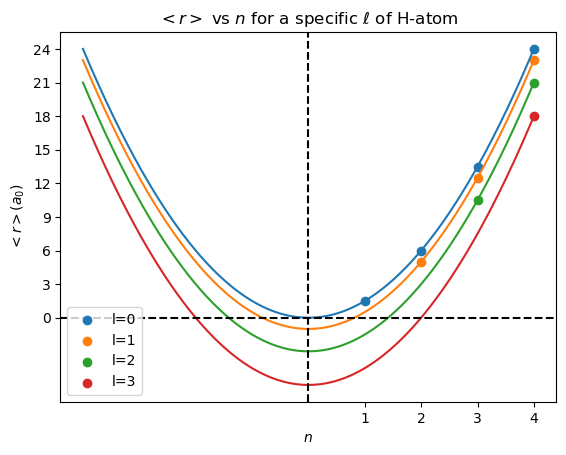}
    \caption{$\langle \hat{r} \rangle$ vs $n$ for a specific $\ell$ of H-atom}
\end{figure}
\begin{figure}[h!]
    \centering
    \includegraphics[width=0.5\textwidth]{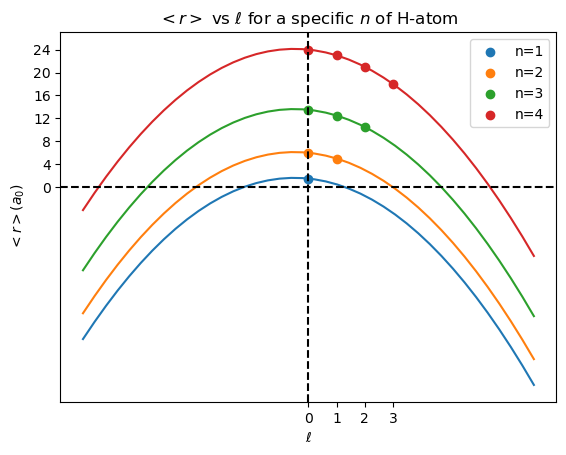}
    \caption{$\langle \hat{r} \rangle$ vs $\ell$ for a specific $n$ of H-atom}
\end{figure}
Fig. (3) is about $\langle \hat{r} \rangle$ for a different orbitals. It's look like an increasing zigzag structure, but for increasing atomic number, the height (vertical width) of zigzag will decrease momentarily i.e. the expectation values of radial position going to be same and for a high atomic number, the height becomes zero, hence for any orbital, the expectation values of radial position will be same like $Z=50$.
\begin{figure}[h!]
    \centering
    \includegraphics[width=0.5\textwidth]{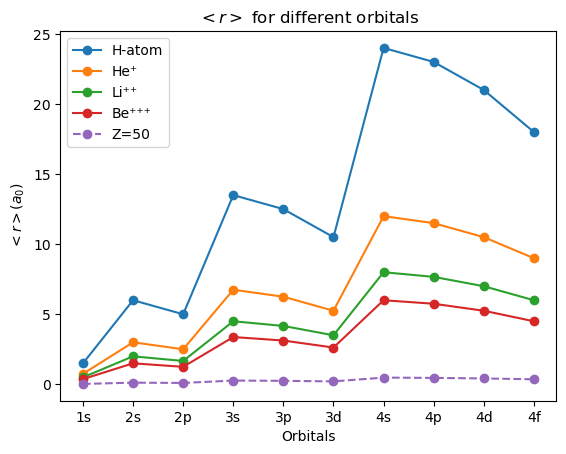}
    \caption{$\langle \hat{r} \rangle$ for different orbitals}
    \label{fig:example}
\end{figure}
\subsubsection{Uncertainty in Radial position}
We have already derived the generalized formula for uncertainty of radial position is given by,
\begin{equation*}
\Delta \hat{r} =\frac{1}{2}\frac{a_0}{Z}\sqrt{n^2(n^2+2)-[\ell(\ell+1)]^2}
\end{equation*}
for all principal quantum numbers ($n$) and azimuthal quantum numbers ($\ell$). If $\ell=0$, then $\Delta \hat{r}=\frac{1}{2}\frac{a_{0}}{Z}\sqrt{n^2(n^2+2)}$ and for ground state wave function ($n=1,\ell=0$), it becomes $\Delta \hat{r}=\frac{\sqrt{3}}{2}\frac{a_{0}}{Z}$. Fig. (4) and (5) are about $\Delta \hat{r}$ vs $n$ for a specific $\ell$ and $\Delta \hat{r}$ vs $\ell$ for a specific $n$. For different orbitals here, Fig. (6) is almost similar for variation of $\langle \hat{r} \rangle$ with different orbitals.
\begin{figure}[h!]
    \centering
    \includegraphics[width=0.5\textwidth]{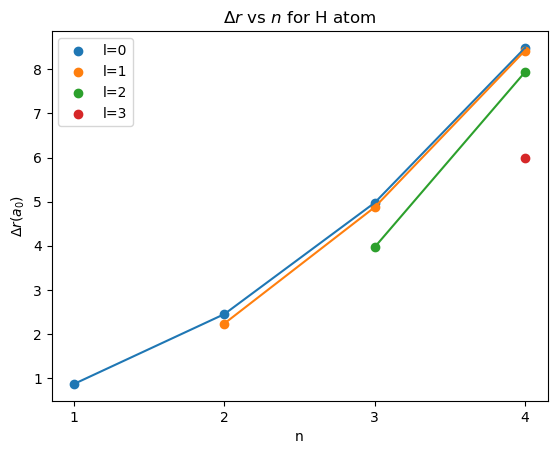}
    \caption{$\Delta \hat{r}$ vs $n$ for a specific $\ell$ of H-atom}
\end{figure}
\begin{figure}[h!]
    \centering
    \includegraphics[width=0.5\textwidth]{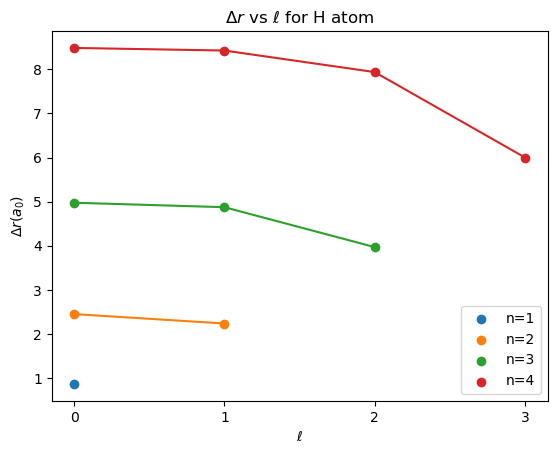}
    \caption{$\Delta \hat{r}$ vs $\ell$ for a specific $n$ of H-atom}
\end{figure}
\begin{figure}[h!]
    \centering
    \includegraphics[width=0.5\textwidth]{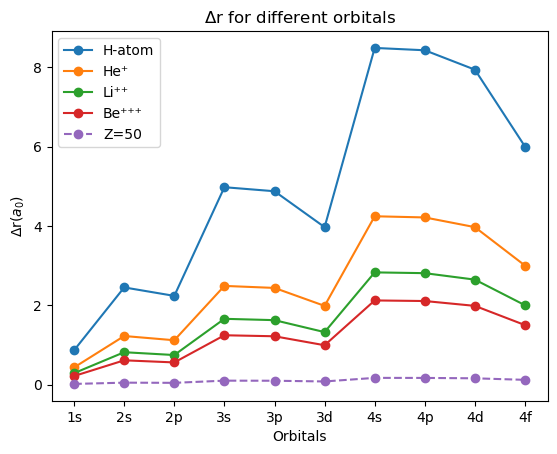}
    \caption{$\Delta \hat{r}$ for different orbitals}
\end{figure}
\subsubsection{Relative dispersion of Radial position}
The relative dispersion or coefficient of variation in the measurement of radial position is defined as, $\sigma_r=\frac{\Delta \hat{r}}{\langle \hat{r} \rangle}$ i.e. the ratio of uncertainty and expectation of radial position.
The relative dispersion is given by
\begin{equation}
    \sigma_r=\frac{\sqrt{n^2(n^2+2)-[\ell(\ell+1)]^2}}{[3n^2-\ell(\ell+1)]}
\end{equation}
It is clear that the relative dispersion is independent of atomic number ($Z$), i.e. for Hydrogen atom or Hydrogen-like ions, the variation of it is same. They are tabulated in Table III. Fig. (7) and (8) are about $\sigma_r$ vs $n$ for a specific $\ell$ and $\sigma_r$ vs $\ell$ for a specific $n$ where Fig. (9) is for different orbitals. Here a fascinating fact is, for $\ell=2$ and $n=3,4$, the relative dispersion of radial position is equal $\left({\sigma}_r=\sqrt{\frac{3}{21}}\right)$.
\begin{figure}[h!]
    \centering
    \includegraphics[width=0.5\textwidth]{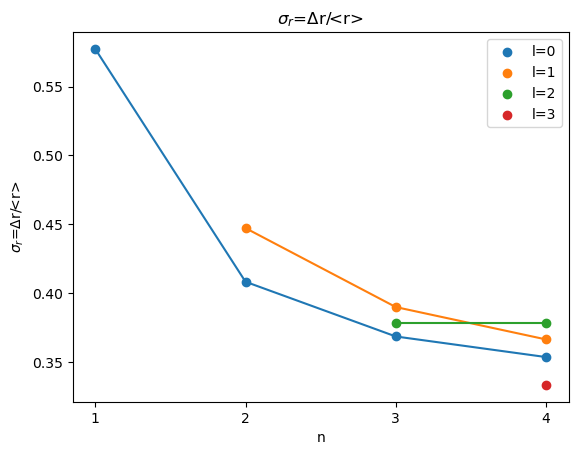}
    \caption{$\sigma_r$ vs $n$ for a specific $\ell$ of H-atom}
\end{figure}
\begin{figure}[h!]
    \centering
    \includegraphics[width=0.5\textwidth]{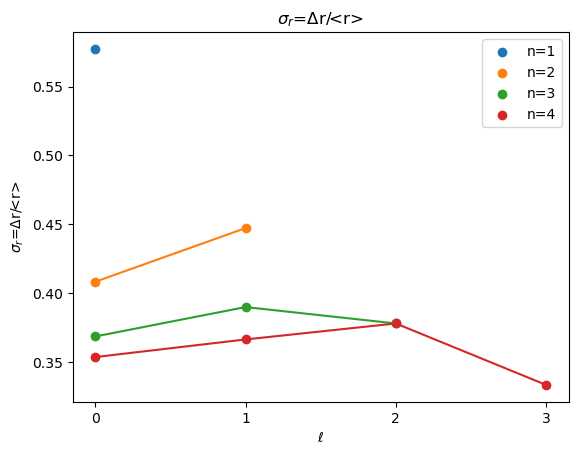}
    \caption{$\sigma_r$ vs $\ell$ for a specific $n$ of H-atom}
\end{figure}
\begin{figure}[h!]
    \centering
    \includegraphics[width=0.5\textwidth]{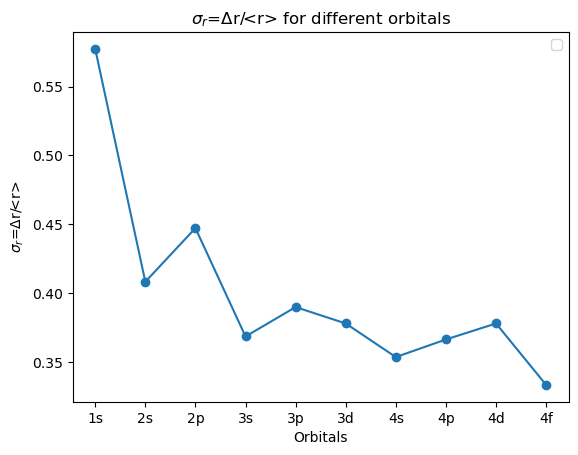}
    \caption{$\sigma_r$ for different orbitals}
\end{figure}
\begin{table}[htb]
    \centering
    \begin{ruledtabular}
    \begin{tabular}{lccc}
        \textrm{Orbital} & 
        \textrm{$\langle \hat{r} \rangle(\frac{a_0}{Z})$} & 
        \textrm{$\Delta \hat{r} (\frac{a_0}{Z})$} & \textrm{$\sigma_r$}\\
        \colrule
        1s & $1.5$  & 0.8660 & 0.5774 \\
        \colrule
2s & $6.0$  & 2.4494 & 0.4082 \\
2p & $5.0$  & 2.2360 & 0.4472 \\
\colrule
3s & $13.5$ & 4.9749 & 0.3685 \\
3p & $12.5$ & 4.8733 & 0.3899 \\
3d & $10.5$ & 3.9686 & 0.3780 \\
\colrule
4s & $24.0$ & 8.4852 & 0.3536 \\
4p & $23.0$ & 8.4261 & 0.3664 \\
4d & $21.0$ & 7.9372 & 0.3780 \\
4f & $18.0$ & 6.0000 & 0.3333 \\
    \end{tabular}
    \end{ruledtabular}
    \caption{Relative dispersion of radial position}
\end{table}
\subsubsection{Expectation value of Radial momentum}
By rigorous calculation, it is able to to that $\langle \hat p_r \rangle =0$, as shown before. Now it is known that the radial probability current is directly proportional to average radial momentum. The radial probability current density in 3d is given by $j_r=\frac{1}{2m}[R^*(r)\hat{p_r}R(r)-R(r)\hat{p_r}R^*(r)]$ where $R(r)$ is the radial wave function. As the radial wave function have no imaginary part i.e. $R^*(r)=R(r)$. Hence, $j_r=0$ and the average radial momentum becomes zero.
\subsubsection{Uncertainty in Radial momentum}
We have already derived the generalized formula for uncertainty of radial momentum is given by,
\begin{equation*}
\Delta \hat p_r=\frac{Z\hbar}{na_0}\sqrt{1-\frac{2\ell(\ell+1)}{n(2\ell+1)}}
\end{equation*}
for all principal quantum numbers ($n$) and azimuthal quantum numbers ($\ell$). If $\ell=0$, then $\Delta \hat p_r=\frac{Z\hbar}{na_0}$ and for ground state wave function ($n=1,\ell=0$), it becomes $\Delta \hat p_r=\frac{Z\hbar}{a_0}$. Fig. (10) and (11) are about $\Delta \hat p_r$ vs $n$ for a specific $\ell$ and $\Delta \hat p_r$ vs $\ell$ for a specific $n$. Fig. (12) is decreasing zigzag structure for variation of $\Delta \hat p_r$ with different orbitals.
\begin{figure}[h!]
    \centering
    \includegraphics[width=0.5\textwidth]{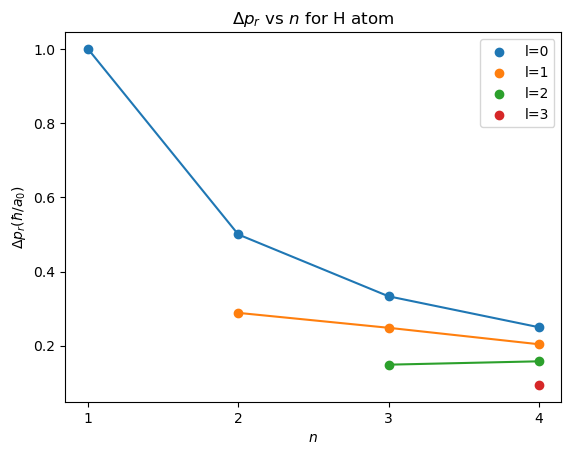}
    \caption{$\Delta \hat p_r$ vs $n$ for a specific $\ell$ of H-atom}
\end{figure}
\begin{figure}[h!]
    \centering
    \includegraphics[width=0.5\textwidth]{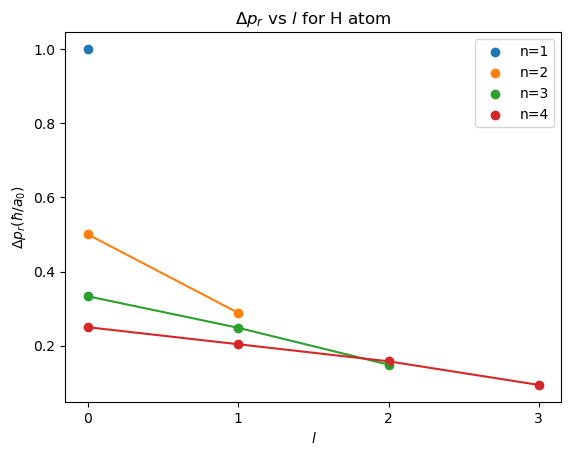}
    \caption{$\Delta \hat p_r$ vs $\ell$ for a specific $n$ of H-atom}
\end{figure}
\begin{figure}[h!]
    \centering
    \includegraphics[width=0.5\textwidth]{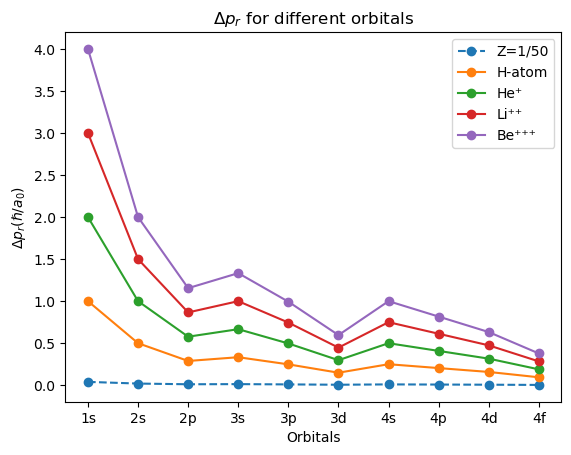}
    \caption{$\Delta \hat p_r$ for different orbitals}
\end{figure}

\subsubsection{Radial uncertainty product}
We have already derived the generalized formula for radial uncertainty for Hydrogen and Hydrogen-like ions,
\begin{equation*}
    \Delta \hat{r} \Delta \hat p_r =\frac{\hbar}{2n}\sqrt{n^2(n^2+2)-[\ell(\ell+1)]^2}\sqrt{1-\frac{2\ell(\ell+1)}{n(2\ell+1)}}
\end{equation*}
Clearly, the radial uncertainty product is independent of Bohr radius and atomic number, it's only related to Planck's constant (reduced). They are tabulated in Table IV. As the range of azimuthal quantum number is given by ${\ell}_{min}=0$ and ${\ell}_{max}=n-1$. For ${\ell}_{min}=0$, $(\Delta \hat{r} \Delta \hat p_r)_{n,0}=\frac{\hbar}{2}\sqrt{n^2+2}$. For ${\ell}_{max}=n-1$, $(\Delta \hat{r} \Delta \hat p_r)_{n,n-1}=\frac{\hbar}{2}\sqrt{\frac{2n+1}{2n-1}}$.  As $n\geq 1$, $\Delta \hat{r} \Delta \hat p_r>\frac{\hbar}{2}$ which is nothing but the Heisenberg's uncertainty principle.
Fig. (13) and (14) are about $\Delta \hat{r} \Delta \hat p_r$ vs $n$ for a specific $\ell$ and $\Delta \hat{r} \Delta \hat p_r$ vs $\ell$ for a specific $n$. Fig. (15) is for variation of $\Delta \hat{r}\Delta \hat p_r$ with different orbitals and all data points are above the $\frac{\hbar}{2}$ line.
\begin{figure}[h!]
    \centering
    \includegraphics[width=0.5\textwidth]{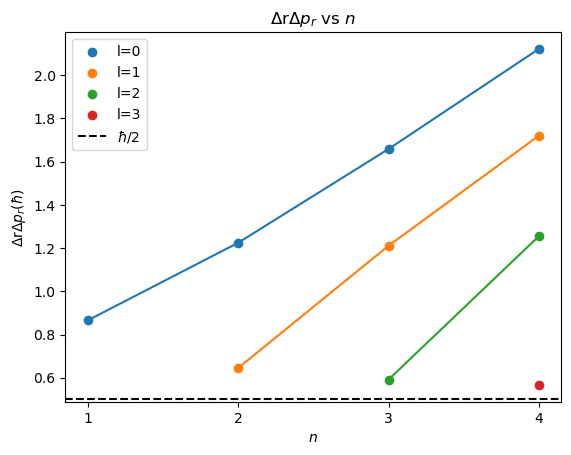}
    \caption{$\Delta \hat{r}\Delta \hat p_r$ vs $n$ for a specific $\ell$ of H-atom}
\end{figure}
\begin{figure}[h!]
    \centering
    \includegraphics[width=0.5\textwidth]{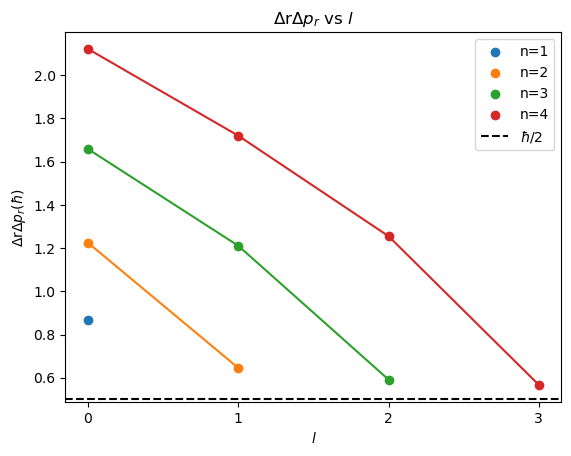}
    \caption{$\Delta \hat{r}\Delta \hat p_r$ vs $\ell$ for a specific $n$ of H-atom}
\end{figure}
\begin{figure}[h!]
    \centering
    \includegraphics[width=0.5\textwidth]{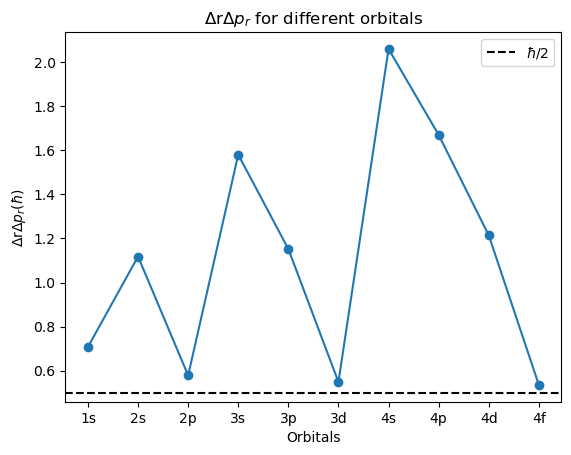}
    \caption{$\Delta \hat{r}\Delta \hat p_r$ for different orbitals}
\end{figure}

\begin{table*}
\begin{ruledtabular}
\begin{tabular}{ccccccc}
Principal quantum number ($n$) & Azimuthal quantum number ($\ell$) & Orbital 
& Radial solution ($R_{n\ell}$) & $\Delta \hat{r} (\frac{a_0}{Z})$ & $\Delta \hat p_r(\frac{Z\hbar}{a_0})$ 
& $\Delta \hat{r} \Delta \hat p_r(\hbar)$ \\ \hline
K-Shell ($n=1$) & $\ell=0$ & 1s & $R_{10}$ & 0.8660 & 1.0000 & 0.8660 \\ \hline
\multirow{2}{*}{L-Shell ($n=2$)} 
 & $\ell=0$ & 2s & $R_{20}$ & 2.4494 & 0.5000 & 1.2247 \\ 
 & $\ell=1$ & 2p & $R_{21}$ & 2.2360 & 0.2886 & 0.6454 \\ \hline
\multirow{3}{*}{M-Shell ($n=3$)} 
 & $\ell=0$ & 3s & $R_{30}$ & 4.9749 & 0.3333 & 1.6583 \\ 
 & $\ell=1$ & 3p & $R_{31}$ & 4.8733 & 0.2484 & 1.2108 \\ 
 & $\ell=2$ & 3d & $R_{32}$ & 3.9686 & 0.1490 & 0.5916 \\ \hline
\multirow{4}{*}{N-Shell ($n=4$)} 
 & $\ell=0$ & 4s & $R_{40}$ & 8.4852 & 0.2500 & 2.1213 \\ 
 & $\ell=1$ & 4p & $R_{41}$ & 8.4261 & 0.2041 & 1.7199 \\ 
 & $\ell=2$ & 4d & $R_{42}$ & 7.9372 & 0.1581 & 1.2549 \\ 
 & $\ell=3$ & 4f & $R_{43}$ & 6.0000 & 0.0944 & 0.5669 \\ 
\end{tabular}
\end{ruledtabular}
\caption{Uncertainties in radial position, momentum and their product}
\end{table*}

\subsubsection{Ground state of Hydrogen atom}
The radial wave function of Hydrogen atom is given by,
\begin{equation*}
    R_{n\ell}(r)=N_{n\ell}e^{-{\frac{r}{na_{0}}}}\left(\frac{2}{na_{0}}r\right)^\ell L^{2\ell+1}_{n-\ell-1}\left({\frac{2}{na_{0}}r}\right)
\end{equation*}
where $N_{n\ell}=\sqrt{{(\frac{2}{na_{0}})}^3\frac{(n-\ell-1)!}{2n(n+\ell)!}}$. For ground state ($n=1,\ell=0$), $N_{10}=2(\frac{1}{a_0})^{\frac{3}{2}}$, $R_{10}(r)=2(\frac{1}{a_0})^{\frac{3}{2}}\,e^{-\frac{r}{a_0}}$ which is plotted in Fig. (16). For ground state, $\langle \hat{r} \rangle=\frac{3a_0}{2}$ and $\Delta \hat{r} =\frac{\sqrt{3}a_0}{2}$.
The radial probability density for ground state, $P_{10}(r)=r^2|R_{10}(r)|^2=4(\frac{1}{a_0})^3r^2e^{-\frac{2r}{a_0}}$.
For most probable radius, $\frac{dP}{dr}=0$ gives $r=0$ (non-acceptable) and $r_{mp}=a_0$. For ground state, Bohr radius is the most probable radius as expected.
\begin{figure}[h!]
    \centering
    \includegraphics[width=0.5\textwidth]{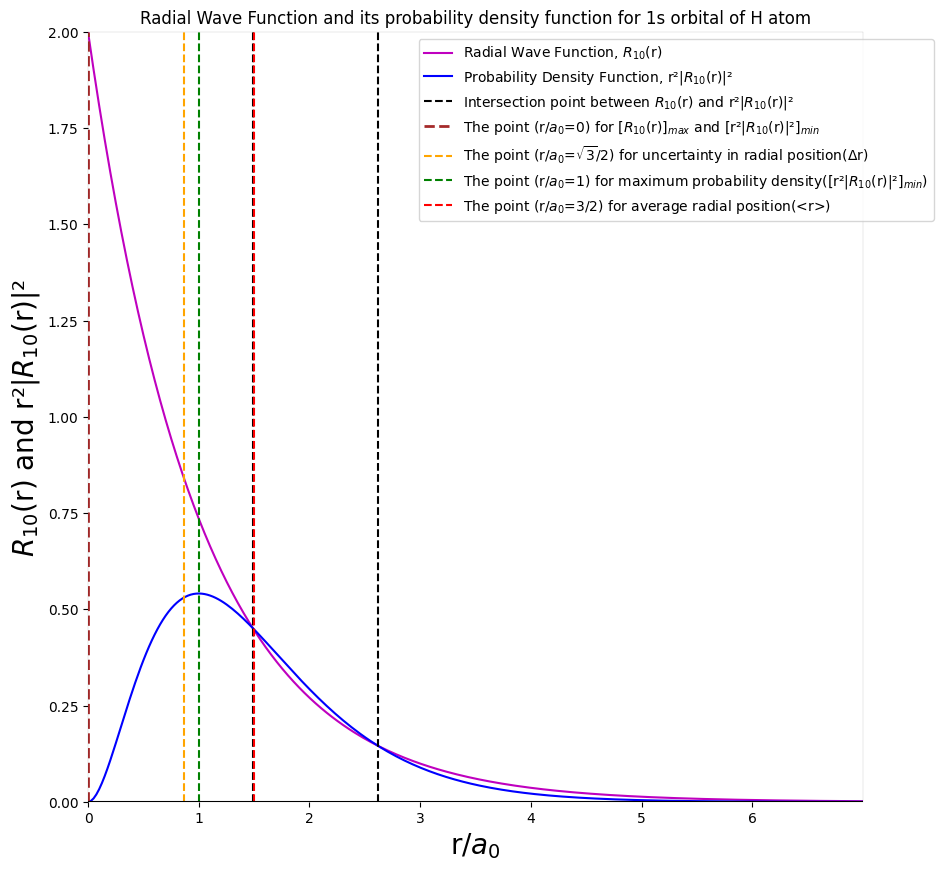}
    \caption{Ground state of Hydrogen atom}
\end{figure}

\section{Infinite Spherical well potential}
\subsection{Radial wave function}
The spherical infinite potential is defined by
\begin{equation}
    V(r) = \begin{cases}
        0, & \text{for } r \leq R \\
        \infty, & \text{for } r > R
    \end{cases}
\end{equation}
where $R$ is the radius of spherical potential well. The wave function outside the spherical potential well is zero according to definition.
And inside the spherical well, the time independent Schrodinger equation is given by,
\begin{equation}
    -\frac{{\hbar}^2}{2m}{\nabla}^2\psi(r,\theta,\phi) = E\psi(r,\theta,\phi)
\end{equation}
The well known radial part (Refs.~\cite{Huang_2017}) is given by,
\begin{equation}
    \frac{d^2R}{dz^2}+\frac{2}{z}\frac{dR}{dz}+\left(1-\frac{\ell(\ell+1)}{z^2}\right)R=0
\end{equation}
where $z=kr$ and $k^2=\frac{2mE}{{\hbar}^2}$ and $\ell$ is the azimuthal quantum number (angular momentum quantum number). The equ. (45) is the so-called Spherical Bessel equation, (Refs.~\cite{Abramowitz_Stegun_1972}) and has the solutions $j_{\ell}(kr)$ and $n_{\ell}(kr)$, where $j_{\ell}(kr)$ and $n_{\ell}(kr)$ are the spherical Bessel and Neumann function of order $\ell$. The spherical Neumann functions $n_{\ell}(kr)$ are discarded, since they are divergent at the center of the sphere, $r= 0$. So the solution of it is given by,
\begin{equation}
    R_{n\ell}(z)=N_{n\ell}j_{\ell}(z)
\end{equation}
where $N_{n\ell}$ is the normalization constant and $j_{\ell}(z)$ is the spherical Bessel function of order $\ell$ is defined by,\\
\begin{equation}
    j_{\ell}(z)=\sqrt{\frac{\pi}{2z}}J_{\ell+\frac{1}{2}}(z)
\end{equation}
where $J_{\ell+\frac{1}{2}}(z)$ is Bessel function of order $\ell$. The important formula for producing spherical Bessel function is known as Rayleigh's formula is defined as
\begin{equation*}
    j_{\ell}(z)=(-z)^{\ell}\left(\frac{1}{z}\frac{d}{dz}\right)^{\ell}\frac{\sin z}{z}
\end{equation*}
Some examples are shown in Table V.
\begin{table}[!ht]
  \centering
  \begin{ruledtabular}
    \begin{tabular}{cc}
      \textrm{$\ell$} & \textrm{Spherical Bessel Function $j_{\ell}(z)$} \\
      \colrule
      0 & $\frac{\sin z}{z}$ \\
      1 & $\frac{\sin z}{z^2} - \frac{\cos z}{z}$ \\
      2 & $\left(\frac{3}{z^3} - \frac{1}{z}\right)\sin z - \frac{3\cos z}{z^2}$ \\
      3 & $\left(\frac{15}{z^4} - \frac{6}{z^2}\right)\sin z - \left(\frac{15}{z^3} - \frac{1}{z}\right)\cos z$ \\
      4 & $\left(\frac{105}{z^5} - \frac{45}{z^3} - \frac{1}{z}\right)\sin z - \left(\frac{105}{z^4} + \frac{10}{z^2}\right)\cos z$ \\
    \end{tabular}
  \end{ruledtabular}
  \caption{Examples of spherical Bessel functions}
\end{table}
In the analogy of one dimensional rigid box potential problem, the wave function is given by $\psi(x)=\sqrt{\frac{2}{a}}\sin kx$ and at the boundary $x=a$, the wave function vanishes i.e., $\sin ka=0$ implies that $ka=n\pi $ or $k=\frac{n\pi}{a}$ where the values of $n$ are 1,2,3,4,.. and the energy eigenvalues are given by $E_n=\frac{n^2{\pi}^2{\hbar}^2}{2ma^2}$. Now for infinite spherical potential well, at the boundary $r=R$, the values of $k$ must be chosen such that $z=ka$ corresponds to the zeros of spherical Bessel functions $j_{\ell}(z)$. Let denote the $n^{th}$ zero of $j_{\ell}(z)$ as $z_{n\ell}$. In this way, $kR=z_{n\ell}$ where the values of $n$ are 1,2,3,4,.... and as a result, the energy of the particle confined in the spherical potential well is given by,
\begin{equation}
    E_{n\ell}=z_{n\ell}^2\frac{{\hbar}^2}{2mR^2}
\end{equation}
The zeroth ($z_{n\ell})$ of spherical Bessel function ($j_{\ell}(z)$) is defined as $j_{\ell}(z_{n\ell})=0$ where $n$ is the place of order of intersection points with z-axis with spherical Bessel function starting from $n=1$ as like for $j_0(z)$ is shown in Fig. (17).
\begin{figure}[h!]
    \centering
    \includegraphics[width=0.5\textwidth]{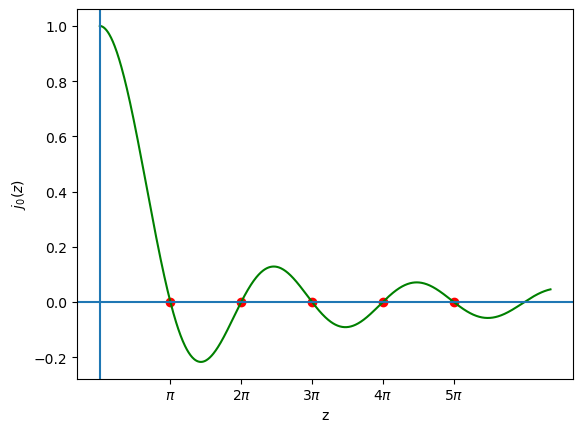}
    \caption{$j_0(z)$ variation w.r.t. $z$}
\end{figure}
Here all possible values of $z_{n\ell}$ and the blue colored is only possible values of $\ell$ for a fixed $n$ are tabulated in Table VI.
\begin{table}[htb]
    \centering
    \begin{ruledtabular}
    \begin{tabular}{lccccc}
        $z_{n\ell}$ & $n=1$ & $n=2$ & $n=3$ & $n=4$ & $n=5$ \\
        \colrule
        $\ell=0$ & \textcolor{blue}{$\pi$} & \textcolor{blue}{$2\pi$} & \textcolor{blue}{$3\pi$} & \textcolor{blue}{$4\pi$} & \textcolor{blue}{$5\pi$} \\
        $\ell=1$ & 4.49341 & \textcolor{blue}{7.72525} & \textcolor{blue}{10.9041} & \textcolor{blue}{14.0662} & \textcolor{blue}{17.2208} \\
        $\ell=2$ & 5.76346 & 9.09501 & \textcolor{blue}{12.3229} & \textcolor{blue}{15.5146} & \textcolor{blue}{18.6890} \\
        $\ell=3$ & 6.98796 & 10.4171 & 13.6980 & \textcolor{blue}{16.9236} & \textcolor{blue}{20.1218} \\
        $\ell=4$ & 8.18256 & 11.7049 & 15.0397 & 18.3013 & \textcolor{blue}{21.5254} \\
    \end{tabular}
    \end{ruledtabular}
    \caption{Values for zeros of spherical Bessel function}
\end{table}
The non-normalized wave function is given by
\begin{equation}
    R_{n\ell}(r)=N_{n\ell}j_{\ell}(kr)=N_{n\ell}j_{\ell}\left(\frac{z_{n\ell}}{R}r\right)
\end{equation}
Let introduce a dimensionless parameter $\rho$ defined as $\rho=\frac{z_{n\ell}}{R}r$. Hence $R_{n\ell}(\rho)=N_{n\ell}j_{\ell}(\rho)$ and let make normalize. The normalization condition for radial wave function is given by $I=\int_0^R r^2|R_{n\ell}|^2=1$. Now this integral is,
\begin{equation}
\begin{alignedat}{2}
    &I=\int_0^R |N_{n\ell}|^2 r^2 \left[j_{\ell}\left(\frac{z_{n\ell}}{R}r\right)\right]^2dr\\
    &=\left(\frac{R}{z_{n\ell}}\right)^3|N_{n\ell}|^2\int_0^{z_{n\ell}}{\rho}^2[j_{\ell}(\rho)]^2d\rho \text{   as $r=\frac{R}{z_{n\ell}}\rho$}\\  
\end{alignedat}
\end{equation}
Now from (Refs.~\cite{bloomfield2017indefiniteintegralssphericalbessel}), the indefinite Bessel squared integral of equ. A(8),
\begin{equation*}
\int x [J_\ell(x)]^2 \, dx = \frac{x^2}{2} \left[ [J_\ell(x)]^2 - J_{\ell-1}(x) J_{\ell+1}(x) \right]
\end{equation*}
Now using equ. (47),
\begin{equation}
    \int_0^x x^2[j_{\ell}(x)]^2dx=\frac{x^3}{2}([j_{\ell}(x)]^2-j_{\ell-1}(x)j_{\ell+1}(x))
\end{equation}
Finally the integral becomes,
\begin{equation}
I=\frac{R^3}{2}|N_{n\ell}|^2([j_{\ell}(z_{n\ell})]^2-j_{\ell-1}(z_{n\ell})j_{\ell+1}(z_{n\ell}))
\end{equation}
As $I=1$, $N_{n\ell}=\sqrt{\frac{2}{R^3}}\frac{1}{\sqrt{[j_{\ell}(z_{n\ell})]^2-j_{\ell-1}(z_{n\ell})j_{\ell+1}(z_{n\ell})}}$\\
Now define $C_{n\ell}=\frac{1}{\sqrt{[j_{\ell}(z_{n\ell})]^2-j_{\ell-1}(z_{n\ell})j_{\ell+1}(z_{n\ell})}}$\\
For $z=z_{n\ell}, \text{ root of $j_{\ell}$ }, j_{\ell}(z_{n\ell})=0$ and $\frac{j_{\ell-1}(z_{n\ell})}{j_{\ell+1}(z_{n\ell})}=-1$, hence $C_{n\ell}=\frac{1}{|j_{\ell-1}(z_{n\ell})|}=\frac{1}{|j_{\ell+1}(z_{n\ell})|}$. The values of $C_{n\ell}$ for corresponding $z_{n\ell}$ are tabulated in Table VII.
\begin{table}[htb]
    \centering
    \begin{ruledtabular}
    \begin{tabular}{lcc}
        \textrm{\((n, \ell)\)} & 
        \textrm{\(z_{n\ell}\)} & 
        \textrm{\(C_{n\ell}\)} \\
        \colrule
        (1,0) & \(3.14159\) & \(3.14159\) \\
        \colrule
        (2,0) & \(6.28319\) & \(6.28319\) \\
        (2,1) & \(7.72525\) & \(7.7897\) \\
        \colrule
        (3,0) & \(9.42478\) & \(9.42478\) \\
        (3,1) & \(10.9041\) & \(10.9498\) \\
        (3,2) & \(12.3229\) & \(12.4464\) \\
        \colrule
        (4,0) & \(12.5664\) & \(12.5664\) \\
        (4,1) & \(14.0662\) & \(14.1017\) \\
        (4,2) & \(15.5146\) & \(15.6122\) \\
        (4,3) & \(16.9236\) & \(17.1046\) \\
        \colrule
        (5,0) & \(15.70796\) & \(15.70796\) \\
        (5,1) & \(17.2208\) & \(17.2498\) \\
        (5,2) & \(18.6890\) & \(18.7698\) \\
        (5,3) & \(20.1218\) & \(20.2731\) \\
        (5,4) & \(21.5254\) & \(21.7633\) \\
    \end{tabular}
    \end{ruledtabular}
    \caption{Values of $C_{n\ell}$ for corresponding $z_{n\ell}$}
\end{table}
\begin{figure}[h!]
    \centering
    \includegraphics[width=0.5\textwidth]{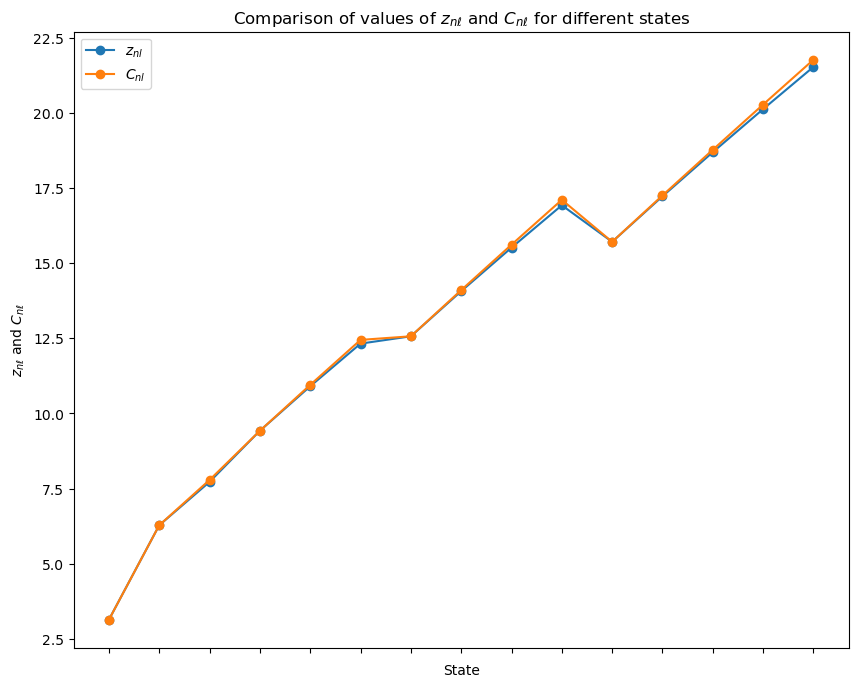}
    \caption{$C_{n\ell}$ and corresponding $z_{n\ell}$ vs state ($n,\ell$)}
\end{figure}
From the Fig. (18), it's clear that $C_{n\ell}=z_{n\ell}$ for $\ell=0$ and the difference is increasing for $\ell$ increasing. Finally got, the normalization constant, $N_{n\ell}=\sqrt{\frac{2}{R^3}}\frac{1}{|j_{\ell+1}(z_{n\ell)}|}$. The normalized radial wave function will be, 
\begin{equation}
    R_{n\ell}(r)=\sqrt{\frac{2}{R^3}}\frac{1}{|j_{\ell+1}(z_{n\ell})|}j_{\ell}\left(\frac{z_{n\ell}}{R}r\right)
\end{equation}

\subsection{Uncertainty in Radial position}
As per equ. (1),
\begin{equation}
\begin{alignedat}{2}
    &\langle \hat{r} \rangle\\
    &=\left(\frac{R}{z_{n\ell}}\right)^4\int_0^{z_{n\ell}}{\rho}^3|R_{n\ell}(\rho)|^2d\rho\\
    &=\frac{2R}{z_{n\ell}^4|j_{\ell+1}(z_{n\ell})|^2}\int_0^{z_{n\ell}}{\rho}^3[j_{\ell}(\rho)]^2d\rho\\
\end{alignedat}
\end{equation}
The evaluation of this integral is little bit harder for by-hand approach, then we have used the Wolfram Mathematica software (Refs.~\cite{mathematica_integration}) and the result is given by
\begin{equation}
\begin{alignedat}{2}
    &\int_0^x x^3[j_{\ell}(x)]^2dx=\frac{1}{4}x^{4+2\ell}\sqrt{\pi}\Gamma(\ell+1)\Gamma(\ell+2)\\
    &{}_{2}F_{3}^{(r)}\left[{\ell+1,\ell+2};{\ell+\frac{3}{2},\ell+3,2\ell+2};-x^2\right]\\
\end{alignedat}
\end{equation}
where ${}_pF_q^{(r)}$ is the regularized hyper-geometric function (Refs.~\cite{wolfram_regularized_hypergeometric}) is related to the generalized hyper-geometric function ${}_pF_q$ by
\begin{equation}
    {}_{p}F_{q}^{(r)}(a_1, \ldots, a_p; b_1, \ldots, b_q; z) = \frac{{}_{p}F_{q}(a_1, \ldots, a_p; b_1, \ldots, b_q; z)}{\Gamma(b_1) \Gamma(b_2) \ldots \Gamma(b_q)}
\end{equation}
So the expectation value of $r$ is,
\begin{equation*}
\begin{alignedat}{2}
    &\langle \hat{r} \rangle\\
    &=\frac{2R}{z_{n\ell}^4|j_{\ell+1}(z_{n\ell})|^2} \frac{1}{4}z_{n\ell}^{4+2\ell}\sqrt{\pi}\Gamma(\ell+1)\Gamma(\ell+2)\\
    &{}_{2}F_{3}^{(r)}\left[{\ell+1,\ell+2};{\ell+\frac{3}{2},\ell+3,2\ell+2};-z_{n\ell}^2\right]\\ 
    &=R\,A(\ell,z_{n\ell})\\
\end{alignedat}
\end{equation*}
where the short-hand notation $A(\ell,z_{n\ell})$ is defined as
\begin{equation}
\begin{alignedat}{2}
    &A(\ell,z_{n\ell})=\frac{z_{n\ell}^{2\ell}}{2|j_{\ell+1}(z_{n\ell})|^2} \frac{\sqrt{\pi}\Gamma(\ell+1)\Gamma(\ell+2)}{\Gamma(\ell+\frac{3}{2})\Gamma(\ell+3)\Gamma(2\ell+2)}\\
    &{}_{2}F_{3}\left[{\ell+1,\ell+2};{\ell+\frac{3}{2},\ell+3,2\ell+2};-z_{n\ell}^2\right]\\  
\end{alignedat}
\end{equation}
Fig. (19) and (20) are about $\langle \hat{r} \rangle$ vs $n$ for a specific $\ell$ and $\langle \hat{r} \rangle$ vs $\ell$ for a specific $n$ and Fig. (21) is for different states.
\begin{figure}[h!]
    \centering
    \includegraphics[width=0.5\textwidth]{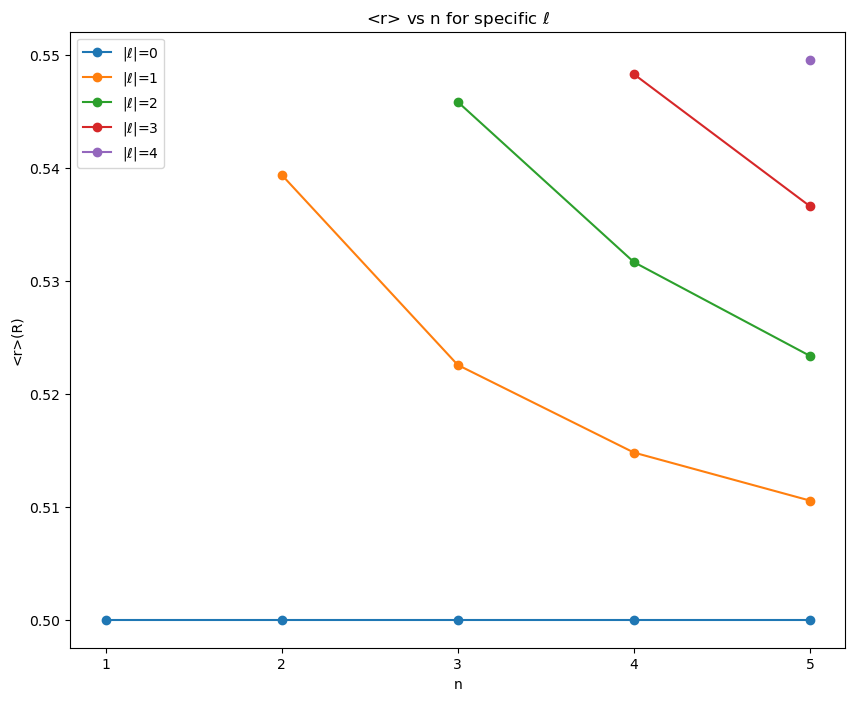}
    \caption{$\langle \hat{r} \rangle$ vs $n$ for a specific $\ell$ of ISW}
\end{figure}
\begin{figure}[h!]
    \centering
    \includegraphics[width=0.5\textwidth]{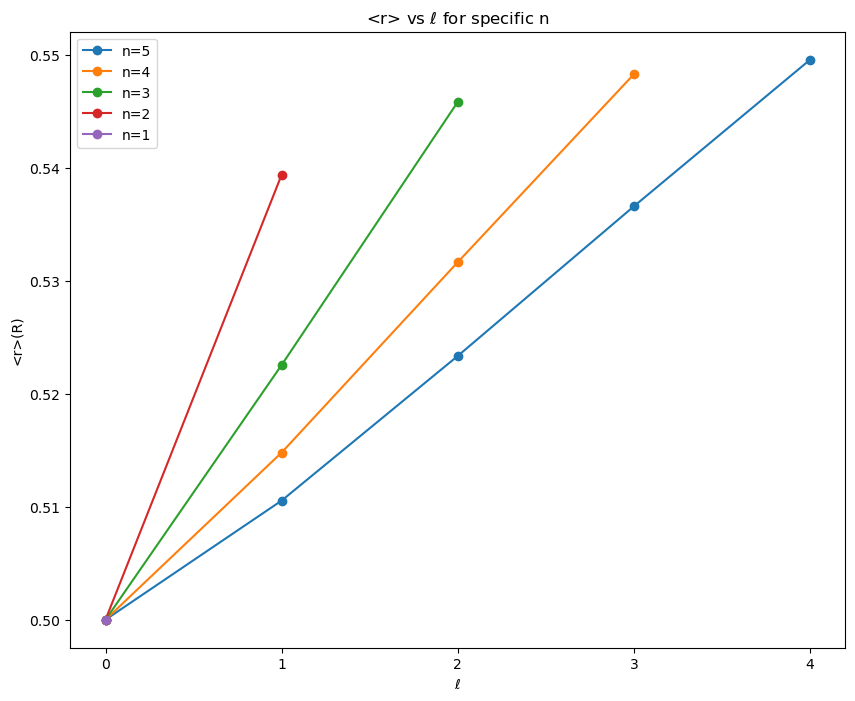}
    \caption{$\langle \hat{r} \rangle$ vs $\ell$ for a specific $n$ of ISW}
\end{figure}
\begin{figure}[h!]
    \centering
    \includegraphics[width=0.5\textwidth]{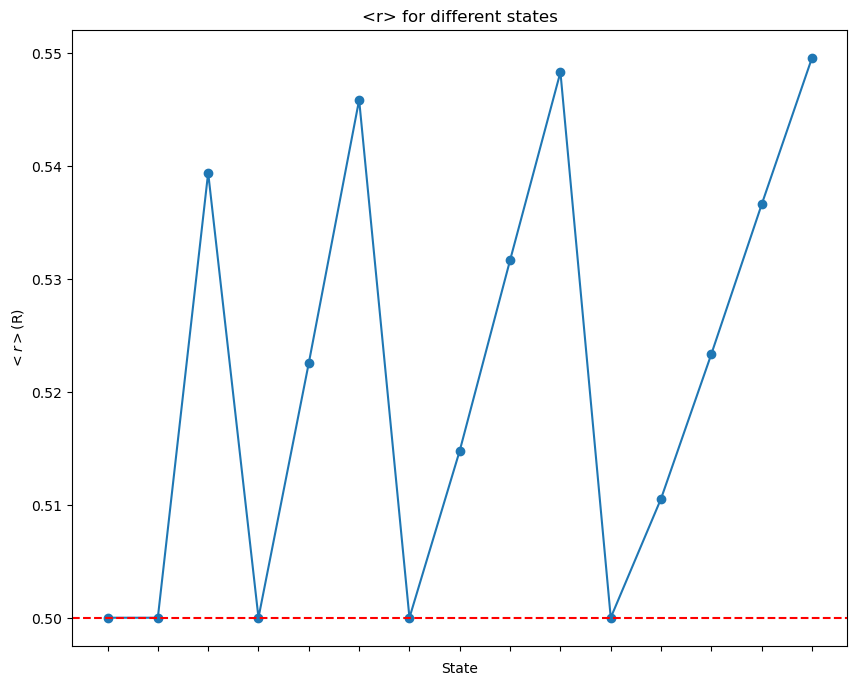}
    \caption{$\langle \hat{r} \rangle$ for different states of ISW}
\end{figure}
Now, our aim is to evaluate $\langle \hat{r}^2 \rangle =0$, as from equ. (2), 
\begin{equation}
\begin{alignedat}{2}
    &\langle \hat{r}^2 \rangle =0\\
    &=\left(\frac{R}{z_{n\ell}}\right)^5\int_0^{z_{n\ell}}{\rho}^4|R_{n\ell}(\rho)|^2d\rho\\
    &=\frac{2R^2}{z_{n\ell}^5|j_{\ell+1}(z_{n\ell})|^2}\int_0^{z_{n\ell}}{\rho}^4[j_{\ell}(\rho)]^2d\rho\\
\end{alignedat}
\end{equation}
Again using Wolfram Mathematica software,
\begin{equation}
\begin{alignedat}{2}
    &\int_0^x x^4[j_{\ell}(x)]^2dx=2^{-2(2+\ell)}x^{5+2\ell}(3+2\ell)\pi \Gamma(2\ell+2)\\
    &{}_{2}F_{3}^{(r)}\left[\ell+1,\ell+\frac{5}{2};\ell+\frac{3}{2},\ell+\frac{7}{2},2\ell+2;-x^2\right]\\ 
\end{alignedat}
\end{equation}
So the expectation value of $r^2$ is,
\begin{equation*}
\begin{alignedat}{2}
    &\langle \hat{r}^2 \rangle =0\\
    &=\frac{R^2 2^{-3-2\ell}z_{n\ell}^{2\ell}}{|j_{\ell+1}(z_{n\ell})|^2}(3+2\ell)\pi \Gamma(2\ell+2)\\
    &{}_{2}F_{3}^{(r)}\left[\ell+1,\ell+\frac{5}{2};\ell+\frac{3}{2},\ell+\frac{7}{2},2\ell+2;-z_{n\ell}^2\right]\\
    &=R^2\,B(\ell,z_{n\ell})\\
\end{alignedat}
\end{equation*}
where the short-hand notation $B(\ell,z_{n\ell})$ is defined as
\begin{equation}
\begin{alignedat}{2}
    &B(\ell,z_{n\ell})=\frac{2^{-3-2\ell}z_{n\ell}^{2\ell}}{|j_{\ell+1}(z_{n\ell})|^2}  \frac{(3+2\ell)\pi}{\Gamma(\ell+\frac{3}{2})\Gamma(\ell+\frac{7}{2})}\\
    &{}_{2}F_{3}\left[\ell+1,\ell+\frac{5}{2};\ell+\frac{3}{2},\ell+\frac{7}{2},2\ell+2;-z_{n\ell}^2\right]\\  
\end{alignedat}
\end{equation}
In terms of $A$ and $B$ the uncertainty in radial position will be,
\begin{equation*}
    \Delta \hat{r}=R\sqrt{B(\ell,z_{n\ell})-A(\ell,z_{n\ell})^2}
\end{equation*}
Fig. (22) and (23) are about $\Delta \hat{r}$ vs $n$ for a specific $\ell$ and $\Delta \hat{r}$ vs $\ell$ for a specific $n$ and Fig. (24) is for different states.
\begin{figure}[h!]
    \centering
    \includegraphics[width=0.5\textwidth]{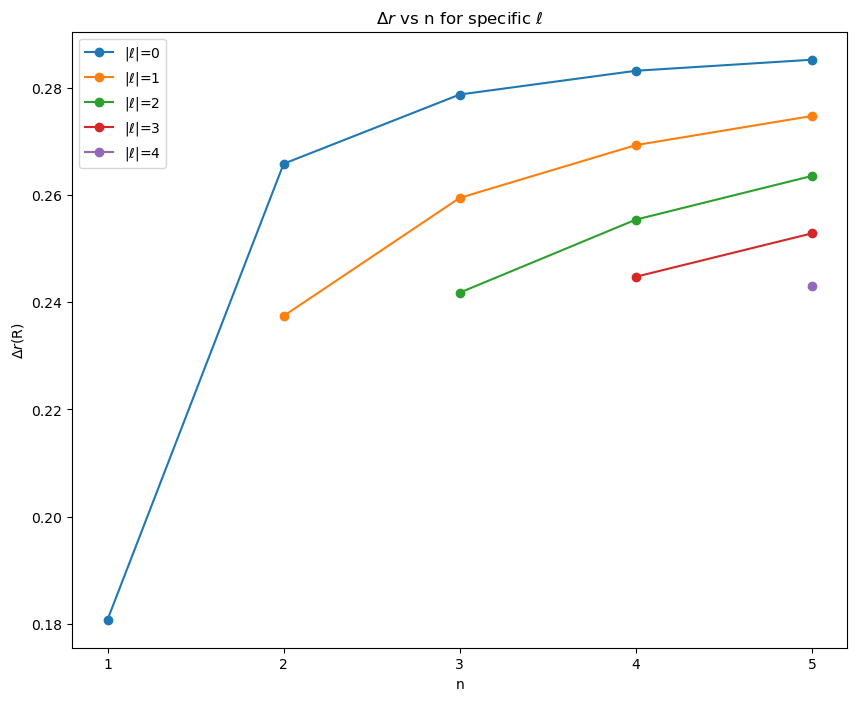}
    \caption{$\Delta \hat{r}$ vs $n$ for a specific $\ell$ of ISW}
\end{figure}
\begin{figure}[h!]
    \centering
    \includegraphics[width=0.5\textwidth]{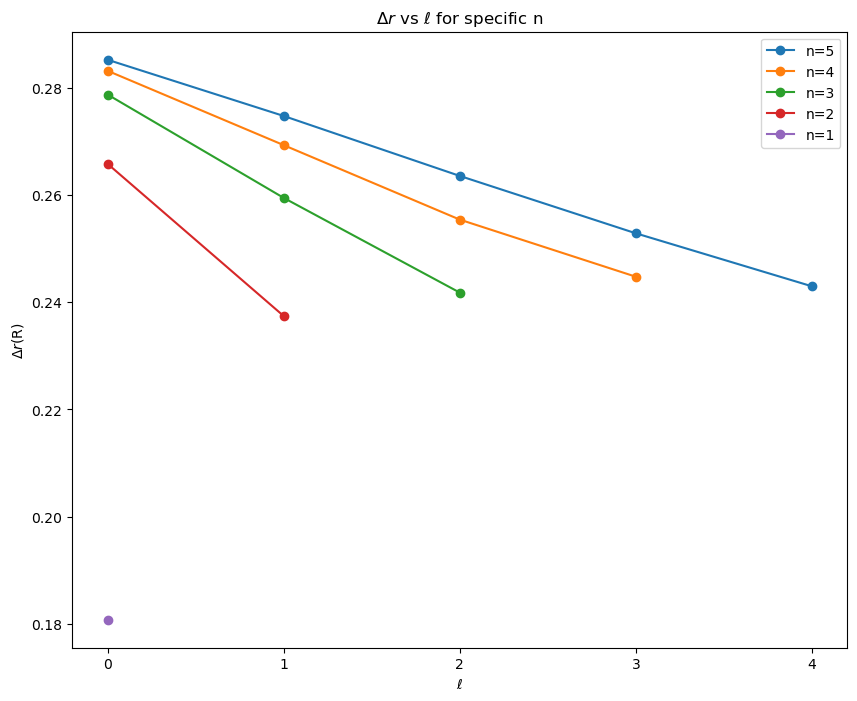}
    \caption{$\Delta \hat{r}$ vs $\ell$ for a specific $n$ of ISW}
\end{figure}
\begin{figure}[h!]
    \centering
    \includegraphics[width=0.5\textwidth]{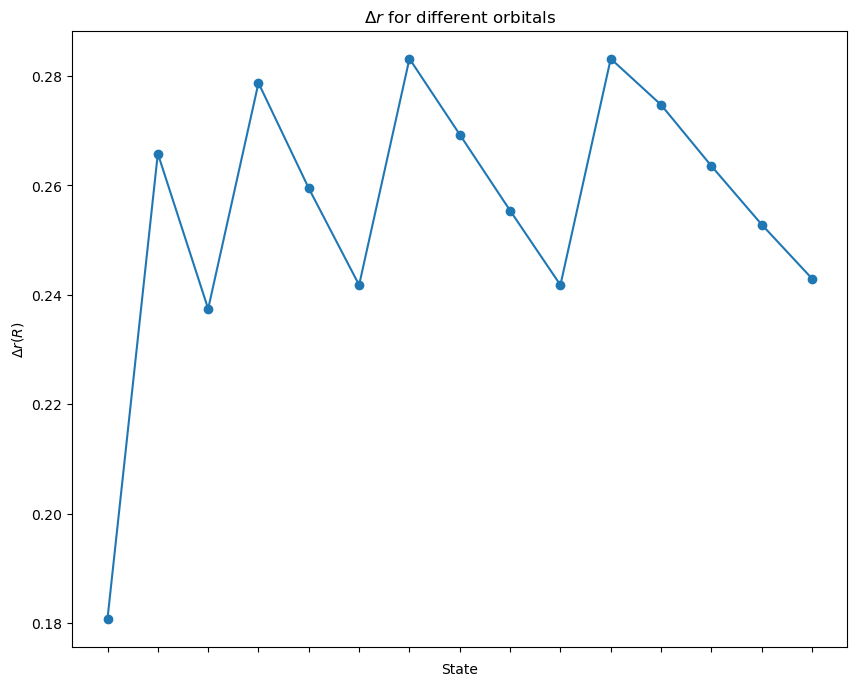}
    \caption{$\Delta \hat{r}$ for different states of ISW}
\end{figure}
We already discuss about the relative dispersion or coefficient of variation in the measurement of radial position is defined as, $\sigma_r=\frac{\Delta \hat{r}}{\langle \hat{r} \rangle}$ i.e. the ratio of uncertainty and expectation of radial position. The plots of $\sigma_r$ follow the trend of corresponding plots of $\Delta \hat{r}$, then we skip those three plots here. We have tabulated the value of $\langle \hat{r} \rangle$, $\Delta \hat{r}$ and $\sigma_r$ for different states in Table VIII.
\begin{table}[htb]
    \centering
    \begin{ruledtabular}
    \begin{tabular}{ccccc}
        \textrm{\((n,\ell)\)} & 
        \textrm{\(z_{n\ell}\)} & 
        \textrm{\(\langle \hat{r} \rangle(R)\)} & 
        \textrm{\(\Delta \hat{r}(R)\)} & 
        \textrm{\(\sigma_r\)} \\
        \colrule
        (1,0) & \(3.14159\) & \(0.50000\) & \(0.18076\) & \(0.36152\) \\
        \colrule
        (2,0) & \(6.28319\) & \(0.50000\) & \(0.26583\) & \(0.53166\) \\
        (2,1) & \(7.72525\) & \(0.53937\) & \(0.23473\) & \(0.44019\) \\
        \colrule
        (3,0) & \(9.42478\) & \(0.50000\) & \(0.27875\) & \(0.55750\) \\
        (3,1) & \(10.9041\) & \(0.53254\) & \(0.25947\) & \(0.49655\) \\
        (3,2) & \(12.3229\) & \(0.54582\) & \(0.24179\) & \(0.44298\) \\
        \colrule
        (4,0) & \(12.5664\) & \(0.50000\) & \(0.28318\) & \(0.56636\) \\
        (4,1) & \(14.0662\) & \(0.51479\) & \(0.26931\) & \(0.52314\) \\
        (4,2) & \(15.5146\) & \(0.53164\) & \(0.25541\) & \(0.48042\) \\
        (4,3) & \(16.9236\) & \(0.54827\) & \(0.24271\) & \(0.44268\) \\
        \colrule
        (5,0) & \(15.70796\) & \(0.50000\) & \(0.28524\) & \(0.57048\) \\
        (5,1) & \(17.2208\) & \(0.51054\) & \(0.27475\) & \(0.53815\) \\
        (5,2) & \(18.6890\) & \(0.52334\) & \(0.26356\) & \(0.50361\) \\
        (5,3) & \(20.1218\) & \(0.53658\) & \(0.25285\) & \(0.47122\) \\
        (5,4) & \(21.5254\) & \(0.54955\) & \(0.24296\) & \(0.44211\) \\
    \end{tabular}
    \end{ruledtabular}
    \caption{ $\langle \hat{r} \rangle$, $\Delta \hat{r}$ and $\sigma_r$ for different states}
\end{table}

\subsection{Uncertainty in Radial momentum}
We see that the radial momentum operator is given by,
\begin{equation*}
    \hat p_r=-i\hbar\frac{1}{r}\frac{\partial}{\partial{r}}[r]=-i\hbar\left(\frac{\partial}{\partial{r}}+\frac{1}{r}\right)
\end{equation*}
As per equ. (7),
\begin{equation*}
\begin{alignedat}{2}
    &\langle \hat{p_{r}} \rangle\\
    &=(-i\hbar)\int_{0}^{R} r^2R_{n\ell}^*(r)\left(\frac{\partial}{\partial{r}}+\frac{1}{r}\right)R_{n\ell}(r) dr\\
    &=(-i\hbar)\left[\int_0^R r^2R_{n\ell}^*(r)dR_{n\ell}(r)+\int_0^R r|R_{n\ell}(r)|^2dr\right]\\
    &=(-i\hbar)\left(\frac{R}{z_{n\ell}}\right)^2 \left[\int_0^{z_{n\ell}}{\rho}^2R_{n\ell}^*(\rho)dR_{n\ell}(\rho)+\int_0^{z_{n\ell}} \rho |R_{n\ell}(\rho)|^2d\rho\right]\\
\end{alignedat}
\end{equation*}
We have the radial wave function is $R_{n\ell}(\rho)=N_{n\ell}j_{\ell}(\rho)$ which gives $dR_{n\ell}(\rho)=N_{n\ell}dj_{\ell}(\rho)$.
\begin{equation*}
\begin{alignedat}{2}
    &\langle \hat p_r \rangle\\
    &=(-i\hbar)\frac{2}{R}\left[\int_0^{z_{n\ell}} {\rho}^2j_{\ell}(\rho)dj_{\ell}(\rho)+\int_0^{z_{n\ell}} \rho [j_{\ell}(\rho)]^2d\rho\right]\\ 
    &=(-i\hbar)\frac{1}{R}\int_0^{z_{n\ell}} d[{\rho}^2 [j_{\ell}(\rho)]^2]=0\text{ for } j_{\ell}(z_{n\ell})=0\\
\end{alignedat}
\end{equation*}
Therefore the expectation value of radial momentum is zero, $\langle \hat p_r \rangle=0$ as expected. Now our aim is $\langle   \hat p_r^2  \rangle$, for that we will use Virial theorem. The total energy of Infinite spherical well, $E_{n\ell m}=E_{n\ell}=z_{n\ell}^2\frac{{\hbar}^2}{2mR^2}$ and the total momentum is defined as $p_{n\ell m}=\sqrt{p_r^2+p_{\ell}^2}$ where $p_r$ is radial momentum and $p_\ell$ is orbital momentum. We know that, orbital angular momentum is defined as $L=p_\ell r=\sqrt{\ell(\ell+1)}\hbar$, then  $p_{\ell}=\frac{\sqrt{\ell(\ell+1)}}{r}\hbar$. Now from the momentum-kinetic energy relation $\langle E_{n\ell m}=T=\frac{p_{n\ell m}^2}{2m}\rangle$ as $V=0$ and $\langle z_{n\ell}^2\frac{{\hbar}^2}{R^2}=p_r^2+\frac{\ell(\ell+1)}{r^2}{\hbar}^2\rangle$ gives
\begin{equation}
    \langle \hat p_r^2  \rangle=z_{n\ell}^2\frac{{\hbar}^2}{R^2}-\ell(\ell+1){\hbar}^2\langle   \frac{1}{\hat{r}^2}  \rangle
\end{equation}
\begin{equation*}
\begin{alignedat}{2}
    &\langle   \frac{1}{\hat{r}^2}  \rangle=\int_0^{R}|R_{n\ell}|^2dr=\frac{R}{z_{n\ell}}\int_0^{z_{n\ell}}|R_{n\ell}(\rho)|^2\,d\rho\\  
    &=\frac{2}{R^2}\frac{1}{z_{n\ell}|j_{\ell+1}(z_{n\ell})|^2}\int_0^{z_{n\ell}}[j_{\ell}(\rho)]^2d\rho
\end{alignedat}
\end{equation*}
Therefore
\begin{equation*}
    \langle   \hat{p_r}^2 \rangle=\frac{{\hbar}^2}{R^2}\left[z_{n\ell}^2-\frac{2\ell(\ell+1)}{z_{n\ell}|j_{\ell+1}(z_{n\ell})|^2}\int_0^{z_{n\ell}}[j_{\ell}(\rho)]^2d\rho\right]
\end{equation*}
where using Wolfram Mathematica, 
\begin{equation}
\begin{alignedat}{2}
    &\int_0^{z_{n\ell}}[j_{\ell}(\rho)]^2d\rho=2^{-2\ell}z_{n\ell}^{1+2l}\pi \frac{1}{(2\ell+1)^3\Gamma(\ell+\frac{1}{2})^2}\\
    &{}_{2}F_{3}\left[\ell+\frac{1}{2},\ell+1;\ell+\frac{3}{2},\ell+\frac{3}{2},2\ell+2;-z_{n\ell}^2\right]\\
\end{alignedat}
\end{equation}
The uncertainty in radial momentum will be,
\begin{equation*}
\Delta \hat p_r=\frac{{\hbar}}{R}z_{n\ell}\sqrt{1-D(\ell,z_{n\ell})}
\end{equation*}
where $D(\ell,z_{n\ell})$ is defined as
\begin{equation}
\begin{alignedat}{2}
    &D(\ell,z_{n\ell})=\frac{2^{-2\ell+1}z_{n\ell}^2}{|j_{\ell+1}(z_{n\ell})|^2} \frac{\pi \ell(\ell+1)}{(2\ell+1)^3\Gamma(\ell+\frac{1}{2})^2}\\
    &{}_{2}F_{3}\left[\ell+\frac{1}{2},\ell+1;\ell+\frac{3}{2},\ell+\frac{3}{2},2\ell+2;-z_{n\ell}^2\right]\\
\end{alignedat}
\end{equation}
Fig. (25) and (26) are about $\Delta \hat p_r$ vs $n$ for a specific $\ell$ and $\Delta \hat p_r$ vs $\ell$ for a specific $n$ and Fig. (27) is for different states.
\begin{figure}[h!]
    \centering
    \includegraphics[width=0.5\textwidth]{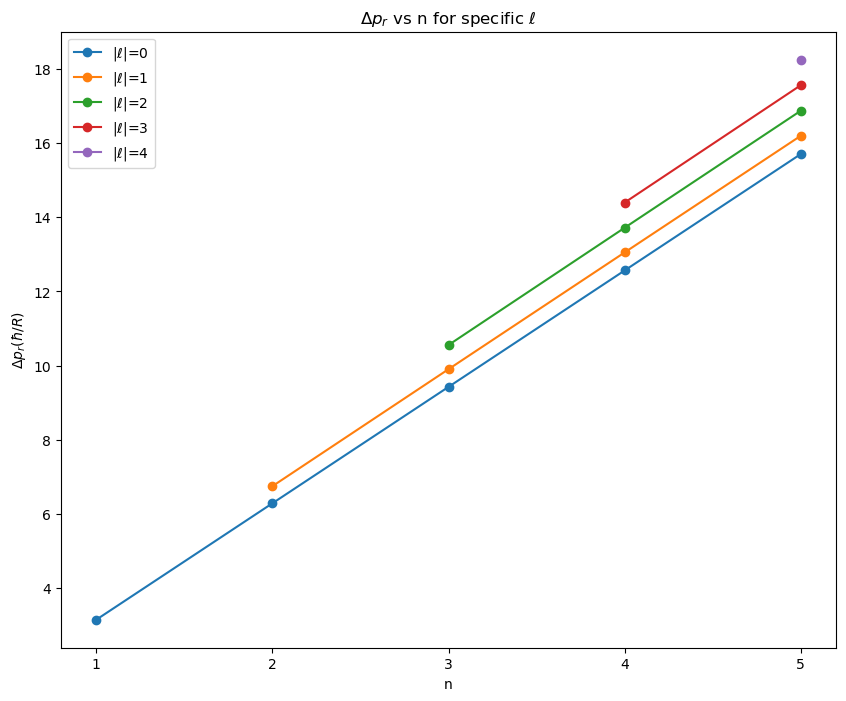}
    \caption{$\Delta \hat p_r$ vs $n$ for a specific $\ell$ of ISW}
\end{figure}
\begin{figure}[h!]
    \centering
    \includegraphics[width=0.5\textwidth]{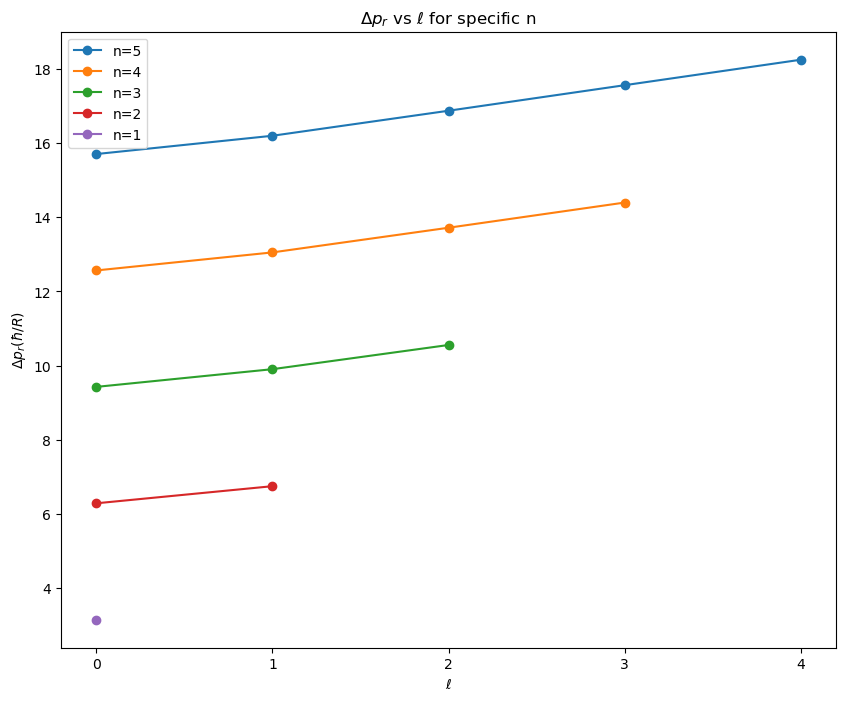}
    \caption{$\Delta \hat p_r$ vs $\ell$ for a specific $n$ of ISW}
\end{figure}
\begin{figure}[h!]
    \centering
    \includegraphics[width=0.5\textwidth]{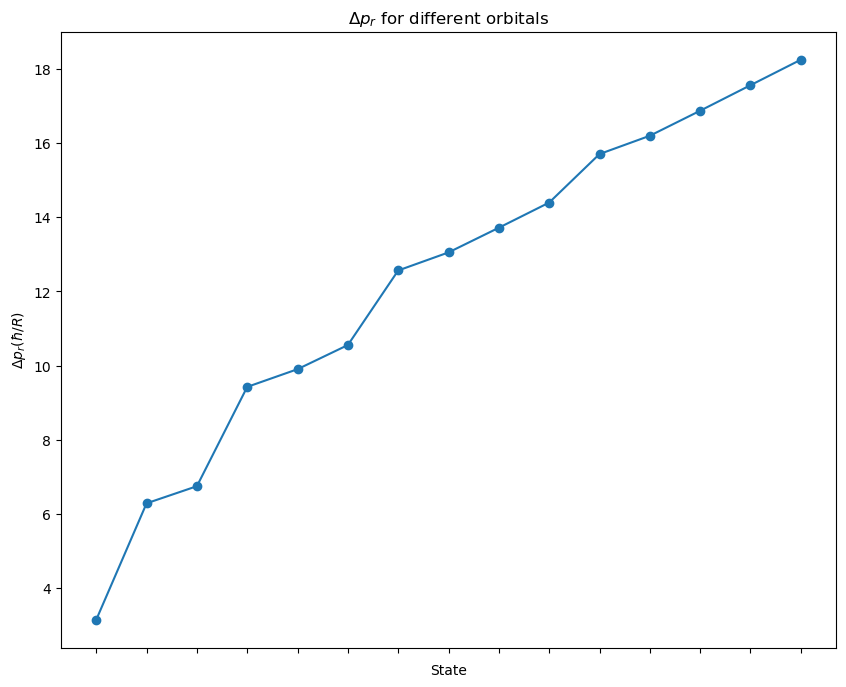}
    \caption{$\Delta \hat p_r$ for different states of ISW}
\end{figure}
\subsection{Radial uncertainty product}
The radial uncertainty product in terms of $A(\ell,z_{n\ell})$, $B(\ell,z_{n\ell})$ and $D(\ell,z_{n\ell})$ is given by
\begin{equation*}
    \Delta \hat{r} \Delta \hat p_r = z_{n\ell} \hbar \sqrt{B(\ell,z_{n\ell})-A(\ell,z_{n\ell})^2} \sqrt{1-D(\ell,z_{n\ell})}
\end{equation*}
where $A(\ell,z_{n\ell})$, $B(\ell,z_{n\ell})$ and $D(\ell,z_{n\ell})$ are defined in equ. (57), equ. (60) and equ. (63) respectively. The uncertainties in radial position and momentum and their product for different states are tabulated in Table IX.
\begin{table}[htb]
    \centering
    \begin{ruledtabular}
    \begin{tabular}{ccccc}
        \textrm{$(n,\ell)$} & \textrm{$z_{n\ell}$} & \textrm{$\Delta \hat{r}(R)$} & \textrm{$\Delta \hat p_r (\frac{\hbar}{R})$} & \textrm{$\Delta \hat{r} \Delta \hat p_r (\hbar)$} \\
        \colrule
        (1,0) & 3.14159 & 0.18076 & 3.14159 & 0.5679 \\
        \colrule
        (2,0) & 6.28319 & 0.26583 & 6.28319 & 1.6703 \\
        (2,1) & 7.72525 & 0.23743 & 6.7449 & 1.6014 \\
        \colrule
        (3,0) & 9.42478 & 0.27875 & 9.42478 & 2.6271 \\
        (3,1) & 10.9041 & 0.25947 & 9.90247 & 2.5694 \\
        (3,2) & 12.3229 & 0.24179 & 10.55537 & 2.5522 \\
        \colrule
        (4,0) & 12.5664 & 0.28318 & 12.56638 & 3.5585 \\
        (4,1) & 14.0662 & 0.26931 & 13.05358 & 3.5154 \\
        (4,2) & 15.5146 & 0.25541 & 13.71948 & 3.5041 \\
        (4,3) & 16.9236 & 0.24271 & 14.39885 & 3.4947 \\
        \colrule
        (5,0) & 15.70796 & 0.28524 & 15.70795 & 4.4805 \\
        (5,1) & 17.2208 & 0.27475 & 16.20148 & 4.4513 \\
        (5,2) & 18.6890 & 0.26356 & 16.87693 & 4.4481 \\
        (5,3) & 20.1218 & 0.25285 & 17.56596 & 4.4415 \\
        (5,4) & 21.5254 & 0.24296 & 18.25349 & 4.4349 \\
    \end{tabular}
    \end{ruledtabular}
    \caption{$\Delta \hat{r}$, $\Delta \hat p_r$ and $\Delta \hat{r}\Delta \hat p_r$ for different states}
\end{table}
Fig. (28) and (29) are about $\Delta \hat{r}\Delta \hat p_r$ vs $n$ for a specific $\ell$ and $\Delta \hat{r}\Delta \hat p_r$ vs $\ell$ for a specific $n$ and Fig. (30) is for different states.
\begin{figure}[h!]
    \centering
    \includegraphics[width=0.5\textwidth]{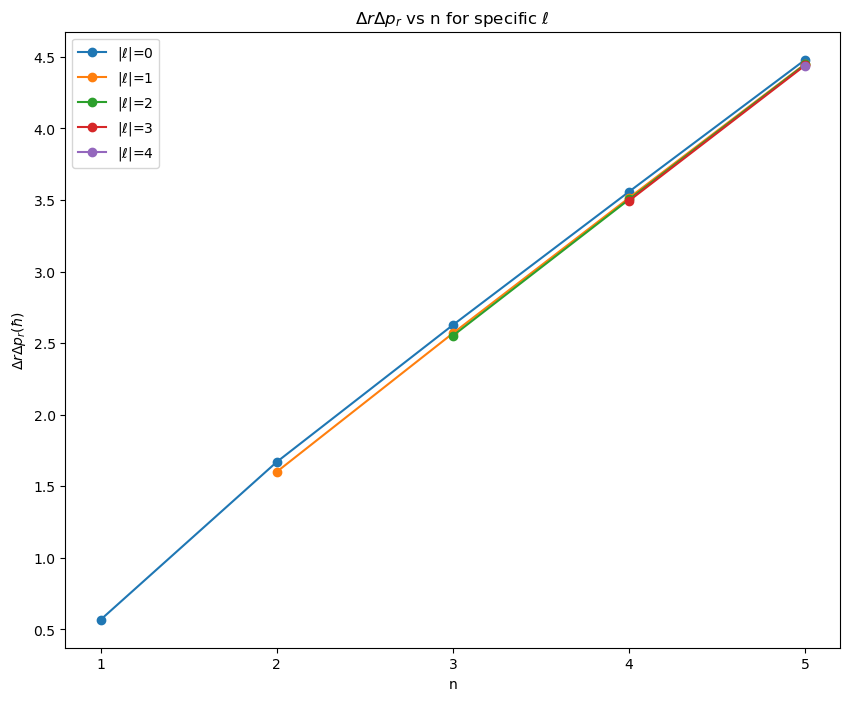}
    \caption{$\Delta \hat{r}\Delta \hat p_r$ vs $n$ for a specific $\ell$ of ISW}
\end{figure}
\begin{figure}[h!]
    \centering
    \includegraphics[width=0.5\textwidth]{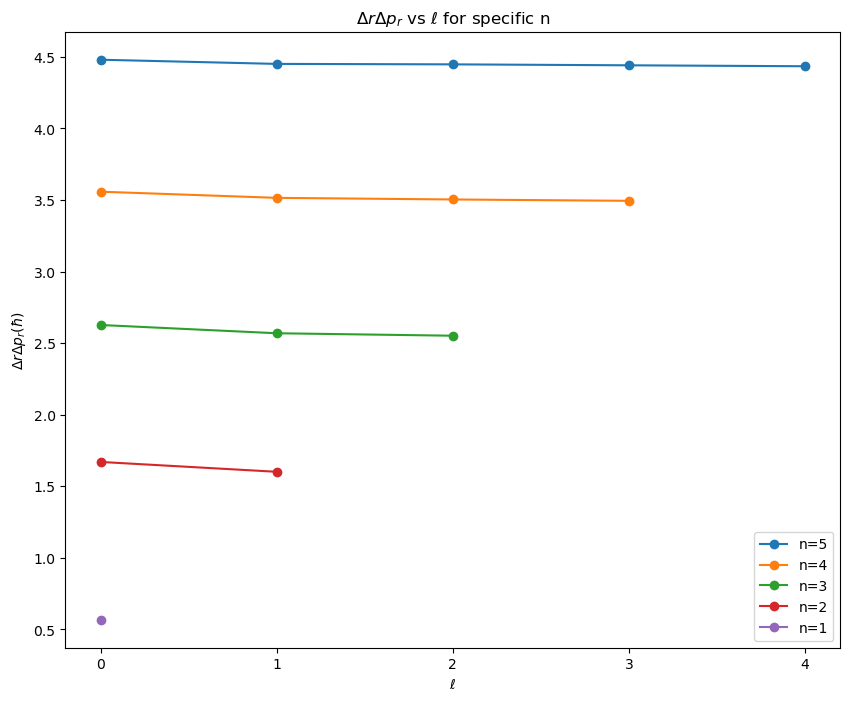}
    \caption{$\Delta \hat{r}\Delta \hat p_r$ vs $\ell$ for a specific $n$ of ISW}
\end{figure}
\begin{figure}[h!]
    \centering
    \includegraphics[width=0.5\textwidth]{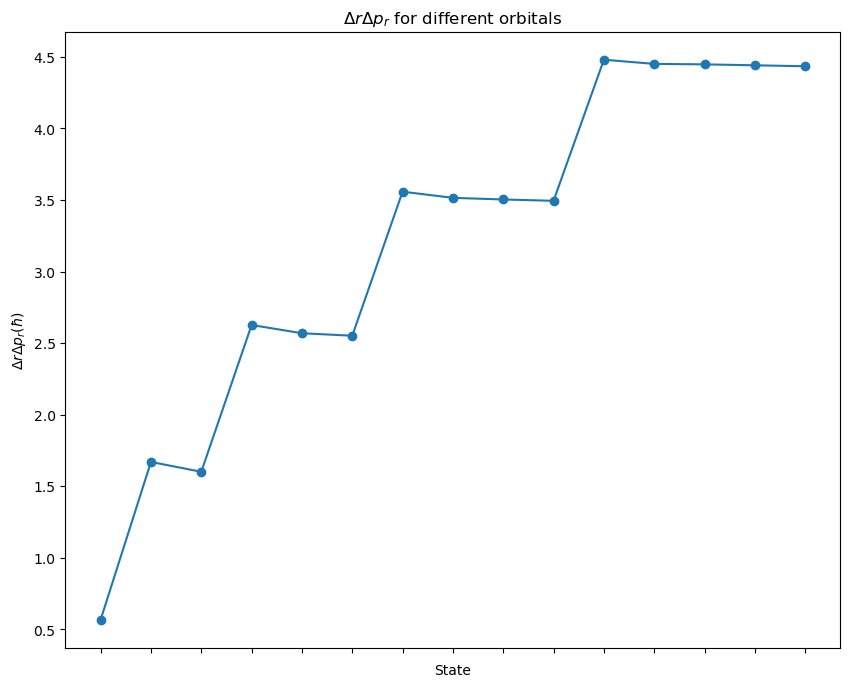}
    \caption{$\Delta \hat{r}\Delta \hat p_r$ for different states of ISW}
\end{figure}
\subsection{Ground state of Infinite Spherical Well}
The radial wave function of Hydrogen atom is given by,
\begin{equation*}
    R_{n\ell}(r)=\sqrt{\frac{2}{R^3}}\frac{1}{|j_{\ell+1}(z_{n\ell})|}j_{\ell}\left(\frac{z_{n\ell}}{R}r\right)
\end{equation*}
For ground state ($n=1,\ell=0$), $R_{10}(r)=\sqrt{\frac{2}{R^3}}\frac{1}{|j_{1}(\pi)|}j_0(\frac{\pi r}{R})$ where $z_{10}=\pi$.
The radial probability density for ground state, $P_{10}(r)=r^2|R_{10}(r)|^2=N_{10}^2r^2[j_0( \frac{\pi r}{R})]^2$. For most probable distance from origin, $\frac{dP}{dr}=0$ gives $j_0( \frac{\pi r}{R})+ \frac{\pi r}{R}j_0^{'}( \frac{\pi r}{R})=0$ gives $r_{mp}=\frac{R}{2}$.
We have already, $\langle \hat{r} \rangle=\frac{R}{2}$ and $\Delta \hat{r}=0.18076R$. Therefore the most probable distance is nothing but the expectation value of radial position. The radial wave function of ground state and its probability density function plots are shown in Fig. (31).
\begin{figure}[h!]
    \centering
    \includegraphics[width=0.5\textwidth]{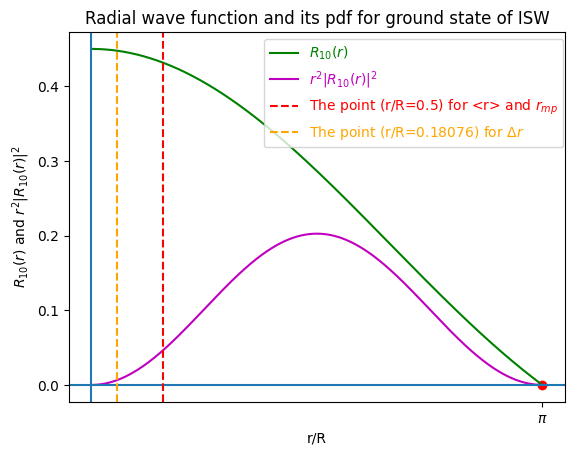}
    \caption{Ground state of ISW}
\end{figure}

\section{Spherical harmonic oscillator}
\subsection{Radial wave function}
The potential of spherical harmonic oscillator is given by 
\begin{equation}
    V(r)=\frac{1}{2}m{\omega}^2r^2
\end{equation}
where $\omega$ is the angular frequency. The time independent Schrodinger equation for this system,
\begin{equation}
    -\frac{{\hbar}^2}{2m}{\nabla}^2\psi(r,\theta,\phi)+\frac{1}{2}m{\omega}^2r^2\psi(r,\theta,\phi) = E\psi(r,\theta,\phi)
\end{equation}
And the corresponding well known radial part (Refs.~\cite{cahaya2022radial}) is given by,
\begin{equation}
    \frac{d^2R}{dr^2}+\frac{2}{r}\frac{dR}{dr}+\left({\alpha}^2r^2-\frac{\ell(\ell+1)}{r^2}+k^2\right)R=0
\end{equation}
where $\alpha=\frac{m\omega}{\hbar}$ and $k=\sqrt{\frac{2mE}{{\hbar}^2}}$ and $\ell$ is azimuthal quantum number which is from effective potential,
\begin{equation}
    V_{eff}(r)=\frac{1}{2}m{\omega}^2r^2+\frac{\ell(\ell+1){\hbar}^2}{2mr^2}
\end{equation}
The solution of equ. (66) is given by,
\begin{equation}
    R_{n_r \ell}(r)=N_{n_r \ell}(\sqrt{\alpha}r)^{\ell}e^{-\frac{\alpha r^2}{2}}L_{n_r}^{\ell+\frac{1}{2}}(\alpha r^2)
\end{equation}
A known fact for 3d harmonic oscillator, $n=2n_r+\ell$ where $n_r$, $n$ and $\ell$ are radial, principal and azimuthal quantum number respectively where the radial quantum number $n_r$ is specifically used for radial solution of Schrodinger equation and numbers of nodes in radial wave function. Meanwhile, the revised radial wave function (Refs.~\cite{cahaya2022radial}) can be written as,
\begin{equation}
     R_{n \ell}(r)=N_{n \ell}(\sqrt{\alpha}r)^{\ell}e^{-\frac{\alpha r^2}{2}}L_{\frac{1}{2}(n-\ell)}^{\ell+\frac{1}{2}}(\alpha r^2)
\end{equation}
Let introduce a dimensionless parameter, $\rho=\sqrt{\alpha}r$ and
\begin{equation}
     R_{n \ell}(\rho)=N_{n \ell}{\rho}^{\ell}e^{-\frac{{\rho}^2}{2}}L_{\frac{1}{2}(n-\ell)}^{\ell+\frac{1}{2}}({\rho}^2)
\end{equation}
where $N_{n\ell}$ is normalization constant, which can be determined by normalization condition and $L_{\frac{1}{2}(n-\ell)}^{\ell+\frac{1}{2}}$ is associated Laguerre function. Here the associated  Laguerre is of non-linear weight ${\rho}^2$, we can replace it by linear weight $\eta={\rho}^2$. Hence the fully revised radial wave function can be written as,
\begin{equation}
    R_{n\ell}(\eta)=N_{n\ell}{\eta}^{\frac{\ell}{2}}e^{-\frac{\eta}{2}}L_{\frac{1}{2}(n-\ell)}^{\ell+\frac{1}{2}}(\eta)
\end{equation}
One important result is $k$ can be expressed as $k=\sqrt{2(2n_r+\ell+\frac{3}{2}})\alpha$ (Refs.~\cite{ucsd3dHO}) that gives energy eigen values as, 
\begin{equation}
    E_{n_r \ell}=\left(2n_r+\ell+\frac{3}{2}\right)\hbar \omega 
\end{equation}
shows that energy eigen values can be also also written as
\begin{equation}
    E_n=\left(n+\frac{3}{2}\right)\hbar\omega
\end{equation}
which is a well known result for 3d harmonic quantum oscillator in Cartesian co-ordinate. Let make normalize the radial wave function. The normalization condition is given by $\int_0^{\infty}P(r)dr=1$ where $P(r)=r^2|R_{n\ell}(r)|^2$.
\begin{equation}
\begin{alignedat}{2}
    &\int_0^{\infty}r^2|R_{n\ell}(r)|^2dr\\
    &=\frac{1}{2\alpha\sqrt{\alpha}}\int_0^{\infty}\sqrt{\eta}|R_{n\ell}(\eta)|^2d\eta \text{ as $\eta=\alpha r^2$}\\
    &=\frac{{\alpha}^{-\frac{3}{2}}}{2}|N_{n\ell}|^2\int_0^{\infty} 
    {\eta}^{\ell+\frac{1}{2}}e^{-\eta}\left[L_{\frac{1}{2}(n-\ell)}^{\ell+\frac{1}{2}}(\eta)\right]^2d\eta\\
    &=\frac{{\alpha}^{-\frac{3}{2}}}{2}|N_{n\ell}|^2 \widetilde{I}
\end{alignedat}
\end{equation}
The orthonormal property of associated Laguerre function is given by as seen earlier in equ. (13),
\begin{equation*}
    \int_{0}^{\infty}z^a e^{-z}L^a_b(z)L^a_c(z)dz=\frac{\Gamma{(a+b+1)}}{\Gamma{(b+1)}}\delta_{bc}
\end{equation*}
After naming $a=\ell+\frac{1}{2}$ and $b=c=\frac{1}{2}(n-\ell)$, the above normalization integral becomes
\begin{equation}
\begin{alignedat}{2}
    &\widetilde{I}=\frac{{\Gamma(a+b+1)}}{\Gamma(b+1)}=\frac{\Gamma(\frac{1}{2}(n+\ell)+\frac{3}{2})}{(\frac{1}{2}(n-\ell))!}
\end{alignedat}
\end{equation}
This gives us the expression of normalization constant, is given by
\begin{equation}
    N_{n\ell}=\sqrt{\frac{2{\alpha}^{\frac{3}{2}}(\frac{1}{2}(n-\ell))!}{\Gamma(\frac{1}{2}(n+\ell)+\frac{3}{2})}}
\end{equation}
We know that for 3d harmonic oscillator problem, the azimuthal quantum number, $\ell$ can take values in way, (Refs.~\cite{Griffiths_2018}) $\ell=n,n-2,n-4,....,{\ell}_{min}$ where ${\ell}_{min}$ is defined as,
\begin{equation}
\ell_{\text{min}} =
\begin{cases}
    1 & \text{for } n \text{ odd} \\
    0 & \text{for } n \text{ even}
\end{cases}
\end{equation}
The examples be like, for $n=6$, $\ell=6,4,2,0$ and for $n=7$, $\ell=7,5,3,1$. From this idea, it's crystal clear that $n+\ell$ or $n-\ell$ is a double positive integer multiple.
It gives the introduction of $p=\frac{1}{2}(n+\ell)$ and $q=\frac{1}{2}(n-\ell)$ is both positive integer including zero and we can re-write the normalization integral as 
\begin{equation}
    \widetilde{I}=\frac{\Gamma\left(p+\frac{3}{2}\right)}{q!}
\end{equation}
For $r$ is positive integer, $\Gamma\left(r+\frac{1}{2}\right)=\frac{(2r)!}{2^r r!}\sqrt{\pi}$, then 
\begin{equation*}
    \Gamma\left(p+\frac{3}{2}\right)=\frac{(2p+2)!}{2^{p+1}(p+1)!}\sqrt{\pi}
\end{equation*}

\subsection{Uncertainty in Radial position}
As per equ. (1),
\begin{equation}
\begin{alignedat}{2}
    &\langle \hat{r} \rangle\\
    &=\frac{1}{2\alpha^2}\int_0^{\infty}\eta|R_{n\ell}(\eta)|^2d\eta\\
    &=\frac{1}{\sqrt{\alpha}}\frac{(\frac{1}{2}(n-\ell))!}{\Gamma(\frac{1}{2}(n+\ell)+\frac{3}{2})}\int_0^{\infty}{\eta}^{\ell+1}e^{-\eta}\left[L_{\frac{1}{2}(n-\ell)}^{\ell+\frac{1}{2}}(\eta)\right]^2d\eta\\
    &=\sqrt{\frac{\hbar}{m\omega}}\widetilde{C}_{n\ell}\widetilde{I}_1\\
\end{alignedat}
\end{equation}
where $\widetilde{C}_{n\ell}\widetilde{I}=1$. The azimuthal quantum number $\ell$ takes values as $\ell=n-2x$ where the possible values of $x$ will be 0,1,2,3..... and the integral $\widetilde{I}_1$ becomes
\begin{equation}
    \widetilde{I}_1(x)=\int_0^{\infty}{\eta}^{n-2x+1}e^{-\eta}\left[L_{x}^{n-2x+\frac{1}{2}}(\eta)\right]^2d\eta
\end{equation}
The evaluation of this general integral is a little awkward using Wolfram Mathematica software, but we can evaluate this integrals for a particular value of $x$.
\begin{equation*}
    \widetilde{I}_1(x=0)=\Gamma(2+n)
\end{equation*}
\begin{equation*}
    \widetilde{I}_1(x=1)=\left(\frac{1}{4}+n\right)\Gamma(n)
\end{equation*}
\begin{equation*}
    \widetilde{I}_1(x=2)=\frac{1}{64}[33+16n(-5+2n)]\Gamma(-2+n)
\end{equation*}
\begin{equation*}
    \widetilde{I}_1(x=3)=\frac{1}{768}[-1965+4n(667+8n(-33+4n))]\Gamma(-4+n)
\end{equation*}
Fig. (32) and (33) are about $\langle \hat{r} \rangle$ vs $n$ for a specific $\ell$ and $\langle \hat{r} \rangle$ vs $\ell$ for a specific $n$ and Fig. (34) is for different states.
\begin{figure}[h!]
    \centering
    \includegraphics[width=0.5\textwidth]{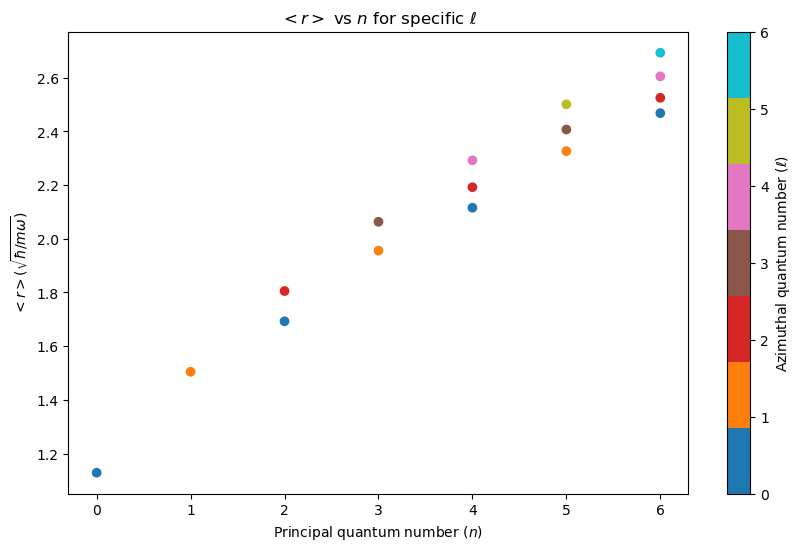}
    \caption{$\langle \hat{r} \rangle$ vs $n$ for a specific $\ell$ of SHO}
\end{figure}
\begin{figure}[h!]
    \centering
    \includegraphics[width=0.5\textwidth]{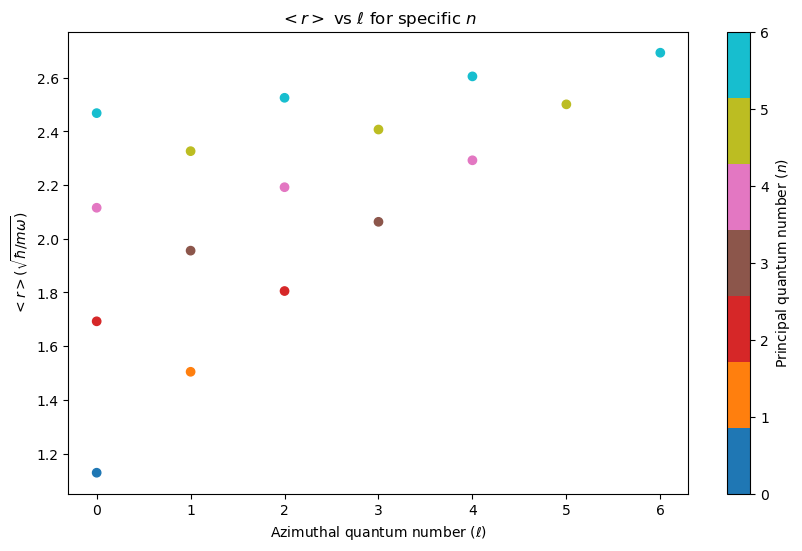}
    \caption{$\langle \hat{r} \rangle$ vs $\ell$ for a specific $n$ of SHO}
\end{figure}
\begin{figure}[h!]
    \centering
    \includegraphics[width=0.5\textwidth]{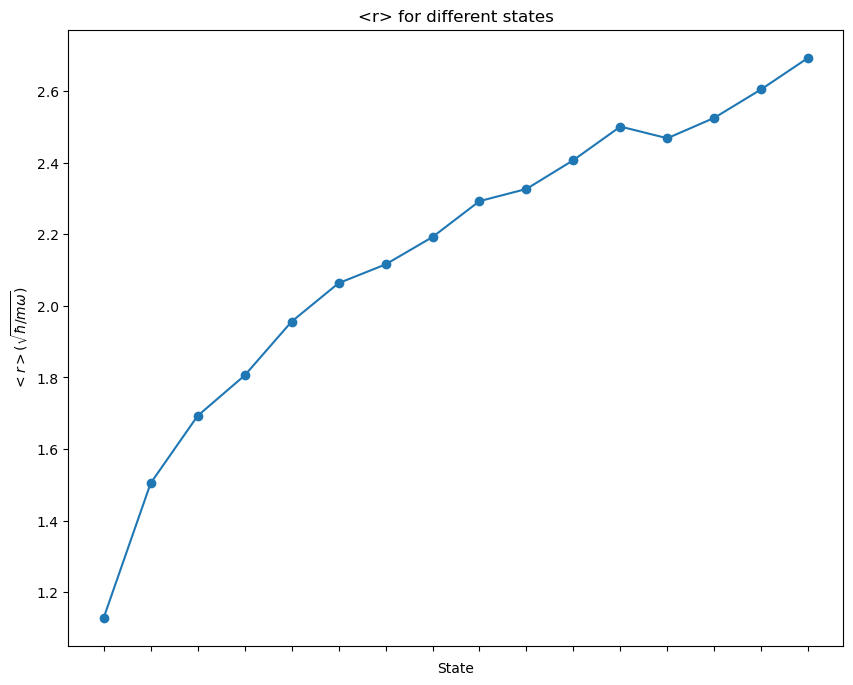}
    \caption{$\langle \hat{r} \rangle$ for different states of SHO}
\end{figure}
Now our aim is to evaluate $\langle \hat{r}^2 \rangle$, as from equ. (2),
\begin{equation}
\begin{alignedat}{2}
    &\langle \hat{r}^2 \rangle\\
    &=\frac{1}{2\alpha^{\frac{5}{2}}}\int_0^{\infty}\eta^{\frac{3}{2}}|R_{n\ell}(\eta)|^2d\eta\\
    &=\frac{1}{2{\alpha}^{\frac{5}{2}}}|N_{n\ell}|^2\int_0^{\infty}\eta.{\eta}^{\ell+\frac{1}{2}}e^{-\eta}\left[L_{\frac{1}{2}(n-\ell)}^{\ell+\frac{1}{2}}(\eta)\right]^2d\eta\\
    &=\frac{1}{2{\alpha}^{\frac{5}{2}}}|N_{n\ell}|^2\widetilde{I}_2
\end{alignedat}
\end{equation}
Using our nomenclature, the integral $\widetilde{I}_2$ becomes,
\begin{equation}
    \widetilde{I}_2=\int_0^{\infty}z.z^ae^{-z}[L_b^a(z)]^2dz
\end{equation}
Using the recursive property of equ. (14),
\begin{equation}
\begin{alignedat}{2}
    &zL^a_b(z)=(a+2b+1)L_b^a(z)\\
    &+\text{other terms contain $L_c^a(z)$ where $c\neq b$}\\  
\end{alignedat}
\end{equation}
Then
\begin{equation*}
\begin{alignedat}{2}
    &\widetilde{I}_2=\int_0^{\infty}(a+2b+1)z^ae^{-z}[L_b^a(z)]^2dz\\
    &=(a+2b+1)\frac{\Gamma(a+b+1)}{\Gamma(b+1)}\\
\end{alignedat}
\end{equation*}
Here $a+2b+1=n+\frac{3}{2}$,
    $a+b+1=\frac{n+\ell+3}{2}$ and $b+1=\frac{1}{2}(n-\ell)+1$, which gives us 
\begin{equation}
    \widetilde{I}_2=\left(n+\frac{3}{2}\right)\widetilde{I}
\end{equation}
and using the relation $|N_{n\ell}|^2{\widetilde{I}=2\alpha^{\frac{3}{2}}}$, we have
\begin{equation}
    \langle \hat{r}^2 \rangle=\frac{1}{\alpha}\left(n+\frac{3}{2}\right)=\left(n+\frac{3}{2}\right)\frac{\hbar}{m\omega}
\end{equation}
There is also a fascinating way to find integral $\langle \hat{r}^2 \rangle$ by energy calculation. The Virial theorem states that for $V(r)\propto r^n$, $\langle T  \rangle=\frac{n}{2}\langle V  \rangle$. For harmonic potential, $V(r)\propto r^2$ and $\langle T  \rangle=\langle V  \rangle$ that states that $\langle E  \rangle=\langle T  \rangle+\langle V  \rangle=2\langle V  \rangle$. For Spherical harmonic oscillator, $V=\frac{1}{2}m{\omega}^2r^2$ and $E=(n+\frac{3}{2})\hbar\omega$. And we have got a direct result of equ. (85). Hence the uncertainty in radial position will be,
\begin{equation}
    \Delta \hat{r}=\sqrt{\frac{\hbar}{m\omega}}\sqrt{\left(n+\frac{3}{2}-\widetilde{C}_{n\ell}^2\widetilde{I}_1^2\right)}
\end{equation}
where
\begin{equation*}
    \widetilde{C}_{n\ell}=\frac{(\frac{1}{2}(n-\ell))!}{\Gamma(\frac{1}{2}(n+\ell)+\frac{3}{2})}
\end{equation*}
and
\begin{equation*}
    \widetilde{I}_1=\int_0^{\infty}{\eta}^{\ell+1}e^{-\eta}\left[L_{\frac{1}{2}(n-\ell)}^{\ell+\frac{1}{2}}(\eta)\right]^2d\eta
\end{equation*}
Fig. (35) and (36) are about $\Delta \hat{r}$ vs $n$ for a specific $\ell$ and $\Delta \hat{r}$ vs $\ell$ for a specific $n$ and Fig. (37) is for different states. We have tabulated the value of $\langle \hat{r} \rangle$, $\Delta \hat{r}$ and $\sigma_r$ for different states in Table X. Fig. (38) and (39) are about $\sigma_r$ vs $n$ for a specific $\ell$ and $\sigma_r$ vs $\ell$ for a specific $n$ and Fig. (40) is for different states.
\begin{figure}[h!]
    \centering
    \includegraphics[width=0.5\textwidth]{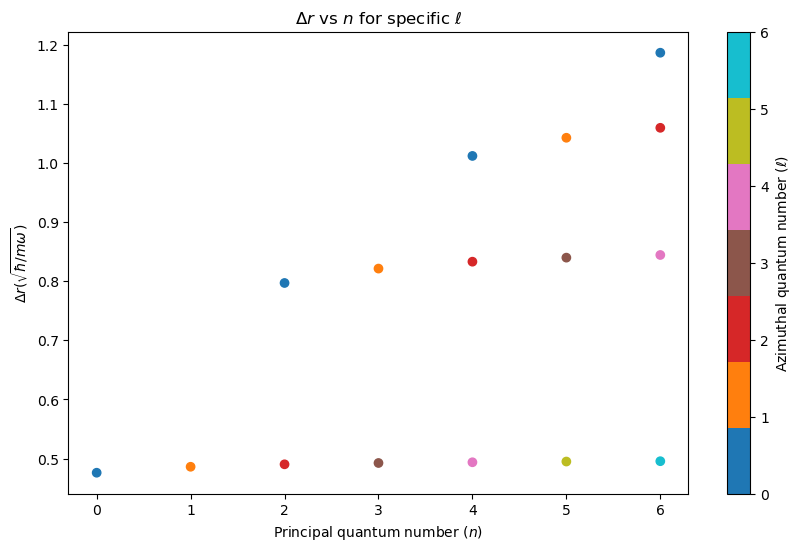}
    \caption{$\Delta \hat{r}$ vs $n$ for a specific $\ell$ of SHO}
\end{figure}
\begin{figure}[h!]
    \centering
    \includegraphics[width=0.5\textwidth]{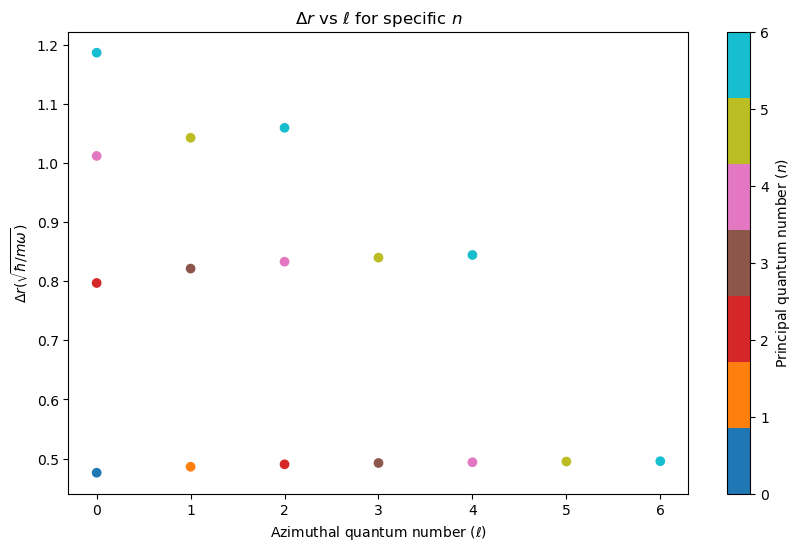}
    \caption{$\Delta \hat{r}$ vs $\ell$ for a specific $n$ of SHO}
\end{figure}
\begin{figure}[h!]
    \centering
    \includegraphics[width=0.5\textwidth]{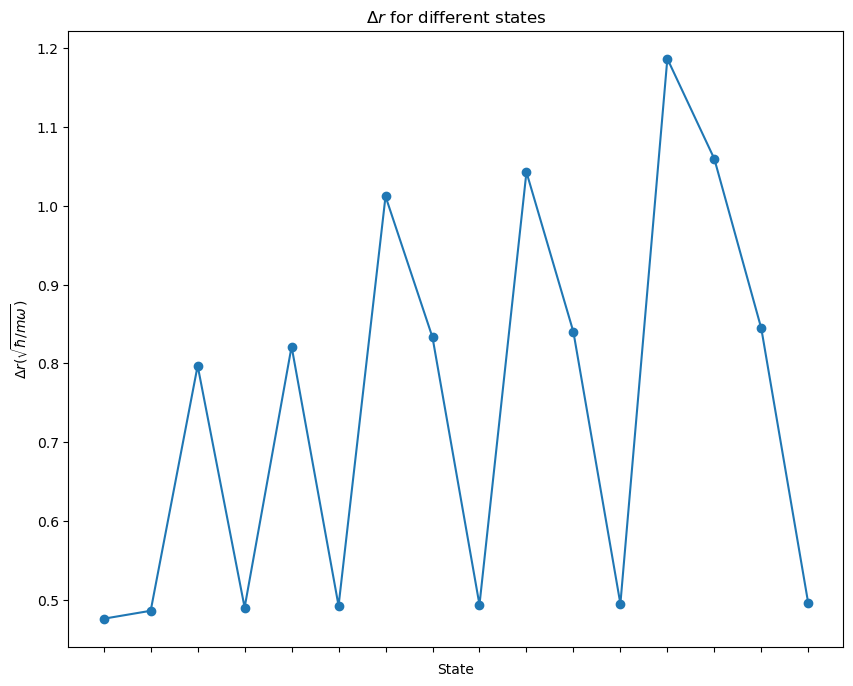}
    \caption{$\Delta \hat{r}$ for different states of SHO}
\end{figure}
\begin{table}[htb]
    \centering
    \begin{ruledtabular}
    \begin{tabular}{lccccc}
        \textrm{ ($n,\ell$)} & 
        \textrm{\(\widetilde{C}_{n\ell}\)} & 
        \textrm{\(\widetilde{I}_1\)} & 
        \textrm{\(\langle \hat{r} \rangle \, (\sqrt{\hbar / m \omega})\)} & 
        \textrm{\(\Delta \hat{r} \, (\sqrt{\hbar / m \omega})\)} & 
        \textrm{\(\sigma_r\)} \\
        \colrule
        \((0,0) \) & \( \frac{2}{\sqrt{\pi}} \) & 1 & \( \frac{2}{\sqrt{\pi}} \) & \( \sqrt{\frac{3}{2} - \frac{4}{\pi}} \) & 0.42201 \\
        \colrule
        \((1,1) \) & \( \frac{4}{3\sqrt{\pi}} \) & 2 & \( \frac{8}{3\sqrt{\pi}} \) & \( \sqrt{\frac{5}{2} - \frac{64}{9\pi}} \) & 0.32321 \\
        \colrule
        \((2,0) \) & \( \frac{4}{3\sqrt{\pi}} \) & \( \frac{9}{4} \) & \( \frac{3}{\sqrt{\pi}} \) & \( \sqrt{\frac{7}{2} - \frac{9}{\pi}} \) & 0.47088 \\
        \( (2,2) \) & \( \frac{8}{15\sqrt{\pi}} \) & 6 & \( \frac{16}{5\sqrt{\pi}} \) & \( \sqrt{\frac{7}{2} - \frac{256}{25\pi}} \) & 0.27163 \\
        \colrule
        \( (3,1) \) & \( \frac{8}{15\sqrt{\pi}} \) & \( \frac{13}{2} \) & \( \frac{52}{15\sqrt{\pi}} \) & \( \sqrt{\frac{9}{2} - \frac{2704}{225\pi}} \) & 0.41994 \\
        \( (3,3) \) & \( \frac{16}{105\sqrt{\pi}} \) & 24 & \( \frac{128}{35\sqrt{\pi}} \) & \( \sqrt{\frac{9}{2} - \frac{16384}{1225\pi}} \) & 0.23876 \\
        \colrule
        \( (4,0) \) & \( \frac{16}{15\sqrt{\pi}} \) & \( \frac{225}{64} \) & \( \frac{15}{4\sqrt{\pi}} \) & \( \sqrt{\frac{11}{2} - \frac{225}{16\pi}} \) & 0.47824 \\
        \( (4,2) \) & \( \frac{16}{105\sqrt{\pi}} \) & \( \frac{51}{2} \) & \( \frac{136}{35\sqrt{\pi}} \) & \( \sqrt{\frac{11}{2} - \frac{18496}{1225\pi}} \) & 0.37997 \\
        \( (4,4) \) & \( \frac{32}{945\sqrt{\pi}} \) & 120 & \( \frac{256}{63\sqrt{\pi}} \) & \( \sqrt{\frac{11}{2} - \frac{65536}{3969\pi}} \) & 0.21542 \\
        \colrule
        \( (5,1) \) & \( \frac{32}{105\sqrt{\pi}} \) & \( \frac{433}{32} \) & \( \frac{433}{105\sqrt{\pi}} \) & \( \sqrt{\frac{13}{2} - \frac{187489}{11025\pi}} \) & 0.44809 \\
        \( (5,3) \) & \( \frac{32}{945\sqrt{\pi}} \) & 126 & \( \frac{64}{15\sqrt{\pi}} \) & \( \sqrt{\frac{13}{2} - \frac{4096}{225\pi}} \) & 0.34888 \\
        \( (5,5) \) & \( \frac{64}{10395\sqrt{\pi}} \) & 720 & \( \frac{1024}{231\sqrt{\pi}} \) & \( \sqrt{\frac{13}{2} - \frac{1048576}{53362\pi}} \) & 0.19796 \\
        \colrule
        \( (6,0) \) & \( \frac{32}{35\sqrt{\pi}} \) & \( \frac{1225}{256} \) & \( \frac{35}{8\sqrt{\pi}} \) & \( \sqrt{\frac{15}{2} - \frac{1225}{64\pi}} \) & 0.48061 \\
        \( (6,2) \) & \( \frac{64}{9455\sqrt{\pi}} \) & \( \frac{2115}{32} \) & \( \frac{94}{21\sqrt{\pi}} \) & \( \sqrt{\frac{15}{2} - \frac{8836}{441\pi}} \) & 0.41948 \\
        \( (6,4) \) & \( \frac{64}{10395\sqrt{\pi}} \) & 750 & \( \frac{3200}{693\sqrt{\pi}} \) & \( \sqrt{\frac{15}{2} - \frac{10240000}{480249\pi}} \) & 0.32409 \\
        \( (6,6) \) & \( \frac{128}{135135\sqrt{\pi}} \) & 5040 & \( \frac{2048}{429\sqrt{\pi}} \) & \( \sqrt{\frac{15}{2} - \frac{4194304}{184041\pi}} \) & 0.18404 \\
    \end{tabular}
    \end{ruledtabular}
    \caption{$\langle \hat{r} \rangle$, $\Delta \hat{r}$, and $\sigma_r$ for different states}
\end{table}
\begin{figure}[h!]
    \centering
    \includegraphics[width=0.5\textwidth]{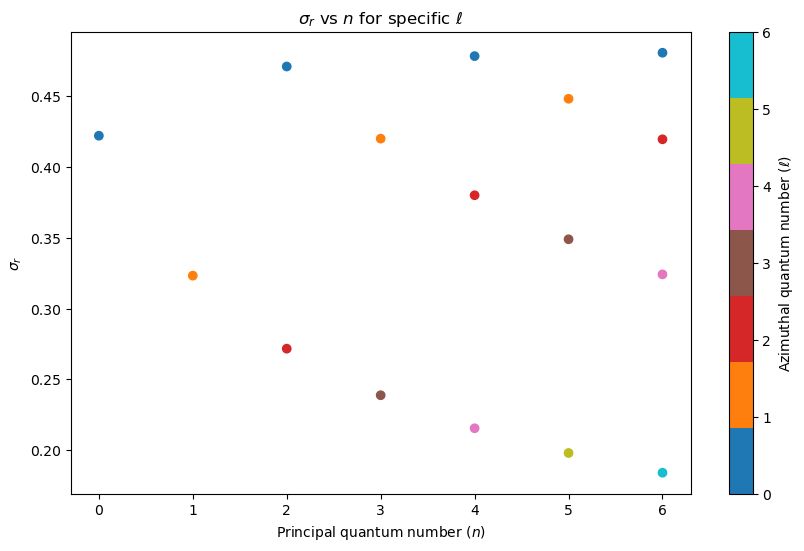}
    \caption{$\sigma_r$ vs $n$ for a specific $\ell$ of SHO}
\end{figure}
\begin{figure}[h!]
    \centering
    \includegraphics[width=0.5\textwidth]{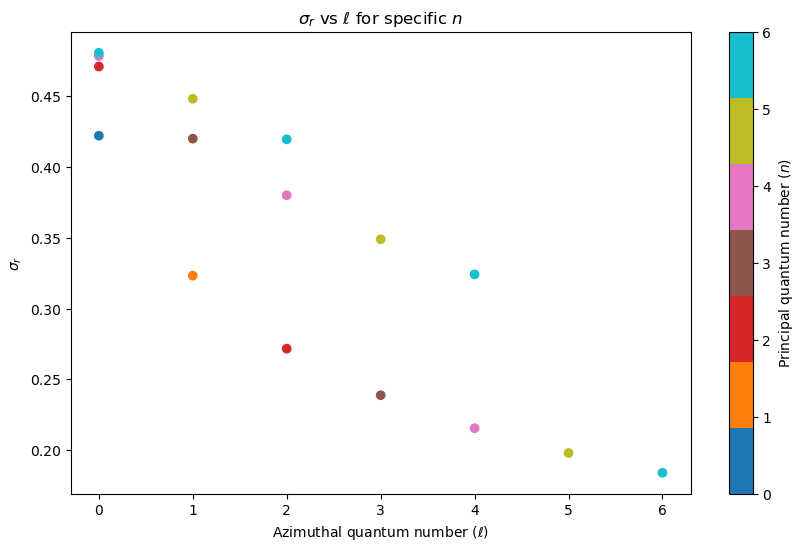}
    \caption{$\sigma_r$ vs $\ell$ for a specific $n$ of SHO}
\end{figure}
\begin{figure}[h!]
    \centering
    \includegraphics[width=0.5\textwidth]{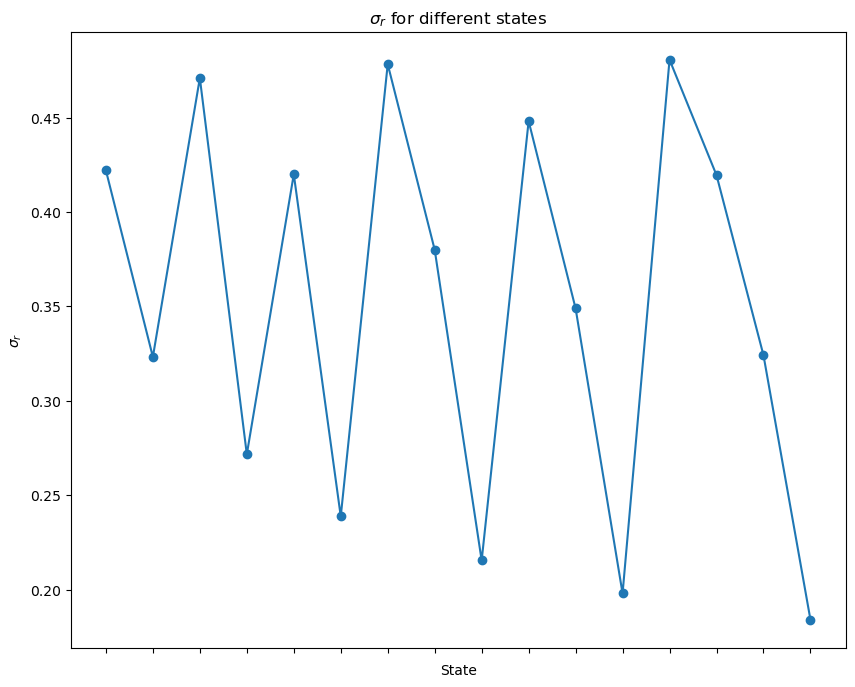}
    \caption{$\sigma_r$ for different states of SHO}
\end{figure}

\subsection{Uncertainty in Radial momentum}
We see that the radial momentum operator is given by,
\begin{equation*}
    \hat{p}_{r}=-i\hbar\frac{1}{r}\frac{\partial}{\partial{r}}[r]=-i\hbar\left(\frac{\partial}{\partial{r}}+\frac{1}{r}\right)
\end{equation*}
As per equ. (7),
\begin{equation}
\begin{alignedat}{2}
    &\langle\hat{p_{r}}\rangle\\
    &=(-i\hbar)\left[\int_0^{\infty} r^2R_{n\ell}^*(r)dR_{n\ell}(r)+\int_0^{\infty} r|R_{n\ell}(r)|^2dr\right]\\
    &=(-i\hbar)[I_2+I_3]\\
\end{alignedat}
\end{equation}
where
\begin{equation}
    \widetilde{I}_2=\frac{1}{\alpha}\int_0^{\infty}\eta R_{n\ell}^*(\eta)dR_{n\ell}(\eta) \text{ where $\eta=\alpha r^2$}
\end{equation}
We see the radial wave function of equ. (71) is given by 
\begin{equation*}
    R_{n\ell}(\eta)=N_{n\ell}{\eta}^{\frac{\ell}{2}}e^{-\frac{\eta}{2}}L_{\frac{1}{2}(n-\ell)}^{\ell+\frac{1}{2}}(\eta)
\end{equation*}
which gives 
\begin{equation}
\begin{alignedat}{2}
     &dR_{n\ell}(\eta)=N_{n\ell}\left[\frac{\ell}{2}{\eta}^{\frac{\ell}{2}-1}e^{-\frac{\eta}{2}}L_{\frac{1}{2}(n-\ell)}^{\ell+\frac{1}{2}}(\eta)\right.\\
     &\left. -\frac{1}{2}{\eta}^{\frac{\ell}{2}}e^{-\frac{\eta}{2}}L_{\frac{1}{2}(n-\ell)}^{\ell+\frac{1}{2}}(\eta)+{\eta}^{\frac{\ell}{2}}e^{-\frac{\eta}{2}}\frac{d}{d\eta}L_{\frac{1}{2}(n-\ell)}^{\ell+\frac{1}{2}}(\eta)\right]d\eta
\end{alignedat}
\end{equation}
Now using the derivative property of associated Laguerre function of equ. (15) and our nomenclature,
\begin{equation}
\begin{alignedat}{2}
    & dR_{n\ell}(\eta)=N_{n\ell}\left[\left(\frac{\ell}{2}+b\right){\eta}^{\frac{\ell}{2}-1}e^{-\frac{\eta}{2}}L_b^a(\eta)-\frac{1}{2}{\eta}^{\frac{\ell}{2}}e^{-\frac{\eta}{2}}L_b^a(\eta)\right.\\
    &\left. -(b+a){\eta}^{\frac{\ell}{2}-1}e^{-\frac{\eta}{2}}L_{b-1}^a(\eta)\right]d\eta\\ 
\end{alignedat}
\end{equation}
\begin{equation}
\begin{alignedat}{2}
    &\widetilde{I}_2=\frac{N_{n\ell}^2}{\alpha}\left[\left(\frac{\ell}{2}+b\right)\int_0^{\infty}{\eta}^{\ell}e^{-\eta}[L_b^a(\eta)]^2d\eta\right.\\
    &-\frac{1}{2}\int_0^{\infty}{\eta}^{\ell+1}e^{-\eta}[L_b^a(\eta)]^2d\eta\\
    &\left. -(b+a)\int_0^{\infty}{\eta}^{\ell}e^{-\eta}L_b^a(\eta)L_{b-1}^a(\eta)d\eta\right]\\  
\end{alignedat}
\end{equation}
Here we have the expanded integral form of the integral $\widetilde{I}_2$ and the another one is
\begin{equation}
   \widetilde{I}_3=\frac{1}{2\alpha}\int_0^{\infty}|R_{n\ell}(\eta)|^2d\eta=\frac{N_{n\ell}^2}{2\alpha}\int_0^{\infty}{\eta}^{\ell}e^{-\eta}[L_b^a(\eta)]^2d\eta
\end{equation}
Hence the expectation value of radial momentum in terms of some integrals is given by
\begin{equation}
\begin{alignedat}{2}
    &\langle \hat p_r \rangle\\
    &=(-i\hbar)\frac{N_{n\ell}^2}{\alpha}\left[\left(\frac{\ell}{2}+b+\frac{1}{2}\right)\int_0^{\infty}{\eta}^{\ell}e^{-\eta}[L_b^a(\eta)]^2d\eta \right.\\
    &-\frac{1}{2}\int_0^{\infty}{\eta}^{\ell+1}e^{-\eta}[L_b^a(\eta)]^2d\eta\\
    &\left. -(b+a)\int_0^{\infty}{\eta}^{\ell}e^{-\eta}L_b^a(\eta)L_{b-1}^a(\eta)d\eta \right]\\
    &=(-i\hbar)\frac{N_{n\ell}^2}{\alpha}[\widetilde{I}_4+\widetilde{I}_5+\widetilde{I}_6]
\end{alignedat}
\end{equation}
For a set of $(n,\ell)$, we can't do the generalized evaluation of these three integrals, but for a particular value of $(n,\ell)$, we have tabulated the results and add them in Table XI. And always their addition results zero. Hence, the expectation value of the radial momentum is zero as expected.
\begin{table}[htb]
    \centering
    \begin{ruledtabular}
    \begin{tabular}{lcccc}
        \textrm{($n,\ell$)} & 
        \textrm{$\widetilde{I}_4$} & 
        \textrm{$\widetilde{I}_5$} & 
        \textrm{$\widetilde{I}_6$} & 
        \textrm{$\widetilde{I}_4 + \widetilde{I}_5 + \widetilde{I}_6$} \\
        \colrule
        $(0,0)$ & $\frac{1}{2}$ & $-\frac{1}{2}$ & 0 & 0 \\
        \colrule
        $(1,1)$ & 1 & -1 & 0 & 0 \\
        \colrule
        $(2,0)$ & $\frac{15}{8}$ & $-\frac{9}{8}$ & $-\frac{6}{8}$ & 0 \\
        $(2,2)$ & 3 & $-3$ & 0 & 0 \\
        \colrule
        $(3,1)$ & $\frac{18}{4}$ & $-\frac{13}{4}$ & $-\frac{5}{4}$ & 0 \\
        $(3,3)$ & 12 & $-12$ & 0 & 0 \\
        \colrule
        $(4,0)$ & $\frac{445}{128}$ & $-\frac{225}{128}$ & $-\frac{220}{128}$ & 0 \\
        $(4,2)$ & $\frac{65}{4}$ & $-\frac{51}{4}$ & $-\frac{14}{4}$ & 0 \\
        $(4,4)$ & 60 & $-60$ & 0 & 0 \\
        \colrule
        $(5,1)$ & $\frac{699}{64}$ & $-\frac{433}{64}$ & $-\frac{266}{64}$ & 0 \\
        $(5,3)$ & $\frac{153}{2}$ & $-\frac{126}{2}$ & $-\frac{27}{2}$ & 0 \\
        $(5,5)$ & 360 & $-360$ & 0 & 0 \\
        \colrule
        $(6,0)$ & $\frac{2667}{512}$ & $-\frac{1225}{512}$ & $-\frac{1442}{512}$ & 0 \\
        $(6,2)$ & $\frac{3087}{64}$ & $-\frac{2115}{64}$ & $-\frac{972}{64}$ & 0 \\
        $(6,4)$ & 441 & $-375$ & $-66$ & 0 \\
        $(6,6)$ & 1520 & $-2520$ & 0 & 0 \\
    \end{tabular}
    \end{ruledtabular}
    \caption{$\langle \hat  p_r \rangle$ for different states}
\end{table}
Now our aim is to evaluate $\langle\hat p_r^2\rangle$, for that we can use energy calculation. The total energy of spherical harmonic oscillator, $E_{n\ell m}=E_{n\ell}=(n+\frac{3}{2})\hbar \omega$ and the total momentum is defined as $p_{n\ell m}=\sqrt{p_r^2+p_{\ell}^2}$ where $p_r$ is radial momentum and $p_\ell$ is orbital momentum is given by $p_{\ell}=\frac{\sqrt{\ell(\ell+1)}}{r}\hbar$. The momentum-kinetic energy relation $\langle E_{n\ell m}=2T=2\frac{p_{n\ell m}^2}{2m}=\frac{p_{n\ell m}^2}{m}\rangle$ as $\langle T=V\rangle$ gives $\langle(n+\frac{3}{2})m\hbar \omega=p_r^2+\frac{\ell(\ell+1}{r^2}{\hbar}^2\rangle$. Hence, 
\begin{equation}
    <p_r^2>=(n+\frac{3}{2})m\hbar \omega-\ell(\ell+1){\hbar}^2\langle\frac{1}{r^2}\rangle
\end{equation}
Now, 
\begin{equation}
\begin{alignedat}{2}
    &\langle\frac{1}{r^2}\rangle\\
    &=\frac{1}{2\sqrt{\alpha}}\int_0^{\infty}\eta^{-\frac{1}{2}}|R_{n\ell}(\eta)|^2d\eta\\
    &=\frac{1}{2\sqrt{\alpha}}\frac{2{\alpha}^{\frac{3}{2}}(\frac{1}{2}(n-\ell))!}{\Gamma(\frac{1}{2}(n+\ell)+\frac{3}{2})}\int_0^{\infty}{\eta}^{\ell-\frac{1}{2}}e^{-\eta}\left[L_{\frac{1}{2}(n-\ell)}^{\ell+\frac{1}{2}}(\eta)\right]^2d\eta\\
    &=\frac{m\omega}{\hbar}\widetilde{C}_{n\ell}\widetilde{I}_7\\
\end{alignedat}
\end{equation}
And finally, 
\begin{equation}
    \langle\hat p_r^2\rangle=m\hbar \omega \left[n+\frac{3}{2}-\ell(\ell+1)\widetilde{C}_{n\ell}\widetilde{I}_7\right]
\end{equation}
The uncertainty in radial momentum will be,
\begin{equation}
   \Delta p_r=\sqrt{m\hbar\omega}\sqrt{\left[n+\frac{3}{2}-\ell(\ell+1)\widetilde{C}_{n\ell}\widetilde{I}_7\right]}
\end{equation}
where
\begin{equation*}
    \widetilde{C}_{n\ell}=\frac{(\frac{1}{2}(n-\ell))!}{\Gamma(\frac{1}{2}(n+\ell)+\frac{3}{2})}
\end{equation*}
and
\begin{equation*}
    \widetilde{I}_7=\int_0^{\infty}{\eta}^{\ell-\frac{1}{2}}e^{-\eta}\left[L_{\frac{1}{2}(n-\ell)}^{\ell+\frac{1}{2}}(\eta)\right]^2d\eta
\end{equation*}
Fig. (41) and (42) are about $\Delta p_r$ vs $n$ for a specific $\ell$ and $\Delta p_r$ vs $\ell$ for a specific $n$ and Fig. (43) is for different states.
\begin{figure}[h!]
    \centering
    \includegraphics[width=0.5\textwidth]{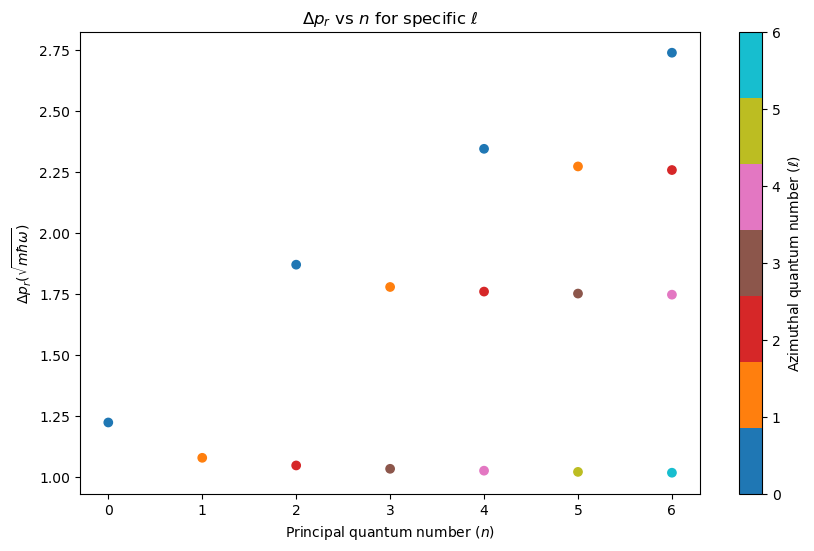}
    \caption{$\Delta p_r$ vs $n$ for a specific $\ell$ of SHO}
\end{figure}
\begin{figure}[h!]
    \centering
    \includegraphics[width=0.5\textwidth]{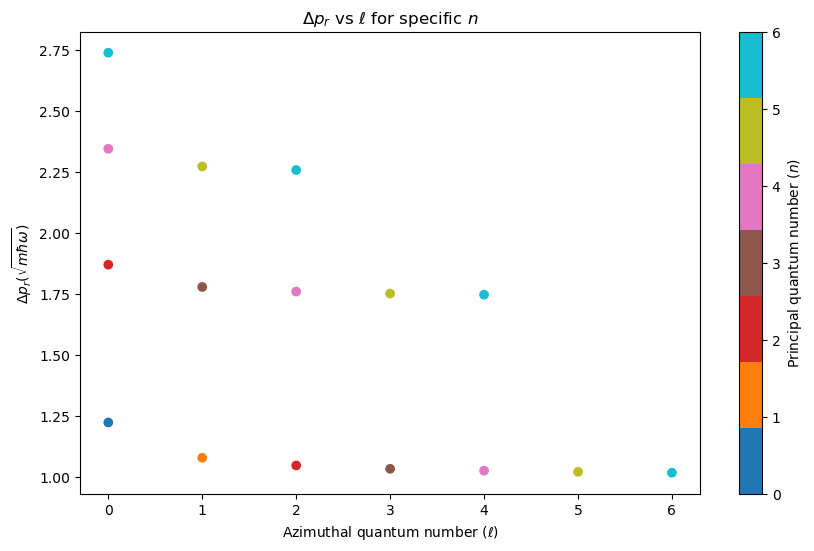}
    \caption{$\Delta p_r$ vs $\ell$ for a specific $n$ of SHO}
\end{figure}
\begin{figure}[h!]
    \centering
    \includegraphics[width=0.5\textwidth]{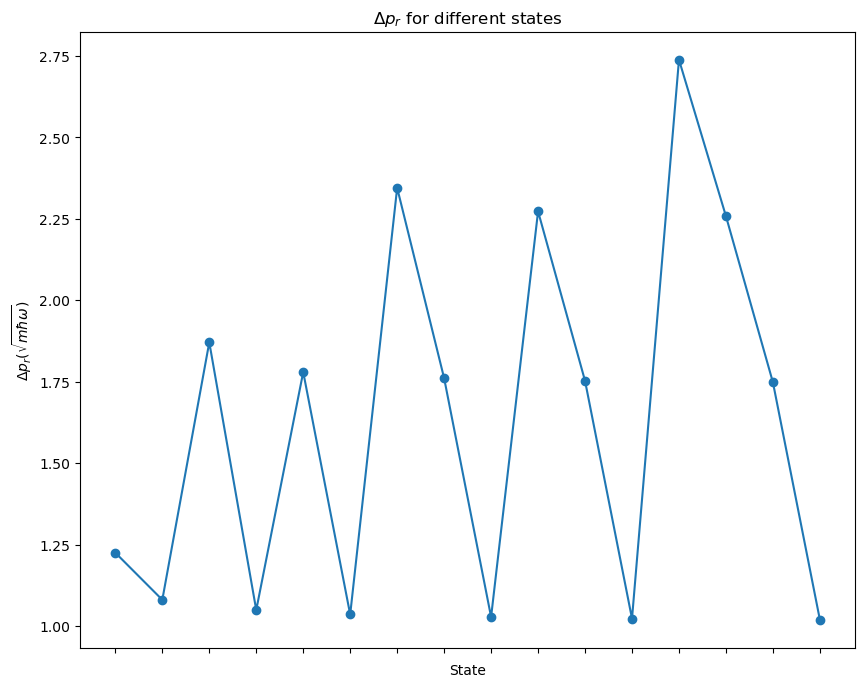}
    \caption{$\Delta p_r$ for different states of SHO}
\end{figure}
\subsection{Radial uncertainty product}
Hence the radial uncertainty product becomes,
\begin{equation}
   \Delta r\Delta p_r=\hbar\sqrt{n+\frac{3}{2}-\widetilde{C}_{n\ell}^2\widetilde{I}_1^2}\sqrt{n+\frac{3}{2}-\ell(\ell+1)\widetilde{C}_{n\ell}\widetilde{I}_7}
\end{equation}
Fig. (44) and (45) are about $\Delta r\Delta p_r$ vs $n$ for a specific $\ell$ and $\Delta r\Delta p_r$ vs $\ell$ for a specific $n$ and Fig. (46) is for different states.
\begin{figure}[h!]
    \centering
    \includegraphics[width=0.5\textwidth]{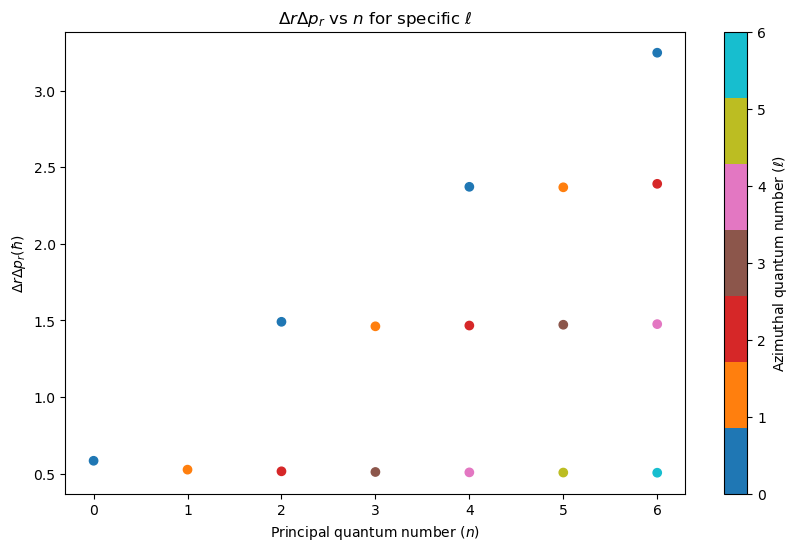}
    \caption{$\Delta r\Delta p_r$ vs $n$ for a specific $\ell$ of SHO}
\end{figure}
\begin{figure}[h!]
    \centering
    \includegraphics[width=0.5\textwidth]{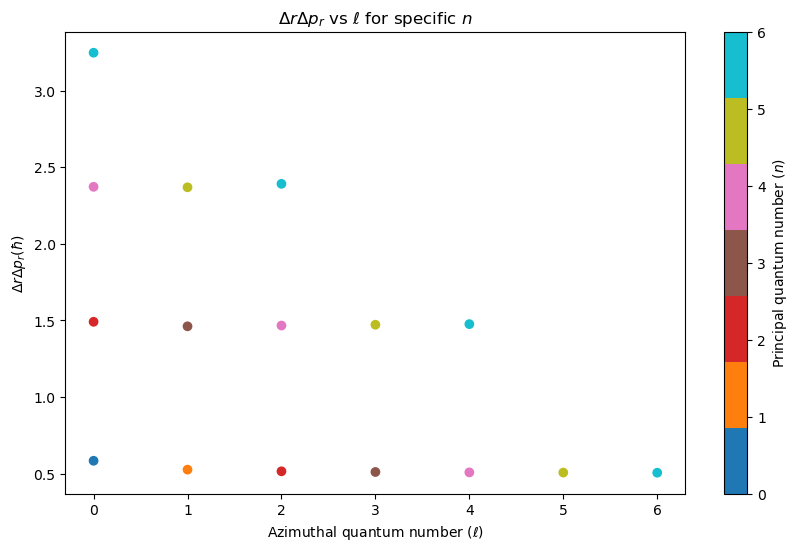}
    \caption{$\Delta r\Delta p_r$ vs $\ell$ for a specific $n$ of SHO}
\end{figure}
\begin{figure}[h!]
    \centering
    \includegraphics[width=0.5\textwidth]{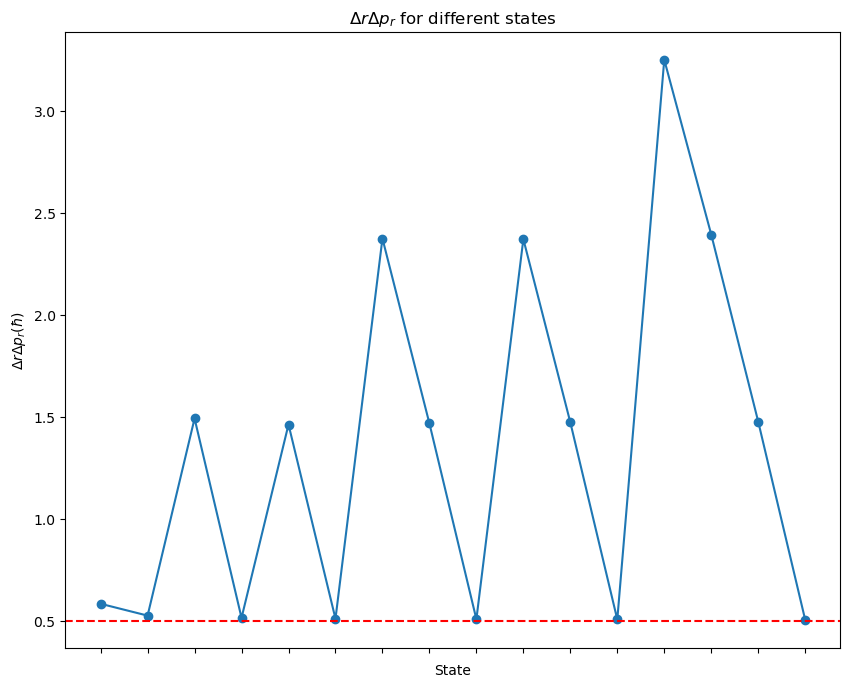}
    \caption{$\Delta r\Delta p_r$ for different states of SHO}
\end{figure}
\begin{table}[htb]
    \centering
    \begin{ruledtabular}
    \begin{tabular}{ccccccc}
        \textrm{($n,\ell$)} & 
        \textrm{$\widetilde{C}_{n\ell}$} & 
        \textrm{$\widetilde{I}_1$} & 
        \textrm{$\widetilde{I}_7$} & 
        \textrm{$\Delta r$ \footnote{in unit of $\sqrt{\hbar / m \omega}$}} & 
        \textrm{$\Delta p_r$\footnote{in unit of $\sqrt{m\hbar\omega}$}} & 
        \textrm{$\Delta r \Delta p_r$\footnote{in unit of $\hbar$}} \\
        \colrule
        $(0,0)$ & $\frac{2}{\sqrt{\pi}}$ & 1 & $\sqrt{\pi}$ & $\sqrt{\frac{3}{2}-\frac{4}{\pi}}$ & $\sqrt{\frac{3}{2}}$ & 0.58321 \\
        \colrule
        $(1,1)$ & $\frac{4}{3\sqrt{\pi}}$ & 2 & $\frac{\sqrt{\pi}}{2}$ & $\sqrt{\frac{5}{2}-\frac{64}{9\pi}}$ & $\sqrt{\frac{7}{6}}$ & 0.52523 \\
        \colrule
        $(2,0)$ & $\frac{4}{3\sqrt{\pi}}$ & $\frac{9}{4}$ & $\frac{3\sqrt{\pi}}{2}$ & $\sqrt{\frac{7}{2}-\frac{9}{\pi}}$ & $\sqrt{\frac{7}{2}}$ & 1.49105 \\
        $(2,2)$ & $\frac{8}{15\sqrt{\pi}}$ & 6 & $\frac{3\sqrt{\pi}}{4}$ & $\sqrt{\frac{7}{2}-\frac{256}{25\pi}}$ & $\sqrt{\frac{11}{10}}$ & 0.51435 \\
        \colrule
        $(3,1)$ & $\frac{8}{15\sqrt{\pi}}$ & $\frac{13}{2}$ & $\frac{5\sqrt{\pi}}{4}$ & $\sqrt{\frac{9}{2}-\frac{2704}{225\pi}}$ & $\sqrt{\frac{19}{6}}$ & 1.46161 \\
        $(3,3)$ & $\frac{16}{105\sqrt{\pi}}$ & 24 & $\frac{15\sqrt{\pi}}{8}$ & $\sqrt{\frac{9}{2}-\frac{16384}{1225\pi}}$ & $\sqrt{\frac{15}{14}}$ & 0.50994 \\
        \colrule
        $(4,0)$ & $\frac{16}{15\sqrt{\pi}}$ & $\frac{225}{64}$ & $\frac{15\sqrt{\pi}}{8}$ & $\sqrt{\frac{11}{2}-\frac{225}{16\pi}}$ & $\sqrt{\frac{11}{2}}$ & 2.37291 \\
        $(4,2)$ & $\frac{16}{105\sqrt{\pi}}$ & $\frac{51}{2}$ & $\frac{21\sqrt{\pi}}{8}$ & $\sqrt{\frac{11}{2}-\frac{18496}{1225\pi}}$ & $\sqrt{\frac{31}{10}}$ & 1.46667 \\
        $(4,4)$ & $\frac{32}{945\sqrt{\pi}}$ & 120 & $\frac{105\sqrt{\pi}}{16}$ & $\sqrt{\frac{11}{2}-\frac{65538}{3969\pi}}$ & $\sqrt{\frac{19}{18}}$ & 0.50758 \\
        \colrule
        $(5,1)$ & $\frac{32}{105\sqrt{\pi}}$ & $\frac{433}{32}$ & $\frac{35\sqrt{\pi}}{16}$ & $\sqrt{\frac{13}{2}-\frac{187489}{11025\pi}}$ & $\sqrt{\frac{31}{6}}$ & 2.36972 \\
        $(5,3)$ & $\frac{32}{945\sqrt{\pi}}$ & 126 & $\frac{135\sqrt{\pi}}{16}$ & $\sqrt{\frac{13}{2}-\frac{4096}{225\pi}}$ & $\sqrt{\frac{43}{14}}$ & 1.47188 \\
        $(5,5)$ & $\frac{64}{10395\sqrt{\pi}}$ & 720 & $\frac{945\sqrt{\pi}}{32}$ & $\sqrt{\frac{13}{2}-\frac{1048576}{53362\pi}}$ & $\sqrt{\frac{23}{22}}$ & 0.50612 \\
        \colrule
        $(6,0)$ & $\frac{32}{35\sqrt{\pi}}$ & $\frac{1225}{256}$ & $\frac{35\sqrt{\pi}}{16}$ & $\sqrt{\frac{15}{2}-\frac{1225}{64\pi}}$ & $\sqrt{\frac{15}{2}}$ & 3.24886 \\
        $(4,2)$ & $\frac{64}{9455\sqrt{\pi}}$ & $\frac{2115}{32}$ & $\frac{189\sqrt{\pi}}{32}$ & $\sqrt{\frac{15}{2}-\frac{8836}{441\pi}}$ & $\sqrt{\frac{51}{10}}$ & 2.39238 \\
        $(6,4)$ & $\frac{64}{10395\sqrt{\pi}}$ & 750 & $\frac{1155\sqrt{\pi}}{32}$ & $\sqrt{\frac{15}{2}-\frac{10240000}{480249\pi}}$ & $\sqrt{\frac{55}{18}}$ & 1.47592 \\
        $(6,6)$ & $\frac{128}{135135\sqrt{\pi}}$ & 5040 & $\frac{10395\sqrt{\pi}}{64}$ & $\sqrt{\frac{15}{2}-\frac{4194304}{184041\pi}}$ & $\sqrt{\frac{27}{26}}$ & 0.50512 \\
    \end{tabular}
    \end{ruledtabular}
    \caption{$\Delta r$, $\Delta p_r$ and  $\Delta r \Delta p_r$ for different states of SHO}
\end{table}
\subsection{Ground state of Spherical harmonic oscillator}
The radial wave function of Spherical harmonic oscillator is given by,
\begin{equation*}
    R_{n\ell}(r)=N_{n\ell}(\sqrt{\alpha}r)^{\ell}e^{-\frac{\alpha r^2}{2}}L_{\frac{1}{2}(n-\ell)}^{\ell+\frac{1}{2}}(\alpha r^2)
\end{equation*}
where $N_{n\ell}=\frac{2{\alpha}^{\frac{3}{2}}(\frac{1}{2}(n-\ell))!}{\Gamma(\frac{1}{2}(n+\ell)+\frac{3}{2})}$ and $\alpha=\frac{m\omega}{\hbar}$. For ground state ($n=0,\ell=0$), $N_{00}=\frac{2{\alpha}^{\frac{3}{2}}}{\sqrt{\pi}}$, $R_{00}(r)=N_{00}e^{-\frac{\alpha r^2}{2}}L_0^{\frac{1}{2}}(\alpha r^2)=N_{00}e^{-\frac{\alpha r^2}{2}}$ as $L_0^{\frac{1}{2}}(x)=1$. The radial probability density for ground state, $P_{00}(r)=r^2|R_{00}(r)|^2=N_{00}^2r^2e^{-\alpha r^2}$. For most probable radius, $\frac{dP}{dr}=0$ gives $r=-\frac{1}{\sqrt{\alpha}}$ (non-acceptable) and $r_{mp}=\frac{1}{\sqrt{\alpha}}=\sqrt{\frac{\hbar}{m\omega}}$. And we have, $\langle r\rangle=\frac{2}{\sqrt{\pi}}\sqrt{\frac{\hbar}{m\omega}}=1.1283\sqrt{\frac{\hbar}{m\omega}}$ and $\Delta r=\sqrt{\frac{3}{2}-\frac{4}{\pi}}\sqrt{\frac{\hbar}{m\omega}}=0.4762\sqrt{\frac{\hbar}{m\omega}}$.
\begin{figure}[h!]
    \centering
    \includegraphics[width=0.5\textwidth]{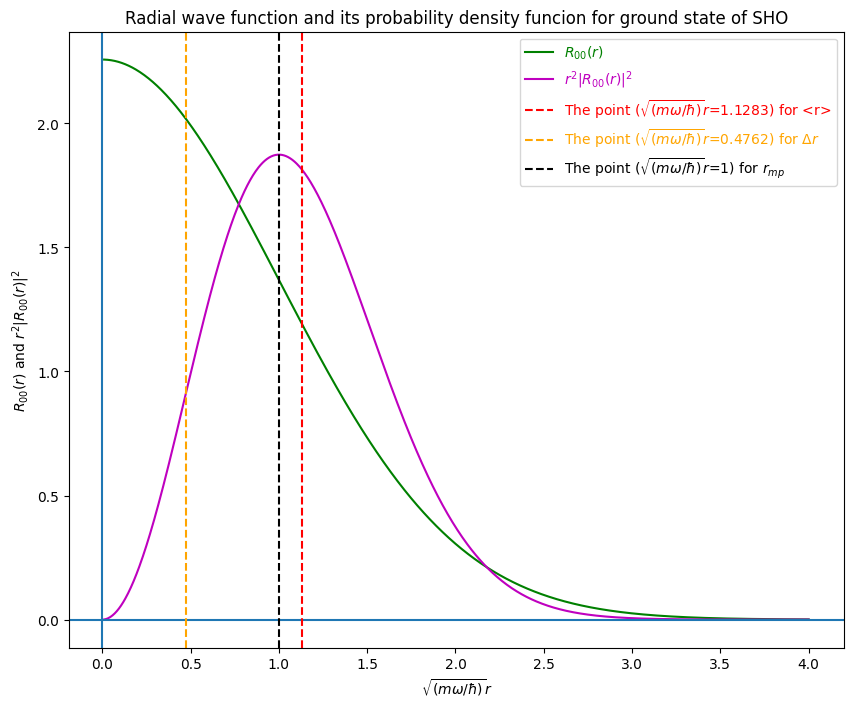}
    \caption{Ground state of SHO}
\end{figure}

\begin{acknowledgments}
I express my heartfelt gratitude to the esteemed Indian Institute of Technology (IIT), Delhi, for providing me with invaluable access to various journal platforms. I extend my sincere appreciation to my esteemed classmate, Molla Suman Rahaman for his unwavering encouragement and support throughout the course of my work. In the realm of computational tools, I am thankful for the indispensable roles played by Wolfram Mathematica and Jupyter Notebooke and Overleaf.
\end{acknowledgments}

\appendix
\section{Parity, degeneracy for SHO}
The degree of degeneracy of H-atom i.e. the total number of possible energy eigen states
$=\Sigma_{\ell=0}^{n-1}(2\ell+1)=n^2$. The electrons have two different type spins, so overall degeneracy $=2n^2$. Meanwhile for infinite spherical well, there are $2\ell +1$ states with different $z$ components of angular momentum for the spherical coordinate states and the parity is $(-1)^\ell$ for the angular momentum states. In case for spherical harmonic oscillator, the number of states at each energy matches exactly and also the parities of the states match. These are tabulated in Table XIII. If we were more calculative, we could verify that the wave functions in spherical coordinates are just linear combinations of the solutions in Cartesian coordinates.
\begin{table}
    \centering
    \begin{ruledtabular}
    \begin{tabular}{ccccccccc}
        \textrm{$n$} & 
        \textrm{$(n_r\ell)$} & 
        \textrm{$d_s$ \footnote{$d_s=\sum (2\ell+1)$, s for spherical}} & 
        \textrm{$p_s$ \footnote{$p_s=(-1)^{\ell}$}} & 
        \textrm{$E_s$ \footnote{$E_s=\left(n+\frac{3}{2}\right)$ in unit of $\hbar\omega$}} & 
        \textrm{$(n_xn_yn_z)$} & 
        \textrm{$d_c$ \footnote{$d_c=\frac{1}{2}(n_x+n_y+n_z)(n_x+n_y+n_z+1)$, c for cartesian}} & 
        \textrm{$p_c$ \footnote{$p_c=(-1)^{n_x+n_y+n_z}$}} & 
        \textrm{$E_c$ \footnote{$E_s=\left(n_x+n_y+n_z+\frac{3}{2}\right)$ in unit of $\hbar\omega$}} \\
        \colrule
        0 & (00) & 1 & + & $\frac{3}{2}$ & (000) & 1 & + & $\frac{3}{2}$ \\
        \colrule
        1 & (01) & 3 & - & $\frac{5}{2}$ & (001,100,010) & 3 & - & $\frac{5}{2}$ \\
        \colrule
        2 & (10) & 6 & + & $\frac{7}{2}$ & (002,200,020) & 6 & + & $\frac{7}{2}$ \\
          & (02) &   &   &               & (011,110,101) &   &   &               \\
        \colrule
        3 & (11) & 10 & - & $\frac{9}{2}$ & (003,300,030) & 10 & - & $\frac{9}{2}$ \\
          & (03) &    &   &               & (210,120,012, &    &   &               \\
          &      &    &   &               & 021,102,201) &    &   &               \\
          &      &    &   &               & (111)         &    &   &               \\  
        \colrule
        4 & (20) & 15 & + & $\frac{11}{2}$ & (004,400,040) & 15 & + & $\frac{11}{2}$ \\
          & (12) &    &   &                & (310,130,301, &    &   &                \\
          & (04) &    &   &                & 103,031,013) &    &   &                \\
          &      &    &   &                & (220,022,202)&    &   &                \\
          &      &    &   &                & (211,121,112) &    &   &                \\
        \colrule
        5 & (21) & 21 & - & $\frac{13}{2}$ & (005,500,050) & 21 & - & $\frac{13}{2}$ \\
          & (13) &    &   &                & (410,140,041,  &    &   &               \\
          & (05) &    &   &                & 014,104,401)  &    &   &               \\
          &      &    &   &                & (320,230,032,  &    &   &               \\
          &      &    &   &                & 023,302,203)  &    &   &               \\
          &      &    &   &                & (311,113,131) &    &   &               \\
          &      &    &   &                & (122,221,121)  &    &   &               \\ 
        \colrule
        6 & (30) & 28 & + & $\frac{15}{2}$ & (006,600,060) & 28 & + & $\frac{15}{2}$ \\
          & (22) &    &   &                & (510,150,015,  &    &   &                \\
          & (14) &    &   &                & 051,501,105)   &    &   &                \\
          & (06) &    &   &                & (420,240,042,  &    &   &                \\
          &      &    &   &                & 024,204,402)  &    &   &                \\
          &      &    &   &                & (330,033,303) &    &   &                \\
          &      &    &   &                & (411,114,141) &    &   &                \\
          &      &    &   &                & (312,321,123,  &    &   &                \\
          &      &    &   &                & 132,231,213)  &    &   &                \\
          &      &    &   &                & (222)          &    &   &                \\
    \end{tabular}
    \end{ruledtabular}
    \caption{Energy (E), parity (p), and degeneracy (d) for SHO}
\end{table}
\section{Spherical polar coordinate}
The coordinates convention is followed by $(r,\theta, \phi)$=(radial distance, polar/zenith angle, azimuthal angle) where $r\in [0,\infty)$ $\theta \in [0,\pi]$ and $\phi \in $. The radius vector is $(x,y,z)=(r\sin\theta\cos\phi,r\sin\theta\sin\phi,r\cos\theta$).
\begin{figure}[h!]
    \centering
    \includegraphics[width=0.5\textwidth]{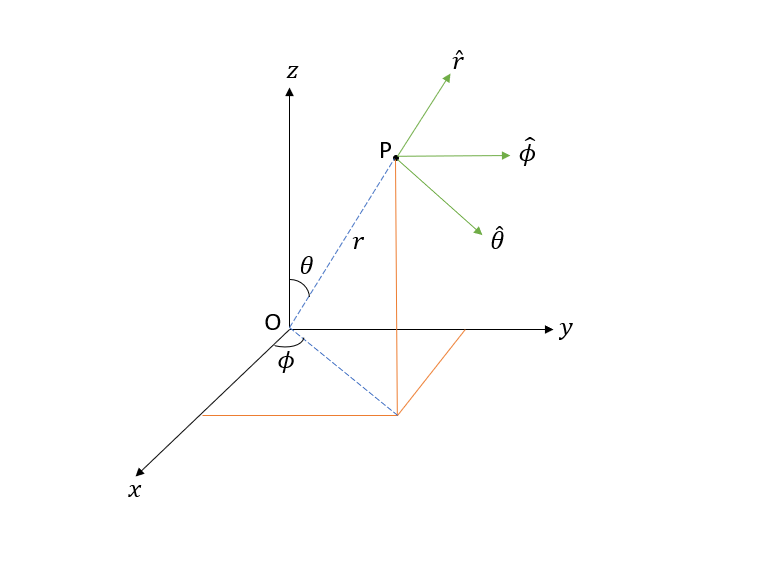}
    \caption{Spherical polar coordinate}
\end{figure}



\nocite{*}
\begin{thebibliography}{99}

\bibitem{khelashvili2022}
Anzor Khelashvili and Teimuraz Nadareishvili,
\emph{Generalized Heisenberg uncertainty relation in spherical coordinates},
International Journal of Modern Physics B, 36(15), 2250072 (2022),
doi:10.1142/S0217979222500722. \url{https://doi.org/10.1142/S0217979222500722}

\bibitem{bracher2011}
Christian Bracher,
\emph{Uncertainty relations for angular momentum eigenstates in two and three spatial dimensions},
American Journal of Physics, 79(3), 313--319 (2011),
doi:10.1119/1.3534840. \url{https://doi.org/10.1119/1.3534840}

\bibitem{aljaber2016}
Sami M. Al-Jaber,
\emph{Uncertainty Relations for Some Central Potentials in N-Dimensional Space},
Applied Mathematics, 7(6), March 2016,
doi:10.4236/am.2016.76047. \url{https://doi.org/10.4236/am.2016.76047}

\bibitem{Dehesa_2021}
J. S. Dehesa,
\emph{Spherical-Symmetry and Spin Effects on the Uncertainty Measures of Multidimensional Quantum Systems with Central Potentials},
Entropy, 23(5), 607 (2021),
doi:10.3390/e23050607. \url{https://doi.org/10.3390/e23050607}

\bibitem{Li_2020}
Jun‐Li Li and Cong‐Feng Qiao,
\emph{The Generalized Uncertainty Principle},
Annalen der Physik, 533(1), November 2020,
doi:10.1002/andp.202000335. \url{http://dx.doi.org/10.1002/andp.202000335}

\bibitem{AlJaber_1998}
Sami M. Al-Jaber,
\emph{Hydrogen Atom in N Dimensions},
International Journal of Theoretical Physics, 37, 1289--1298 (1998),
doi:10.1023/A:1026679921970. \url{https://doi.org/10.1023/A:1026679921970}

\bibitem{Paz_2001}
Gil Paz,
\emph{On the connection between the radial momentum operator and the Hamiltonian in n dimensions},
European Journal of Physics, 22(4), 337 (2001),
doi:10.1088/0143-0807/22/4/308. \url{https://arxiv.org/abs/quant-ph/0009046}

\bibitem{Supriadi_2019}
B. Supriadi, A. Harijanto, M. Maulana, Z. R. Ridlo, W. D. Wisesa, and A. Nurdiniaya,
\emph{The function of the radial wave of a hydrogen atom in the principal quantum numbers (n) 4 and 5},
Journal of Physics: Conference Series, 1211, 012052 (2019),
doi:10.1088/1742-6596/1211/1/012052.

\bibitem{Bransden_Joachain_1989}
B. Bransden and C. Joachain,
\emph{Introduction to Quantum Mechanics},
Wiley, New York, 1989.

\bibitem{Abramowitz_Stegun_1972}
M. Abramowitz and I. A. Stegun,
\emph{Handbook of Mathematical Functions with Formulas, Graphs, and Mathematical Tables},
9th printing, Dover, New York, 1972. Chapter 22: Orthogonal Polynomials, 771--802.

\bibitem{Griffiths_2018}
David J. Griffiths and Darrell F. Schroeter,
\emph{Introduction to Quantum Mechanics},
3rd ed., Cambridge University Press, Cambridge, 2018.

\bibitem{Zettili_2013}
Nouredine Zettili,
\emph{Quantum Mechanics: Concepts and Applications},
2nd ed., Wiley, Hoboken, NJ, 2013.

\bibitem{Huang_2017}
Young-Sea Huang, Kung-Te Wu, and Chun-Hsien Wu,
\emph{The right way to solve the infinite spherical well in quantum mechanics},
ResearchGate, 2017, In Progress. \url{https://doi.org/10.13140/RG.2.2.18172.44162}

\bibitem{bloomfield2017indefiniteintegralssphericalbessel}
Jolyon K. Bloomfield, Stephen H. P. Face, and Zander Moss,
\emph{Indefinite Integrals of Spherical Bessel Functions},
arXiv:1703.06428 [math.CA], 2017. \url{https://arxiv.org/abs/1703.06428}

\bibitem{mathematica_integration}
Wolfram Research,
\emph{Integration Solution in Mathematica}, 2025. \url{https://www.wolfram.com/mathematica/}

\bibitem{wolfram_regularized_hypergeometric}
Wolfram Research,
\emph{Regularized Hypergeometric Function}, 2025. \url{https://mathworld.wolfram.com/RegularizedHypergeometricFunction.html}

\bibitem{cahaya2022radial}
Adam Badra Cahaya,
\emph{Radial Wave Function of 2D and 3D Quantum Harmonic Oscillator},
AL-FIZIYA JOURNAL OF MATERIALS SCIENCE, GEOPHYSICS, INSTRUMENTATION AND THEORETICAL PHYSICS, 5(2), 2022. \url{https://example.com/your-article-link}

\bibitem{ucsd3dHO}
UC San Diego,
\emph{3D Symmetric Harmonic Oscillator in Spherical Coordinates}, n.d. \url{https://quantummechanics.ucsd.edu/ph130a/130_notes/node244.html}

\bibitem{Kuo2005}
Cheng-Deng Kuo,
\emph{The uncertainties in radial position and radial momentum of an electron in the non-relativistic hydrogen-like atom},
Annals of Physics, 316(2), 431--439 (2005),
doi:10.1016/j.aop.2004.09.005.

\bibitem{Patil2007}
S.H. Patil and K.D. Sen,
\emph{Uncertainty relations for modified isotropic harmonic oscillator and Coulomb potentials},
Physics Letters A, 362(2-3), 109--114 (2007),
doi:10.1016/j.physleta.2007.01.021.

\bibitem{Pauling1935}
L. Pauling and E.B. Wilson Jr.,
\emph{Introduction to Quantum Mechanics With Applications to Chemistry},
McGraw-Hill, New York, 1935.

\end{thebibliography}
\end{document}